\documentclass[11pt,a4paper]{article}

\usepackage{jheppub}

\usepackage{multirow, graphicx,amssymb,url,mathrsfs,amsmath}
\allowdisplaybreaks[4]
\usepackage{wrapfig,boxedminipage,setspace,subfigure,epsfig}
\usepackage{amsxtra,amstext,latexsym,dsfont,amsfonts}
\usepackage{color,eucal}
\usepackage[dvipsnames,table]{xcolor}
\usepackage{float}
\usepackage{tikz}
\usepackage{slashed,comment}
\usepackage{kotex}
\usepackage{tensor}
\usepackage{pifont}
    
\usepackage{makecell}          
\usepackage{booktabs}          
\usepackage{amsmath}           
\usepackage{tabularx}  
\usepackage{indentfirst}

\newtheorem{theorem}{Theorem}[section]

\newcommand{\eg}{{\it e.g.}}
\newcommand{\ie}{{\it i.e.}}

\title{Upper bound of holographic entanglement entropy combinations}

\author[a]{Xin-Xiang Ju,}
\author[a,b]{Ya-Wen Sun}
\author[a]{and Yang Zhao}

\emailAdd{juxinxiang21@mails.ucas.ac.cn}
\emailAdd{yawen.sun@ucas.ac.cn}
\emailAdd{zhaoyang20a@mails.ucas.ac.cn}

\affiliation[a]{School of Physical Sciences, University of Chinese Academy of Sciences, Zhongguancun east road 80, Beijing 100190, China}
\affiliation[b]{Kavli Institute for Theoretical Sciences, University of Chinese Academy of Sciences, Beijing 100049, China}

\abstract{In this work, we develop a systematic formalism to evaluate the upper bound of a large family of holographic entanglement entropy combinations when fixing \( n \) subsystems and fine-tuning one other subsystem. The upper bound configurations and values of these entropy combinations can be derived and classified. The upper bound of these entropy combinations reveals holographic $n+1$-partite entanglement that $n$ fixed subsystems participate in. In AdS\(_3\)/CFT\(_2\), AdS\(_4\)/CFT\(_3\), and even higher-dimensional holography, one can, in principle, find different formulas of upper bound values, reflecting the fundamental difference in entanglement structure in different dimensions.
}

\begin{document}
\maketitle


\section{Introduction} \label{sec1}

\noindent Quantum entanglement has played a significant role in holography \cite{Maldacena:1997re,Maldacena_2013,VanRaamsdonk:2010pw} since the proposal of the Ryu–Takayanagi formula \cite{Ryu_2006}, which evaluates the bipartite entanglement entropy of the boundary subsystem via the area of the bulk minimal surface. Partitioning the boundary system into multiple parties leads to the study of multipartite entanglement structure in holography, an intriguing area with many recent developments \cite{Iizuka:2025ioc,Iizuka:2025bcc,Yuan:2024yfg,Umemoto:2018jpc,Balasubramanian:2024ysu,Gadde:2023zzj,Penington:2022dhr,Akers:2019gcv,Ju:2023dzo,Basak:2024uwc}. 
Usually, people study multipartite entanglement structures in holography by first defining multipartite entanglement measures and applying them in holography, \eg\ those on multi‑entropy \cite{Gadde:2022cqi,Gadde:2023zzj}. However, the study of multipartite entanglement measures is a formidable area in quantum information theory \cite{Walter:2016lgl,Guo_2020,bengtsson2016brief}, since no calculable measure is perfect and no perfect measure is calculable \cite{Li_2018,Guo_2020}. To introduce these measures in holography, one must further accept that very few have a computable holographic correspondence.

Choosing not to face those problems directly, we do not introduce any fancy new measures in this work. Instead, we use computable but imperfect measures: linear combinations of holographic entanglement entropy of {multiple subsystems} \cite{Hernandez-Cuenca:2023iqh}. In general, we cannot quantify the amount of quantum entanglement from their values except when they reach their upper or lower bounds. For example, the mutual information \(I(A\!:\!B)\) is an imperfect bipartite entanglement measure when \(\rho_{AB}\) is mixed, since it may be nonzero for a separable \(\rho_{AB}\). However, when it reaches its upper bound \(2\min(S_A,S_B)\), we can conclude that \(A\) (\(B\)) contributes all its degrees of freedom to the entanglement with \(B\) (\(A\)).  {In \cite{Ju:2024hba,Ju:2024kuc}, we have evaluated the upper bound of holographic $n$ partite information and concluded that there exists a huge amount multipartite entanglement in holography when the upper bound is reached.} 

The core idea of this work is to evaluate the upper bounds of more general combinations of holographic entanglement entropy with $n$ regions fixed and one other region arbitrarily chosen. In this way, we are detecting the multipartite entanglement that these $n$ subregions participate in the $n+1$ body system. This reflects how a less-partite density matrix (sub-state) actually is embedded in the multipartite state.
Combining this goal with our chosen measures, the path of our study becomes clear: first, we fix several regions (less-partite sub-state); then we carefully embed them into a more partite state so that the entropy combination of the multipartite state reaches an extreme value. We may then assert that those fixed regions participate in multipartite entanglement with another fine‑tuned region, as measured by the entropy combination.

In this work, we develop a systematic procedure to evaluate the upper bound. This procedure involves two steps as follows: 
\begin{enumerate}
    \item  Determine the (dis)connectivity of each entanglement wedge in the upper bound configuration via a so-called lamp diagram. 
    \item Utilize a gap region classification method to calculate the exact upper bound for general configurations of $n$ fixed regions. 
\end{enumerate}

For general combinations of entanglement entropy, we could expect that they have their separate information theoretical upper bounds, which are the upper bounds when all density matrices of the subsystems could be any quantum system. Since holographic states form a subset of quantum states, the upper bounds of general combinations of holographic entanglement entropy may not reach this information theoretical upper bound, and we always have
\[
\textit{Holographic Upper Bound} \leq \textit{Information Theoretical Upper Bound}.
\]
When this inequality is saturated, the multipartite state is in a certain type of multipartite entanglement measured by the entropy combination, while distinctive features of holographic states compared to other random states are obscured; the upper bound of $n$-partite information which we investigated in \cite{Ju:2024hba,Ju:2024kuc} is basically this case. In the cases where the inequality is not saturated, three possible reasons for this non-saturation can be identified:
\begin{itemize}
    \item[I.] The intrinsic nature of holographic states.
    \item[II.] The specific geometrical configuration of the fixed regions.
    \item[III.] The dimensionality of the holographic theory.
\end{itemize}

\textbf{Item I.} It has been thoroughly explored in the holographic entropy cone program \cite{Bao:2015bfa,Hubeny:2018trv,Hubeny:2018ijt,He:2019ttu,HernandezCuenca:2019wgh,He:2020xuo,Avis:2021xnz,Fadel:2021urx,Bao:2024azn}, where authors discover and investigate numerous entropy inequalities that hold only in holography, thus revealing the special nature of holographic states beyond general quantum states. For example, the monogamy of mutual information \cite{Hayden_2013} (i.e., \(I_3 \leq 0\)) implies that the holographic upper bound of \(I_3\) is zero \cite{Ju:2023tvo}, even though quantum states with positive \(I_3\) (such as four-partite GHZ states) can exist. In this sense, our work shares the same motivation as the entropy cone program, although the fixed-region setup results in upper bounds that are expressed differently from conventional holographic entropy inequalities.

\textbf{Item II.} Non-saturation may also result from the geometrical configuration of the fixed regions. For instance, in AdS\(_4\)/CFT\(_3\) it has been observed \cite{Ju:2024kuc} that two convex spatial subregions, when far apart on the boundary, can have all their degrees of freedom contribute to tripartite entanglement with another region, whereas two concave regions cannot. Similarly, in the multi-mouth wormhole scenario considered here, if the fixed regions lie on different boundaries, the upper bound of the entropy combinations is generally not saturated. These observations point to intricate multipartite entanglement structures that depend on the shapes and locations of the regions involved.

\textbf{Item III.} Furthermore, the non-saturation can be fundamentally linked to the dimensionality of the holographic theory, an aspect independent of the specific geometrical configuration. For example, as also demonstrated in \cite{Ju:2024hba}, the upper bound of \(I_4\) obtained by fixing \(A\), \(B\), and \(C\) and fine-tuning \(E\) is always finite in AdS\(_3\)/CFT\(_2\), but may become infinite in higher-dimensional holographic theories. In this paper, we employ the four-color theorem to further distinguish the entanglement structure in AdS\(_4\)/CFT\(_3\) from that in even higher dimensions. Overall, the formalism built in this work can reveal the difference of entanglement structures in different dimensional holography. To our knowledge, these results cannot be derived by any other method, including the holographic entropy cone program.

In this work, while the upper bounds of some of the general combinations of holographic entanglement entropy could reach the information theoretical upper bound, for most of them, the upper bound is lower than that, as could be shown from the fact that the upper bound depends on dimensionality. As we will show in the work, we will analyze the effects of the geometrical configurations and dimensionality in the non-saturation of some of the examples. {This work could serve as a substantial and systematic generalization of \cite{Ju:2024hba,Ju:2024kuc}, from evaluating the upper bound of $n-$partite information to considering large classes of entropy combinations.}

{We organize this paper as follows. In Section \ref{sec2}, we identify two large classes of entropy combinations where we can claim the (dis)connectivity of all entanglement wedges in their upper bound configurations. In Section \ref{sec3}, we evaluate the upper bound of the entropy combinations under those (dis)connectivity conditions, and point out the difference of the upper bound in different dimensions. Combining those two sections, we can thoroughly evaluate the upper bound of an entropy combination. The result is given by the linear combinations of the entropy of fixed regions and the gap regions in between them. We give specific examples in Section \ref{sec4}. We conclude and discuss our results in Section \ref{sec5}.}

\section{Entanglement wedge connectivity in upper bound configurations} \label{sec2}

\noindent To study the holographic multipartite entanglement structure in AdS\(_3\)/CFT\(_2\) and higher dimensions, the upper bound of \( n \)-partite information (\( I_{n+1} \)) for fixed \( n \) boundary regions has been studied in our previous work \cite{Ju:2024hba}. By tuning the $n+1$-th region, we could always find an upper bound configuration of these $n+1$ regions, where the amount of entanglement that the $n+1$-th subregion participates with other $n$ subregions diverges, while the entanglement that the $n+1$-th subregion participates with {any} fewer subregions vanishes. Therefore, the upper bound value of \( I_{n+1} \) in this configuration reflects a large amount of global multipartite entanglement. {We name this upper bound configuration the holographic exclusive global multipartite entanglement configuration (HEGMEC), indicating its purely multipartite entanglement nature.}  The existence of this upper bound configuration reveals the special feature in the multipartite entanglement structure of holographic systems. In particular, we found that even when any two subsystems lack bipartite correlations, they still strongly participate in tripartite entanglement with an additional fine-tuned region, and the divergence behavior of \( I_4 \) in different dimensions reveals fundamental differences of four-partite entanglement structures in different dimensions. 

Our approach involves first proving a disconnectivity condition that constrains the connectivity\footnote{Note that in the whole paper, by connectivity, we refer to the connectivity of the entanglement wedges corresponding to those boundary regions. This is equivalent to nonzero mutual information between the two resulting subsystems when arbitrarily bipartitioning these regions.} of the entanglement wedges of a series of boundary regions at the maximum configuration, and then using this constraint to upper-bound the entanglement quantities via “fake” larger RT surfaces corresponding to incorrect connectivity, thereby achieving configurations where \( {I_{n+1}} \) reaches its information-theoretical upper bound. In this paper, we extend this method to a broader class of holographic entanglement entropy combinations beyond \( {I_{n+1}} \) and derive universal upper bounds for those combinations that apply to general configurations of the fixed \( n \) regions on the boundary.

{The holographic entanglement entropy combinations we are studying in this paper involve $n+1$ subsystems for any $n\geq 2$. There are infinite many combinations of holographic entanglement entropy, and some of their upper bounds are not interesting, \eg, \( S_{AE}+S_{E} \), as it is unbalanced\footnote{Here ``balanced" refers to the requirement that in the combination, for each individual party, the sum of the coefficients of every term which includes that party equals zero \cite{He:2020xuo}. An important issue is that we are only considering combinations of HEE that contain \(E\), so terms like \(S_{AB}\) are irrelevant and completely ignored. Thus, by the term balance, we mean that when considering only terms that contain \(E\), region \(A\) is balanced in those terms. For example, in our context, a combination like \(S_{AE}+S_{BE}-2S_{E}-S_{AB}\) is \textbf{not} balanced, while \(S_{AE}+S_{BE}+S_{CE}-2S_{E}-S_{ABCE}\) and \(S_{AE}+S_{BE}-S_{E}-S_{ABE}+S_{AB}\) \textbf{are} balanced.} with respect to \( E \), which means that one can simply make region \( E \) the union of as many intervals as possible so that both $S_{AE}$ and \( S_E \) have many UV divergent terms, making the upper bound as large as possible. Investigating combinations like these will not yield non-trivial and meaningful results.
Furthermore, not all combinations are calculable using the disconnectivity method mentioned above; for example, for the six-partite information \( I_6(A:B:C:D:E:F) \), we do not know its exact general upper bound except in some special \( ABCDF \) configurations \cite{Ju:2024kuc}, where we can construct \( E \) such that its holographic upper bound coincides with its information-theoretical upper bound.} 

In this section, we build a formalism that answers the following question: for which entropy combinations could we fix the (dis)connectivity properties of the related entanglement wedges in the maximum configuration and how do we fix them? To accomplish this goal, we introduce a so-called ``lamp diagram,'' which is a hypercube that marks the connectivity of all entanglement wedges for any given configuration. We could also read the entropy combination from the diagram and one such diagram corresponds to the connectivity of the upper bound configuration. We could find this particular diagram via the rules of lamp diagram that will be introduced in this section. Finally, we accomplish this procedure for two large classes of entropy combinations.

\subsection{The lamp diagram}\label{sec2.1}
\noindent Let us introduce the following notation, which is adopted throughout the whole paper. Capital letters $A$, $B$, $C$... are used to denote the $n$ regions we keep fixed that appear in the entropy combinations. It should be noted that in this work, we are interested in general cases where $A$, $B$, $C$... are no longer restricted to single, simply-connected intervals. For example, $A$ could consist of multiple intervals that interlace with (or enclose, depending on the specific spacetime dimension) the components of $B$ or $C$ on the boundary. On the other hand, $E$ is the $n+1$-th boundary subregion residing in the gap intervals between $A$, $B$, $C$.... Note that $E$ could also be a union of multiple subregions and we need to adjust them to alter the value of given entropy combinations to seek for maximums.

Directly evaluating the upper bound of a combination of holographic entanglement entropy with \( n \) regions fixed and one region \( E \) arbitrarily chosen is a formidable task, because \( E \) could consist of multiple subregions or could even be very wiggly in higher dimensions in the upper-bound configuration.
In \cite{Ju:2024hba}, we develop a procedure to evaluate the upper bound of tripartite and four-partite information with two (three) regions fixed and one region arbitrarily chosen. Here, we extract the abstract idea from that procedure, which can be employed to find the upper bound of other families of combinations:
\begin{itemize}
    \item I. Find and prove the connectivity of each entanglement wedge that contains \( E \) in the upper-bound configuration (this section).
    \item II. Directly evaluate the upper bound via the constraints of those (dis)connectivity conditions of the entanglement wedges (Section \ref{sec3}).
\end{itemize}

{Note that we will implement I in this section and II in the next section for more general entropy combinations.}
For \( I_3(A:B:E) \) and \( I_4(A:B:C:E) \) studied in \cite{Ju:2024hba}, those (dis)connectivity conditions are
\begin{equation}\label{I3I4dis}
\begin{aligned}
    I_3(A:B:E): \,\,\,\, &EW(E), EW(AE), EW(BE) ~\text{disconnected}, ~EW(ABE) ~\text{connected}.\\
    I_4(A:B:C:E): \,\,\,\, &EW(E), EW(AE), EW(BE), EW(CE), EW(ABE),\\ &EW(ACE), EW(BCE) \,~\text{disconnected}, ~EW(ABCE) \,~\text{connected}.
\end{aligned}
\end{equation}
It is worth noting that all connectivity refers to the connectivity between region $E$ and the rest in the corresponding entanglement wedges. For example, \( EW(ABE) \) being disconnected means that the mutual information \( I(E:AB) \) vanishes, but the mutual information \( I(A:B) \) might not vanish. A special condition is the (dis)connectivity of \( EW(E) \), which stipulates (dis)connectivity between different intervals within \( EW(E) \) itself.
Another important issue is that, due to strong subadditivity, \( I(E:AB)\geq I(E:A) \); therefore, if \( EW(ABE) \) is disconnected, then both \( EW(AE) \) and \( EW(BE) \) must be disconnected \cite{Hernandez-Cuenca:2019jpv}.

\begin{figure}[H]
	\centering
	\includegraphics[scale=0.4]{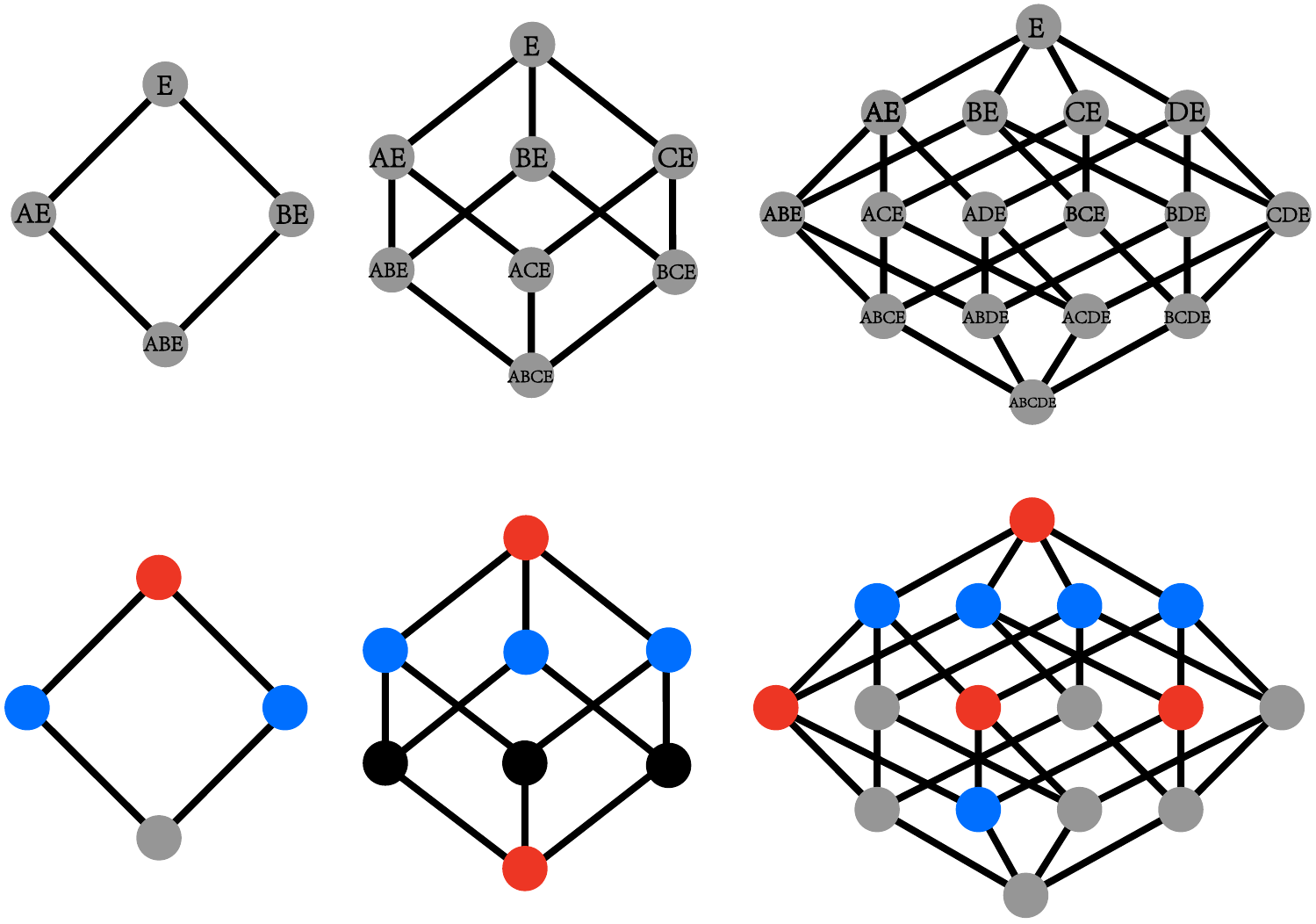}
	\caption{Examples of lamp diagrams for entropy combinations with fixed $n=2,3,4$ subregions (left, middle, right). In the all-grey figures on the first row, all our interested entanglement wedges are connected, while on the second row some of them become disconnected. Specifically, {the grey (other colored) dots of the three diagrams, as labeled, represent the respective connectivity (disconnectivity) of the entanglement wedge of every term in {$I_{n+1}$} that contains $E$. We denote the change of the color from grey to other colors (red, blue or black) as ``illuminating or lighting a lamp". Meanwhile, the lines display the relationship between these connectivity conditions: lighting a lamp requires the illumination of all lamps above it which connect with it by lines.}}\label{lampdiagram}
\end{figure}

 We can draw a so-called ``lamp diagram" to {denote}  all those connectivity conditions and their relationships as follows, as shown in Figure \ref{lampdiagram}.
It can be observed that these diagrams are hypercube-shaped. There are only two elements: dots (lamps) and lines. A dot represents a connectivity condition of the corresponding entanglement wedge\footnote{Note that we only care about the terms in the combination that contains $E$ (we will always omit other terms like $S_{AB}$ {as they stay unchanged when tuning $E$}), so we only care about the connectivity of the entanglement wedges that contains $E$ (the connectivity of other entanglement wedges will not be denoted as dots in this diagram).}. If it is grey, as shown in the figure, the entanglement wedge is connected; otherwise (if the lamp is lit), the entanglement wedge is disconnected. Due to strong subadditivity as discussed above, the necessary condition to light a lamp is to ensure that all the lamps above it, which are connected to it by lines, are also lit. For example, if one wants to light the lamp on the last line, one must ensure that all other lamps are lit; \ie, if \( I(E:ABCD) \) vanishes, then all mutual information such as \( I(E:ABC) \) and \( I(E:CD) \) must vanish.

If a lamp is lit, there are three possible colors for this lamp: red, blue, and black. The specific color is determined by the sign in front of the corresponding entanglement entropy term in the entropy combination. Specifically, if \(S_{AE}\) has a positive sign in the entropy combination, we draw the corresponding lamp, which represents the connectivity of \(EW(AE)\), in blue when it is lit. If \(S_{AE}\) has a negative sign, we then draw it in red\footnote{One might notice that the color designation here is opposite to that in the RT surfaces, where red is used to denote positive terms. The crucial reason behind this is discussed in Section \ref{sec2.2}.}, and if \(S_{AE}\) does not appear in the combination, we then draw it in black. {The absolute value of the coefficient in front of the entropy term is indicated within each dot, e.g. $2, 3$, etc.. We omit this notation when the coefficient is $1$.} Note that we can read a unique entropy combination from a lamp diagram with all lamps lit on. In other words, a lamp diagram with all lamps lit on is in one-to-one correspondence with a unique entropy combination{, and at the same time it indicates a given configuration with fixed disconnectivity conditions for all related entanglement wedges}. In the next subsection, we will use the lamp diagram to present the procedure for finding the upper bound of general entropy combinations.


\subsection{Disconnectivity condition and its proof}\label{sec2.2}


\noindent With the lamp diagram boosting the efficiency of the search for the upper bounds, we would like to extend our study to more general entropy combinations with the aim of revealing more entanglement structures. Before we begin, it should be noted that our analysis is restricted to entropy combinations in which every term contains $E$, for example 3-CMI: $I_3(A:B:C|E)=-2S_E+S_{AE}+S_{BE}+S_{CE}-S_{ABCE}$. Moreover, one can obtain upper bounds of more complicated entropy combinations containing a considerable number of entropy terms in further generalization, such as $n$-CMI or the examples presented in Section \ref{sec4.2}. 

In the upper-bound configuration, the (dis)connectivity of each entanglement wedge can be {fixed}, like in the configurations of (\ref{I3I4dis}).
This dramatically reduces the work required to identify upper-bound configurations because it is sufficient to analyze only those configurations in which the entanglement wedges satisfy these connectivity conditions. 
To prove this, we must show that \textit{if a configuration of \(E\) yields entanglement wedges that do not satisfy the required (dis)connectivity conditions, then there always exists another configuration \(E'\) that does satisfy them, with the value of the entropy combination being no less than that of the original configuration $E$.} 
In other words, once this result is established, evaluating the upper bound under the required (dis)connectivity conditions alone is justified. 

To prove this statement, we classify the entropy combinations into two types based on the structure of the lamp diagram: the CMI type and the $I_4$ type. The difference between them lies in that during the tuning process of the $n+1$-th subregion, the CMI type entropy combinations display only one local maximum configuration, while the $I_4$ type entropy combinations feature two local maxima of subregion configurations, one of which can be shown to be {not the true maximum}. Eventually, configurations corresponding to the global maximum in both cases can be verified to satisfy the {required} disconnectivity condition. 
We will prove the (dis)connectivity {requirements} for each type separately. 

Before going to the detailed analysis in these two cases, let us first determine the connectivity of \(EW(E)\) and \(EW(AB...E)\), where $AB...E$ refers to all $n+1$ subregions, as the (dis)connectivity conditions for these two entanglement wedges do not depend on the type and is a general feature of all balanced entropy combinations. 

The (dis)connectivity conditions for \(EW(E)\) and \(EW(AB...E)\) in the upper bound configurations of every balanced combination should be as follows.
\begin{theorem}
    For a balanced combination of holographic entanglement entropy, {the connectivity of \(EW(E)\) and \(EW(AB...E)\) could always be stipulated} in its upper bound configuration as follows: \footnote{Here we have not fixed the sign in any term in the entropy combination. In fact, any balanced combination and its opposite could share the same connectivity for $EW(E)$ and $EW(AB...E)$ as for any configuration we could find another configuration with disconnected $EW(E)$ and connected $EW(AB...E)$ with an equal value of the entropy combination.}
    \begin{equation*}
        EW(E) \,\, \text{being totally \textbf{disconnected};}\quad EW(AB...E) \,\, \text{being totally \textbf{connected}.}
    \end{equation*}
\end{theorem}

\noindent \paragraph{Proof.}
\begin{itemize}
    \item \textbf{For $EW(E)$.} If $EW(E)$ is not totally disconnected—for example, suppose $E_i$ and $E_j$ are connected in $EW(E)$—split the subregion $E_i$ into three parts, $E_{i1}$, $E_{i2}$, and a gap region between them. Enlarge the gap starting from zero width, and let $E'$ be the region obtained from $E$ by removing this gap. Consider $EW(E')$. For sufficiently small gap width, $E_{i1}$ and $E_{i2}$ remain connected in $EW(E')$. As the gap grows, the entropy combination (with $E$ replaced by $E'$) remains unchanged because it is balanced: the area contribution from the RT surface associated with the small gap cancels out. This continues until a phase transition of $EW(E')$ occurs. At that transition, one of $E_{i1}$ or $E_{i2}$ must become disconnected from $E_j$ (and therefore also from the other). If the midpoint of the gap is chosen sufficiently far toward the left (right) end where $E_{i1}$ ($E_{i2}$) lies, then increasing the gap first makes $E_{i1}$ ($E_{i2}$) very small and it disconnects from $E_j$ first. By continuity, there exists a choice of midpoint for which $E_{i1}$ and $E_{i2}$ disconnect from $E_j$ simultaneously. At this critical point, $E_{i1}$, $E_{i2}$, and $E_j$ are mutually disconnected in $EW(E')$, while the entropy combination remains unchanged. Repeating this procedure, every connected component in $EW(E)$ can be made disconnected in some $EW(E')$.

\item \textbf{For $EW(AB...E)$.} If \(EW(AB...E)\) is not totally connected, \ie, if \(S_{AB...E}\neq \sum_i S_{gap_i}\), where the summation is with respect to all the gap regions \(gap_i\) between different intervals of \(E\), \(A\), \(B\), etc., one can simply add new sub-regions \(E_i\) {in any gap region between the original $A,B,E$} as part of \(E\) and enlarge it from zero size. At first, as $E_i$ is so small, it will disconnect with every region in $EW(AB...E)$. Then as we enlarge $E$, the value of the combination will remain unchanged as $E$ is balanced until it connects with one region on its left (right) side in $EW(AB...E)$. If the middle point of $E_i$ is too far left in the gap region, it will connect with the region on the left side first, and vice versa. 
As a result, one can always fine-tune the center of \(E\) so that \(E\) connects with both sides in \(EW(AB...E)\) simultaneously. 
In the entire process,  $E_i$ is like the glue that connects two disconnected parts of $EW(AB...E)$, and we can add as many \(E_i\) as needed until all regions are connected in \(EW(AB...E)\). During the whole process, the value of the entropy combination will not change due to the balance condition of the entropy combination respective to $E$. 
\end{itemize}
Note that the structures of the proofs for $EW(E)$ and $EW(AB\cdots E)$ are quite similar. The similarity is not accidental. Let $O=(AB\cdots E)^{c}$ be the complement\footnote{Throughout the paper, we denote the complement of boundary subregion $X$ by $X^c$.}. Replacing every holographic entanglement entropy involving $E$ by its purifier involving $O$ produces an entropy combination that is balanced with respect to $O$. Furthermore,
\[
EW(AB\cdots E)\ \text{fully connected}\ \Longleftrightarrow\ EW(O)\ \text{totally disconnected}.
\]
Thus the two arguments are complementary reformulations of the same statement.

\subsubsection{CMI type (dis)connectivity condition}

\noindent In this section, to analyze the upper bound of general combinations of holographic entanglement entropy of $n+1$  subsystems with $n$ subsystems fixed, we determine the connectivity conditions for the upper bound configurations. This allows us to focus on only one specific configuration for every combination when evaluating the upper bound, thereby greatly simplifying our analysis.

As we have already shown above, \(EW(E)\) and \(EW(AB...E)\) are required to be disconnected and connected respectively. Now we determine the connectivity of all other entanglement wedges that contain \(E\) in the combination {e.g., terms like \(EW(AE)\) and \(EW(BE)\)}. {First we need to show that for this CMI type of holographic entropy combinations, connectivity analysis is an efficient method to find the upper bound configuration, i.e. we could always reach the upper bound configuration by {analyzing and fixing} (dis)connectivity conditions on all {subsystems} containing $E$. }
Equivalently, we want to prove that if one configuration of \(E\) does not satisfy the required (dis)connectivity condition, one can always find another configuration of \(E\) that satisfies the condition, with the value of the combination being no less than that of the former one. 

To achieve this, we first assume that there exists at least one ``bad" interval (or ``bad" subregion in higher dimensions) inside \(E\) that does not satisfy {the required} (dis)connectivity condition for at least one of the entanglement wedges containing $E$. Our goal is to find an \(E'\) replacing \(E\) that satisfies that (dis)connectivity condition, {e.g.} (\ref{I3I4dis}), and to ensure that after the replacement, the value of the combination does not decrease.

The core process of finding the region \(E'\) is to split that ``bad" interval into three parts: a new gap region in the middle and the other two subregions that belong to \(E'\). In other words, \(E'\) is the region obtained by deleting one gap region inside the ``bad" subregion from \(E\), analogous to the proof of $EW(E)$'s disconnectivity.

Following this procedure, we present an explicit example in Appendix~A to illustrate how the splitting process is used to evaluate the maximal value of the conditional mutual information \(I(A\!:\!B\mid E)\). This procedure was studied in~\cite{Ju:2024hba}. Furthermore, in Appendix A, we recast the entire process in the language of lamp diagrams and formulate the general rules of comparing a lamp diagram with its successor obtained by lighting one additional lamp (the post–splitting diagram). As each diagram serves as a stipulation of the connectivity conditions of all entanglement wedges, comparing those diagrams to find the maximum one will help us find the rightful stipulation in the upper bound configuration. The rule is as follows:

\begin{itemize}
    \item \textbf{Rules of comparing lamp diagrams via splitting process.}
        \textit{If the number of red lamps is greater than the number of blue lamps in the current lamp diagram, then the splitting process to the next step will increase the value of the entropy combination; if the number of blue lamps is greater than the number of red lamps in the current diagram, then the splitting process to the next step will decrease the value of the entropy combination; if the numbers are equal, the next step of the splitting process will not change the value of the entropy combination.}
\end{itemize}

\begin{figure}[H]
	\centering
	\includegraphics[scale=0.4]{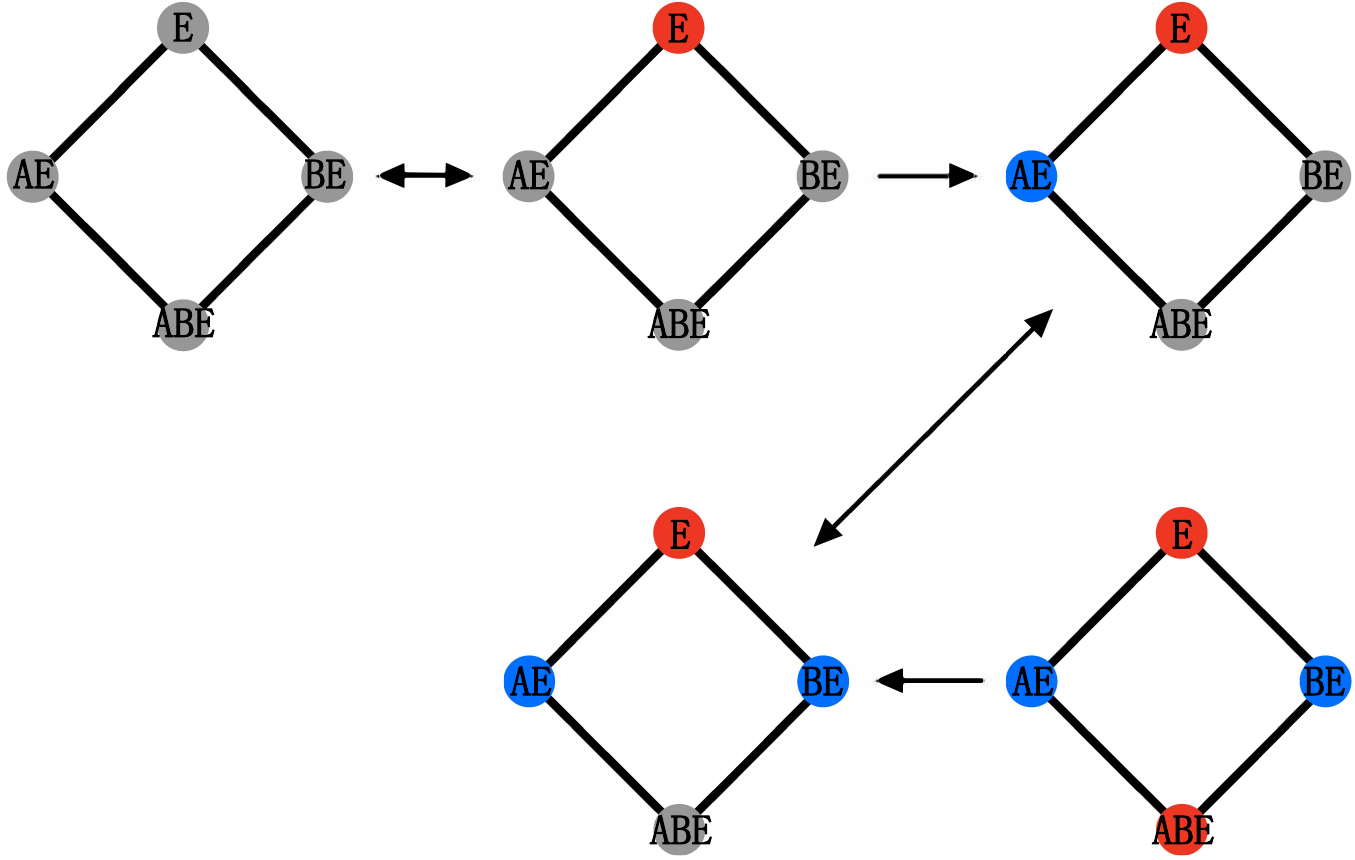}
	\caption{Lamp diagram to {obtain} the disconnectivity condition for the CMI $I(A:B|E)$, which is an explicit example of the CMI type entropy combination. Detailed discussion is presented in Appendix A.} \label{CMIlamp}
\end{figure}

It should be first noted that this is a general conclusion which applies to any entropy combination and its corresponding lamp diagram. Consequently, utilizing this newly found rule, we can simply draw all the arrows in Figure \ref{CMIlamp}, showing the directions of increasing the value of the entropy combination. In this specific example of CMI $I(A:B|E)$, we find that the maximum configuration is represented by the last graph on the first line and the first graph on the second line, as inferred from the directions of the arrows. In principle, choosing any one of these configurations and stipulating the corresponding connectivity condition would give us the upper bound configuration; however, practically, the last graph on the first line is not unique because we can switch \(A\) and \(B\). If we choose that graph, then some of the intervals of \(E\) might be connected with \(A\) in \(EW(AE)\) while others might be connected with \(B\) in \(EW(BE)\). Evaluating the maximum value is still a tedious task. As a result, we should choose the first graph on the second line as the rightful choice to stipulate the connectivity condition of the entanglement wedges so that finding the upper bound value would be easier. The resulting configuration will exactly match the (dis)connectivity condition in (\ref{I3I4dis}) for the CMI $I(A:B|E)$.

{In conclusion, given an arbitrary entropy combination, the basic procedure of obtaining the maximum configuration through lamp diagrams can be summarized as follows. First, a sequence of lamp diagrams should be drawn in order-- from the diagram where all lamps are grey to the diagram where the lamps are fully lit, along with all possible diagrams in between. Then, we can draw arrows (or double-headed arrows) that connect them showing the direction that increases the value of the entropy combination. Eventually we can obtain a figure that resembles Figure \ref{CMIlamp}, through which the maximum configuration can be concluded. It should also be noted that for CMI type combinations, when the maximum configuration consists of multiple diagrams, we should choose the one which is permutation symmetric with respect to the $n$ fixed parties.} We will present more examples in Section \ref{sec4} to realize this procedure.

\subsubsection{$I_4$ type (dis)connectivity condition}

\noindent In the splitting process used to reach the maximum configurations, we can always arrive at the maximum lamp diagram(s) following the direction of the arrows. However, there exist entropy combinations for which the arrows lead to more than one maximum configurations that are not directly connected by an arrow, making it unclear which one represents the true maximum. Among these, \(I_4\) serves as an example, as its splitting process features two maximums—one of which is trivial. The analysis of \(I_4\) therefore provides insight into the upper bounds of a broad class of entropy combinations (which we define as \(I_4\)-type entropy combinations) that exhibit an additional trivial maximum. Readers who are not interested in the details of the proof may proceed directly to the end of this section.

\begin{figure}[H]
	\centering
	\includegraphics[scale=0.6]{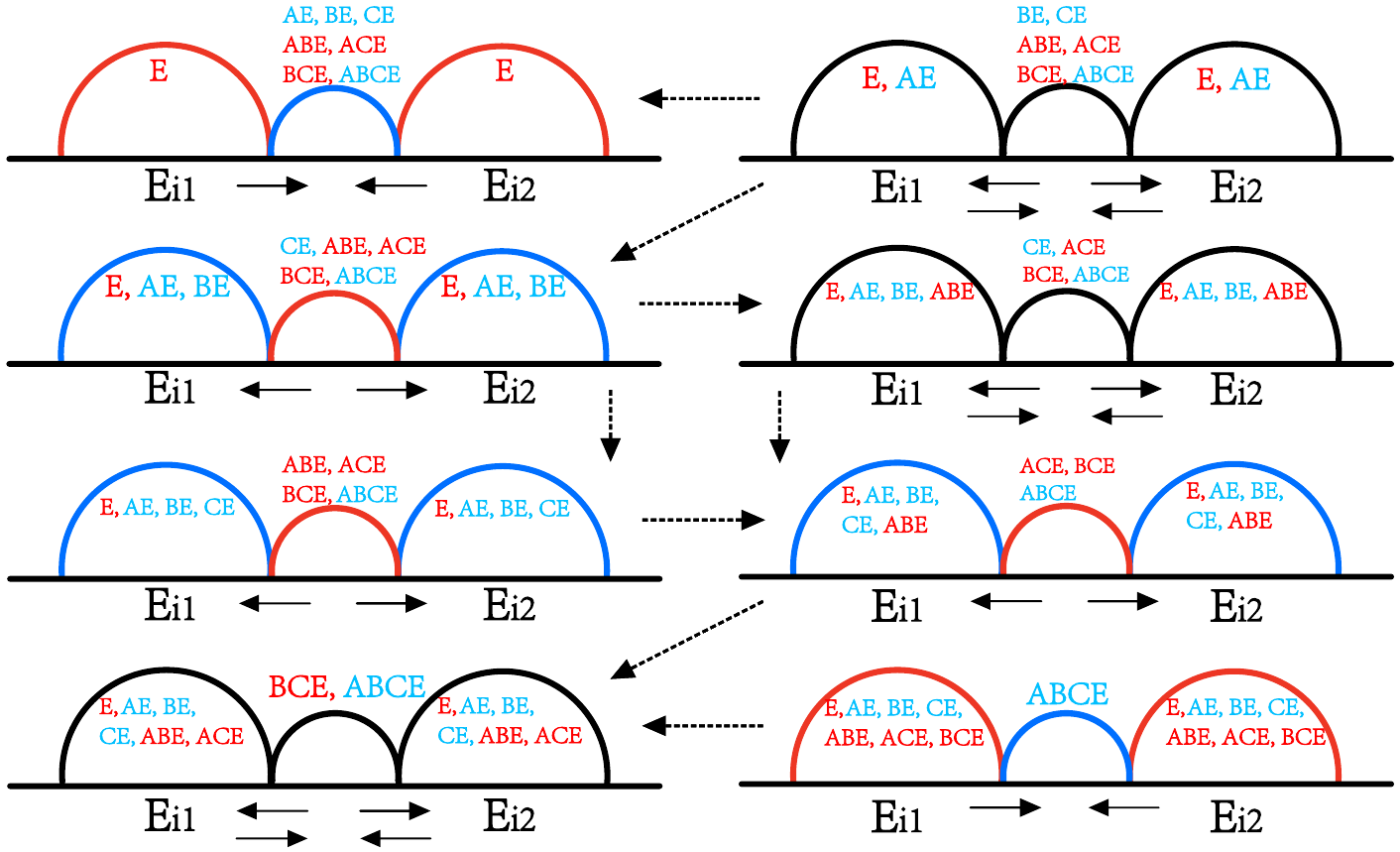}
	\caption{Splitting process of $E_i$ in the search of the maximum configuration for  \(I_4(A:B:C:E)\). {The red (blue) colors represent the overall positivity (negativity) of the geodesic in \( I_4 \). The dashed arrows between each diagram represent the direction of increasing \( I_4 \).} Details could be found in \cite{Ju:2024kuc}.}\label{DISI4}
\end{figure}

\begin{figure}[H]
	\centering
	\includegraphics[scale=0.5]{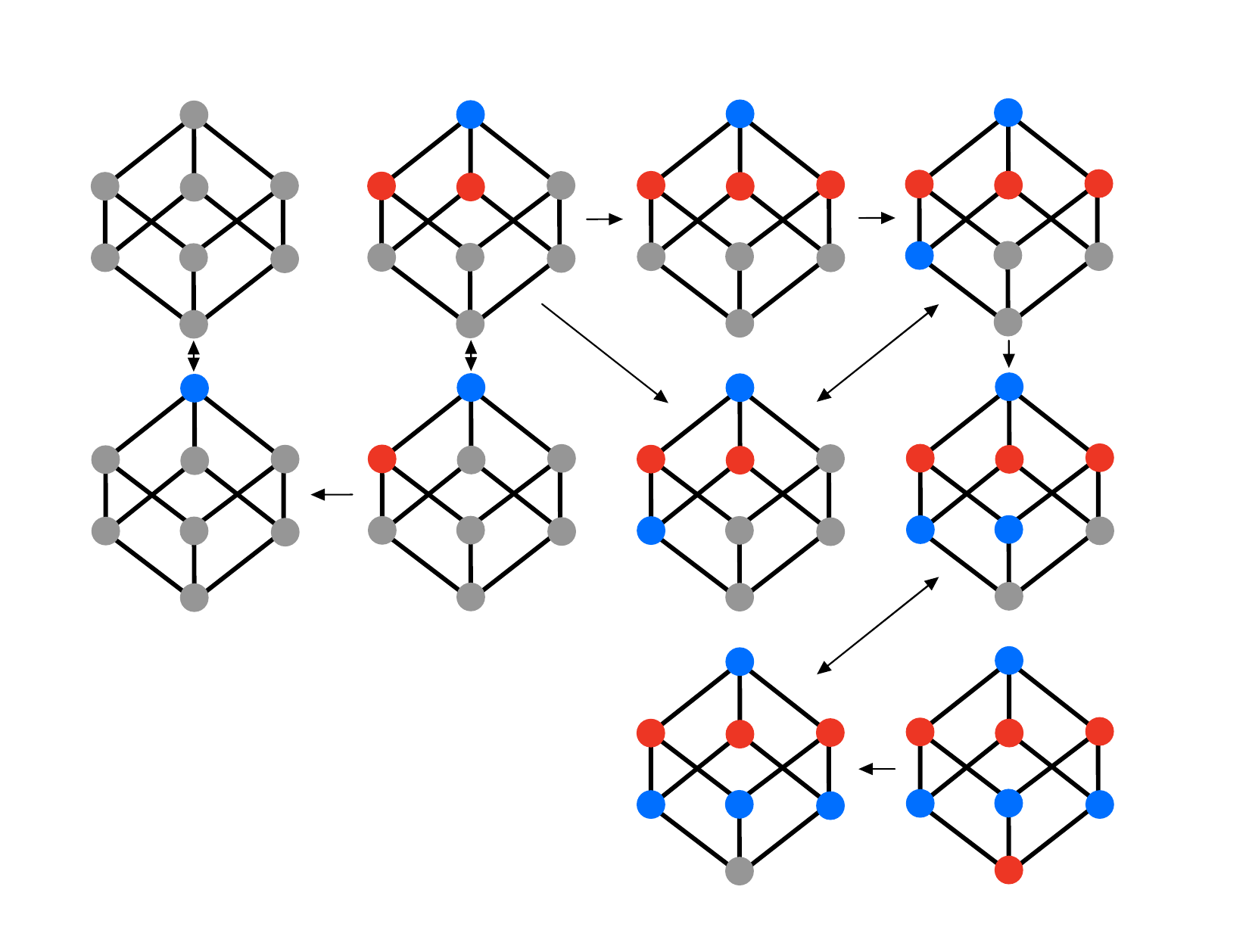}
	\caption{Lamp diagram to prove the \(I_4\) version disconnectivity condition. From now on, we omit the name of the entanglement wedge that each dot represents as the pattern is always the same as in Figure \ref{lampdiagram}.}\label{I4lamp}
\end{figure}

We now analyze the maximum configuration using the splitting process of lamp diagrams for this type, taking \(I_4(A:B:C:E):=I_3(A:B:C)+I_3(A:B:E)-I_3(A:B:CE)\) as an example. Let us review the process of finding the (dis)connectivity condition in the maximum configuration for the \(I_4\) case using the language of the lamp diagram. The {process of fixing} the (dis)connectivity condition at the maximum configuration (see \ref{I3I4dis}) in \cite{Ju:2024kuc} is presented in Figure \ref{DISI4}. Translating it into the language of the lamp diagram, we obtain Figure \ref{I4lamp}, which shows all possible lamp diagrams connected by arrows between adjacent diagrams. After drawing all the arrows properly, we find that there exist two possible maximum configurations: the all-connected configuration (the very first figure) and the configuration corresponding to the (dis)connectivity condition (the second-to-last figure) as given in (\ref{I3I4dis}). These two configurations are not directly connected by an arrow, so we cannot immediately determine which one represents the larger value of \(I_4\).

Note that when we perform the splitting process, we analyze each individual interval \(E_i\) within the region \(E\) and obtain maximum configurations in which its (dis)connectivity in all entanglement wedges is fixed. When two maximum configurations exist, it means that for each interval \(E_i\) there is a choice between these two (dis)connectivity configurations. Consequently, different intervals within \(E\) might adopt different (dis)connectivity configurations corresponding to the two diagrams. For example, some intervals in \(E\) may satisfy (\ref{I3I4dis}), while others may remain connected to all regions in all entanglement wedges, corresponding to the very first diagram. This multiplicity of possibilities makes evaluating the upper bound quite difficult.

However, through specific examples as follows, we find that the existence of an interval $E_i$ which is connected in all entanglement wedges (corresponding to the very first diagram) will result in an even smaller $I_4$, compared to the case where this $E_i$ is deleted from $E$. The simplest case is when $E$ is the complement of regions $ABC$. Through simple evaluation, the four partite information is 
\begin{equation}
    I_4(A:B:C:(ABC^c))=2I_3(A:B:C)\leq0,
\end{equation}
indicating that a large enough $E$ (being the complement of $ABC$, yielding the first diagram) is on the contrary resulting a negative $I_4$. 
Then if we only take $E$ as one gap region in between $A,B,C$, or a single interval inside those gap regions, through other specific evaluation, it will also decrease $I_4$. This fact makes us wonder if we can actually prove that the very first diagram cannot be the upper bound configuration. 

Fortunately, through tedious technical details and the introduction of some other concepts like the O-version diagram and the reverse procedure of the splitting process as will be explained later, we have managed to prove that the first maximum configuration is trivial and cannot be the true maximum configuration for any interval; only the second-to-last configuration is the rightful one for the (dis)connectivity of the entanglement wedges.

 Before further analysis, it is important to note that our discussion is not limited solely to \(I_4\), but applies more generally to a wide class of entropy combinations whose maximum (dis)connectivity configurations can be summarized by the two diagrams described above. These types of entropy combinations are not uncommon, and we will present additional examples in Section \ref{sec4}. 
 
 The justification for eliminating the first maximum configuration is provided by the following theorem. We provide its systematic proof in Appendix B. 

\begin{theorem}
    \textbf{\(I_4\) type disconnectivity theorem.} For some entropy combinations at general $n$, there are two choices of maximum (dis)connectivity configurations in the lamp diagrams: one is the case where all \(2^n\) entanglement wedges are connected, and the other is that \(2^n-1\) entanglement wedges are disconnected while the entanglement wedge of all $n+1$ subregions is connected. The all-connected configuration cannot be the true maximum configuration, and one can directly stipulate the disconnectivity of \(2^n-1\) entanglement wedges and the connectivity of the entanglement wedge of the $n+1$ subregions as the (dis)connectivity condition in the upper-bound configuration.
\end{theorem}

Let us summarize the process for testing whether an entropy combination falls into one of the two types discussed above, so that we can identify the corresponding upper-bound configurations with fixed (dis)connectivity conditions for all entanglement wedges. First, we draw a lamp diagram with all lights turned on along with one where all lights are grey, and then we draw all possible diagrams and arrows between them as we did in Figures \ref{CMIlamp} and \ref{I4lamp}. Next, we test whether the maximum configuration is unique. If it is unique (which corresponds to the CMI type), we can directly {fix} the (dis)connectivity condition. If the maximum configuration is not unique—in most cases, we cannot stipulate the (dis)connectivity condition—but if there are two maximal diagrams, one with all lights turned off and the other with \(2^n-1\) lights turned on (which corresponds to the \(I_4\) type), we can directly omit the first diagram and adopt the diagram with \(2^n-1\) lights turned on.

{Interestingly, we find that in both types of the combination, the upper bound configuration exhibits the following properties: $EW(AB...E)$ fully connects while other EWs of subsystems appearing in the combination are fully disconnected. In those configurations, entanglement which involves less partite subsystems vanishes while entanglement which involves $n+1$ partite prevails. In this sense, we can impart the CMI and $I_4$ type entropy combinations the following special meaning: they serves as genuine multipartite entanglement measures in holography when they reach their upper bound.}

\section{Derive the upper bound: classification of gap regions} \label{sec3}
\noindent In the previous section, we introduced a formalism that identifies the (dis)connectivity condition of maximum configurations in two general types of entropy combinations. The next step is to directly derive the exact upper bound value of an entropy combination at its maximum configuration under this (dis)connectivity condition. This upper bound value should be expressed as a function determined by the fixed \(n\) subregions as well as all the gap regions among these \(n\) subregions.

However, even with the (dis)connectivity condition in place, it is still challenging to extract the upper bound value for the most general configurations of the fixed subregions. This difficulty arises because, under the disconnectivity condition, there are still infinite many possibilities for choosing the intervals that constitute the subregion \(E\), making it hard to obtain an exact value for the upper bound of the entropy combination.

To address this problem, we employ explicit inequalities that constrain the upper bound based on the principle that the minimal RT surface is always smaller than the non-minimal (or fake) ones. Achieving this first requires a universal classification of all the gap regions according to their adjacent subregions, in order to derive the explicit constraints from fake RT surfaces. Therefore, in this section, we propose a universal gap region classification method to derive the upper bound of the combinations which satisfy the (dis)connectivity condition in the last section. When \( n\geq3 \), we find that the non-existence of a gap region adjacent to three regions simultaneously in AdS\(_3\)/CFT\(_2\) results in a fundamental difference in the four-partite entanglement structure from those in higher dimensions. When \( n\geq4 \), the lack of the existence of certain combinations of different gap regions constrained by the famous four-colored theorem results in further differences in the five-partite entanglement structure in holographic \(2+1\)d CFT and even higher dimensional ones.

\subsection{General formalism for gap region classification}
\noindent The (dis)connectivity condition at the maximum configuration can greatly help simplify the explicit expression of the entropy combination. Note that in the (dis)connectivity conditions for both types of the entropy combinations, \(E\) is disconnected in all entanglement wedges except the entanglement wedge of the union of all regions \(EW(AB\ldots E)\). As a result, mutual information like \(I(E:A)\) and \(I(E:B)\) vanishes, \ie, \(S_{AE}=S_A+S_E\). Since \(EW(AB\ldots E)\) is connected, the entanglement entropy of the union of all regions will simply be the summation of the entanglement entropies of all gap regions among \(E, A, B,\ldots\), written as \(\sum_i S_{gap_i}\). At the same time, since \(EW(E)\) is totally disconnected, 
\[
S_E=\sum_i S_{E_{i}},
\]
which is the summation of the entanglement entropy of each interval inside \(E\). As a result, for example, \(I_4(A:B:C:E)\) can be simplified as follows
\begin{equation}\label{I4def}
\begin{aligned}
    I_4 &= S_E - S_{AE} - S_{BE} - S_{CE} + S_{ABE} + S_{ACE} + S_{BCE} - S_{ABCE} + I_3(A:B:C)\\
    &= S_E - S_{A} - S_{E} - S_{B} - S_{E} - S_{C} - S_{E} + S_{AB} + S_{E} + S_{AC} + S_{E} + S_{BC} + S_{E}\\
    &\quad - S_{ABCE} + I_3(A:B:C)\\
    &= S_{ABC} + \sum_i S_{E_{i}} - \sum_i S_{gap_i}.
\end{aligned}
\end{equation}

One can observe that after this simplification, only the term \(\sum_i S_{E_{i}} - \sum_i S_{gap_i}\) matters, while other terms (such as \(S_{ABC}\) in the \(I_4\) case) are constants that will not change when we tune \(E\). In fact, this simplified form could be examined to be still valid for general entropy combinations. More explicitly, as long as the general entropy combination is balanced with respect to \(E\), the expression for the entropy combination will always be simplified into a term of the form 
\[
\text{coeff}~\Bigl(\sum_i S_{E_{i}} - \sum_i S_{gap_i}\Bigr)
\]
plus a constant under the required (dis)connectivity condition, where ``\(\text{coeff}\) " is the negative of the coefficient of \(S_{AB\ldots E}\) in the combination. Note that ``coeff " must be positive because $S_{AB...E}$ must have a negative sign in the entropy combinations so that the entropy combination could belong to the CMI or $I_4$ type.

Therefore, to find the exact upper bound values of general entropy combinations of the CMI or $I_4$ type with the help of the (dis)connectivity conditions, we only need to evaluate the upper bound value of \(\sum_i S_{E_{i}} - \sum_i S_{gap_i}\). The core idea is that the (dis)connectivity condition stipulates the disconnectivity of the entanglement wedges, preventing \(\sum_i S_{E_{i}}\) from being too large and \(\sum_i S_{gap_i}\) from being too small. As a consequence, the disconnectivity condition constrains the value of the combination from being too large, and in principle provides an upper bound value for the combination when tuning $E$.

To be more precise, {such an upper bound is explicitly obtained by requiring the area of the real RT surface that encloses any disconnected entanglement wedge as stipulated in the (dis)connectivity condition for the upper bound configuration to be smaller than all fake ones of the corresponding subregion. This idea is explained in detail as follows.}
Normally, when we say that an entanglement wedge (e.g., \(EW(AE)\)) is disconnected, it means that any connected minimal surface has a larger area compared to the real disconnected RT surfaces. 

Here the fake connected RT surfaces include partially connected cases, \ie, $A$ could be connected with some of the intervals inside $E$ in $EW(AE)$. This results in too many possible configurations for fake RT surfaces because region \(E\) can be placed in all those gap regions among \(A, B, C,\ldots\) and $A$ could connect with any one (or many) of the \(E_i\)'s inside one (or many) of the gap regions. As a result, we can obtain many inequalities for the upper bound value because any such partially connected configuration will have a “fake” RT surface whose area is larger than that of the real disconnected RT surface. Which inequality should we choose in order to obtain the tightest upper bound for \(\sum_i S_{E_{i}} - \sum_i S_{gap_i}\)?


To analyze these faked RT surfaces in detail, the initial step involves categorizing all gap regions based on their adjacent fixed subregions. For each single gap region, we denote it as \(g_{nAB\ldots}\) where \(n\) represents the number of regions adjacent to it, and \(AB\ldots\) indicates the adjacent regions. {Note that each fixed subregion $A, B, C...$ could have a large amount of disconnected intervals and the intervals of these $n$ fixed subregions could array in an arbitrary order, so there could be gap regions adjacent to many different combinations of subregions.} For example, {in 1+1 boundary dimensions,} when we evaluate the upper bound of \(I_4(A:B:C:E)\), all possible gap regions among \(A, B, C\) where we could place \(E\) are as follows
\begin{equation}\label{Gcha_n3d2}
    \begin{aligned}
        &g_{1A},\quad g_{1B},\quad g_{1C},\\
        &g_{2AB},\quad g_{2AC},\quad g_{2BC}.
    \end{aligned}
\end{equation}
It is worth noting that in \(1+1\)d CFT, any gap region has only two endpoints and can be adjacent to at most two fixed subregions; therefore, \(g_{3ABC}\) exists only in higher dimensions.

After labeling each gap region in this manner, the task of identifying the partially connected fake RT surface that yields the tightest upper bound is greatly simplified when we further require that the intervals of \(E\) within the same gap region exhibit uniform (dis)connectivity across all entanglement wedges. In the configuration of the fake RT surface, this allows us to divide all gap regions into two groups: in the first group \(G\), all the intervals of \(E\) {inside these gap regions} are connected in every entanglement wedge; in the second group, the intervals of \(E\) are disconnected. This classification significantly reduces the complexity of finding the optimal configuration for the fake RT surface.


{The area of the fake RT surface sets an upper bound for the real RT surface for each entanglement wedge including $E$, resulting in the following inequality.} We still take $EW(AE)$ as an example. Due to the disconnectivity condition of $EW(AE)$, we can write down the inequality as follows
\begin{equation}\label{ExIn}
    S_A + \sum_{i\subset G} S_{E_i}+ \sum_{i\not\subset G} S_{E_i} \leq S_{AG} + \sum_{i\subset G}  S_{gap_i}+ \sum_{i\not\subset G} S_{E_i} ,
\end{equation}
where the LHS is the real RT surface homologous to $AE$, while the RHS is the fake RT surface with intervals of $E$ inside the gap regions \(G\) connected in $AE$ and the rest disconnected in $AE$. \(\sum_{i\subset G} S_{E_i}\) is the summation of the entanglement entropy of the intervals of $E$ inside \(G\), and \(\sum_{i\subset G} S_{gap_i}\) is the summation of the entanglement entropy of the small gap regions among \(E\) and \(A, B, C\) inside $G$\footnote{An important issue should be clarified regarding inequality (\ref{ExIn}). Notice that we have ignored the contribution of  the small gap region \(gap_{EB}\) between \(E\) and (say) \(B\) or \(C\) {in the right-hand side terms $S_{AG}$ and $\sum_{i\subset G}  S_{gap_i}$}. Assuming \(G = g_{2AB}\) {for simplicity} and taking \(gap_{BE}\) into consideration, we should actually rewrite inequality (\ref{ExIn}) as follows:
\begin{equation}
    S_A + \sum_{i\subset g_{2AB}} S_{E_i} \leq S_{Ag_{2AB}/gap_{BE}} + \sum_{i\subset g_{2AB}} S_{gap_i} - S_{gap_{BE}}.
\end{equation}
One can easily check that the right-hand side increases when the length of \(gap_{BE}\) tends to zero. When it is negligible (its RT surface is close to the UV cutoff \(\epsilon\)), the inequality reduces to (\ref{ExIn}). However, we cannot view this inequality as the tighter one, because the existence of a large \(gap_{BE}\) is not always necessary when choosing \(E\); \ie, we can artificially choose a smaller \(gap_{BE}\) that results in a larger value of the combination, so that the upper bound with a large \(gap_{BE}\) is not valid anymore. Practically, for a configuration with a \(gap_{BE}\) that is not negligible, we can always construct another configuration with a smaller \(gap_{BE}\), with the value of the combination being no less than that of the former one.

Specifically, we can always add a small interval \(E_{\text{small}}\) within \(gap_{BE}\) and enlarge it until a phase transition occurs such that \(E_{\text{small}}\) becomes connected in \(EW(AB\ldots E)\). During this process, the value of the combination remains unchanged, and after the process all intervals within \(E\) still satisfy the (dis)connectivity condition. The only change is that the new \(gap_{BE_{\text{small}}}\) is a subset of the original \(gap_{BE}\). As a result, we can repeat this process until \(gap_{BE}\) becomes completely negligible, at which point inequality (\ref{ExIn}) is the appropriate inequality to use in the following calculations. }.

{In summary, for each disconnected entanglement wedge containing $E$, e.g. $EW(AE)$, {from the constraint that its fake RT surface is larger than the real one,} we could write an inequality constraining the upper bound of part of the terms in the entropy combination. By appropriately choosing $G$ for each entanglement wedge, we could sum all the inequalities to obtain an inequality for the exact total expression of the entropy combination, which gives a valid upper bound for the combination. However, different choices of $G$ for the wedges can lead to inequalities that differ in tightness. Therefore, by analyzing the difference of the valid inequalities utilising CMI and UV divergence, we can eventually obtain the tightest inequality.}

\subsection{An explicit example: upper bound of $I_4$}\label{sec3.2}



\noindent To show an explicit example for this procedure, let us still take $I_4(A:B:C:E)$ in AdS$_3$/CFT$_2$ as an example. {Note that in our previous work \cite{Ju:2024hba}, the upper bound for $I_4(A:B:C:E)$ in a specific case has been derived, where $A$, $B$ and $C$ are single boundary intervals in AdS$_3$/CFT$_2$, and $E$ is arbitrarily chosen. However, when considering another geometry, which is the three-mouth wormhole with $A$ $B$ and $C$ single intervals residing on the three mouths respectively, we obtain an upper bound whose form is distinct from that in the former case (detailed discussion of these cases will be momentarily provided in this section). Therefore, we aim to give a universal upper bound for $I_4$ that applies to the most general cases where we have any finite number of $A$, $B$ and $C$s randomly distributed on the boundary. We hope that such a bound can unify the results of the two examples above.}

\begin{figure}[H]
	\centering
	\includegraphics[scale=0.6]{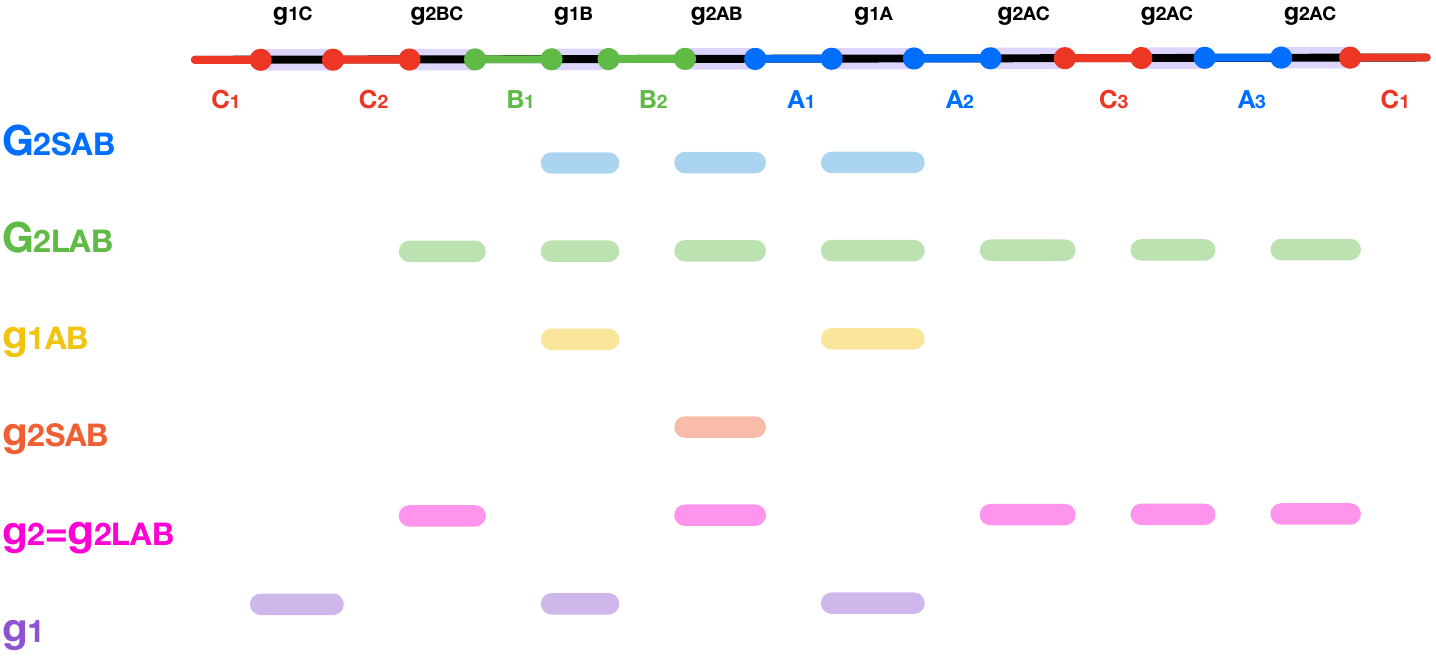}
		\caption{{Illustration of the unions of gap regions considered in this section, displayed as colored strips in an arbitrarily chosen configuration of $A$, $B$ and $C$ as an example. The first five rows show the gap combination choices in (\ref{I4gapgroup}) for $AB$, while the last two rows display our notation of $g_1$ and $g_2$.}}\label{Gdisplay}
\end{figure}

To begin with, {according to the precise disconnectivity of $EW(ABE)$, $EW(BCE)$ and $EW(ACE)$ for the maximum configuration required by (\ref{I3I4dis}), we must first consider the ``fake"  and ``real" RT surfaces by replacing $A$ in (\ref{ExIn}) with $AB$, $BC$ and $AC$.} Then, it is essential to introduce the following gap groups to place $E$ as the union of single gaps in (\ref{Gcha_n3d2}) (taking $AB$ for instance), as shown in Figure \ref{Gdisplay}:
\begin{equation}\label{I4gapgroup}
\begin{aligned}
    &G_{2_{S}AB}=g_{1A}\cup g_{1B}\cup g_{2{AB}}\\
    &G_{2_{L}AB}=g_{1A}\cup g_{1B}\cup g_{2{AB}}\cup g_{2{AC}}\cup g_{2{BC}}\\
    &g_{1AB}=g_{1A}\cup g_{1B}\\
    &g_{2_{S}AB}= g_{2{AB}}\\
    &g_{2_{L}AB}= g_{2{AB}}\cup g_{2{AC}}\cup g_{2{BC}}.
\end{aligned}
\end{equation}
{Here capital letter $G$ refers to the unions containing single gaps that are adjacent to different numbers of parties, while $g$ indicates that only gaps adjacent to the same number of parties are included. Moreover, $L$($S$) represents the inclusion(exclusion) of the $g_{2...}$ that are adjacent to only one of $A$ and $B$, which are $g_{2{BC}}$ and $g_{2{AC}}$ in this example.}

Before further discussion, it should be noted that for each entanglement wedge, we are inclined to compare all possibilities of the choice of $G$ and pick the one corresponding to the tightest constraint. Meanwhile, this aim of finding the tightest bound can also provide certain restrictions that simplify the choice of $G$. First of all, in the maximum configuration, though $EW(AE)$, $EW(BE)$ and $EW(CE)$ are also demanded to be disconnected, they are not at critical points of entanglement wedge phase transitions. In these cases, the ``real" RT surface is always smaller than the ``fake" one, and the inequalities they give cannot be saturated and are not the tightest. 

Moreover, when we study the $G$ to place $E$ (without loss of generality, take $AB$ as an example), we always consider gap combinations that are closest to $AB$, in which it is harder for strips of $E$ to disconnect with $AB$ in $EW(ABE)$, and the inequalities thus obtained are tighter. In other words, when we split and shrink every strip of $E$, those far from $AB$ tend to break from the connected $EW(ABE)$ earlier than those close to $AB$. Therefore, the saturation of inequalities for the far gaps, which states the exact RT surface phase transition for $AB$ and $E$ in those gaps, is not possible for our maximum configuration where $AB$ and the entire $E$ undergo a precise RT surface phase transition.

For convenience we also define 
\begin{equation}
\begin{aligned}
    &g_{1}=g_{1A}\cup g_{1B}\cup g_{1C}\\
    &g_{2}= g_{2{AB}}\cup g_{2{AC}}\cup g_{2{BC}}=g_{2_{L}AB}=g_{2_{L}AC}=g_{2_{L}BC},
\end{aligned}
\end{equation}
{which are the unions of all single gaps adjacent to one and two parties respectively. } {Thus, the process of upper-bounding $\sum_i S_{E_{i}}-\sum_i S_{gap_i}$ by the area of a fake RT surface according to the disconnectivity condition can be carried out in all the gap groups defined in (\ref{I4gapgroup}) along with those with subscripts $BC$ and $AC$. Meanwhile, our goal is to derive general bounds satisfying the permutation symmetry across $A$, $B$ and $C$, and consequently, our final bound of $\sum_i S_{E_{i}}-\sum_i S_{gap_i}$ must consist of summations of (\ref{ExIn}) -shaped inequalities that contain gaps with subscript $AB$, $AC$ and $BC$. For example, with all $G_{2_{S}}$-type gap groups, namely $G_{2_{S}AB}$, $G_{2_{S}BC}$ and $G_{2_{S}AC}$ we have}


\begin{equation}\label{C32G2SI4}
    \begin{aligned}
        & S_{AB}+\sum_{i\subset G_{2_{S}AB}} S_{E_i} \le\sum_{i\subset G_{2_{S}AB}} S_{gaps_i}+S_{ABG_{2_{S}AB}},\\
        & S_{AC}+\sum_{i\subset G_{2_{S}AC}} S_{E_i} \le\sum_{i\subset G_{2_{S}AC}} S_{gaps_i}+S_{ACG_{2_{S}AC}},\\
        & S_{BC}+\sum_{i\subset G_{2_{S}BC}} S_{E_i} \le\sum_{i\subset G_{2_{S}BC}} S_{gaps_i}+S_{BCG_{2_{S}BC}}.\\
    \end{aligned}
\end{equation}

{As the summation of the above three, (\ref{G2SI4}) gives a permutation symmetric bound that covers every single gap defined in (\ref{Gcha_n3d2}). 
Thus, we can obtain five permutation symmetric inequalities(\ding{172}\ding{173}\ding{174}\ding{175}\ding{176}) with regard to the five types of gap groups in (\ref{I4gapgroup}). Each inequality is a combination of restrictions on the area of the minimal RT surface of $AB$, $BC$ and $AC$ along with the $E$s residing in their respective gap groups.} {It should also be noted that, just as the case for (\ref{ExIn}), for each one of $AB$, $BC$ and $AC$, the contribution from the intervals of $E$ outside the respective $G$ is canceled on both sides.}

For $G_{2_{S}}$ we have
\begin{equation}\label{G2SI4}
    \text{\ding{172}} G_{2_{S}}:2\sum_{i \subset g_1} (S_{E_i}-S_{gaps_i})+\sum_{i \subset g_2}(S_{E_i}-S_{gaps_i}) \le \sum_{C_3^2} (S_{ABG_{2_{S}AB}}-S_{AB}),
\end{equation}
where the notation $C_3^2$ on the right-hand side means substituting \(AC\) and \(BC\) for \(AB\) during the summation{, \ie{}, $C_3^2=(AB,AC,BC)$}. It should be noticed, however, that the \(g_1\) and \(g_2\) terms on the left-hand side have different coefficients. Therefore, to obtain the general bound for \(\sum_i S_{E_{i}}-\sum_i S_{gap_i}\), (\ref{G2SI4}) must be paired with {some other inequality chosen from those of the remaining four gap groups in (\ref{I4gapgroup}) given below:} 
\begin{equation}\label{other4I4}
    \begin{aligned}
        &\text{\ding{173}} G_{2_{L}}:2\sum_{i \subset g_1} (S_{E_i}-S_{gaps_i})+3\sum_{i \subset g_2}(S_{E_i}-S_{gaps_i}) \le \sum_{C_3^2} (S_{ABG_{2_{L}AB}}-S_{AB}),\\
        &\text{\ding{174}} g_{1}:2\sum_{i \subset g_1} (S_{E_i}-S_{gaps_i}) \le \sum_{C_3^2} (S_{ABg_{1AB}}-S_{AB}),\\
        &\text{\ding{175}} g_{2_{S}}:\sum_{i \subset g_2}(S_{E_i}-S_{gaps_i}) \le \sum_{C_3^2} (S_{ABg_{2_{S}AB}}-S_{AB}),\\
        &\text{\ding{176}} g_{2_{L}}:3\sum_{i \subset g_2}(S_{E_i}-S_{gaps_i}) \le \sum_{C_3^2} (S_{ABg_{2_{L}AB}}-S_{AB}).\\
    \end{aligned}
\end{equation}
Thus, bounds for \(\sum_i S_{E_{i}}-\sum_i S_{gap_i}\) with correct coefficients can be provided from the following combinations 
\begin{align}\label{combinationI4}
    &\text{\ding{172}}\text{\ding{173}}: \sum (S_{E_i}-S_{gaps_i}) \le \sum_{C_3^2} \Bigl(\frac{1}{4}S_{ABG_{2_{L}AB}}+\frac{1}{4}S_{ABG_{2_{S}AB}}-\frac{1}{2}S_{AB}\Bigr),\notag\\
    &\text{\ding{172}}\text{\ding{175}}: \sum (S_{E_i}-S_{gaps_i}) \le \sum_{C_3^2} \Bigl(\frac{1}{2}S_{ABg_{2_{S}AB}}+\frac{1}{2}S_{ABG_{2_{S}AB}}-S_{AB}\Bigr),\notag\\
    &\text{\ding{172}}\text{\ding{176}}: \sum (S_{E_i}-S_{gaps_i}) \le \sum_{C_3^2} \Bigl(\frac{1}{6}S_{ABg_{2_{L}AB}}+\frac{1}{2}S_{ABG_{2_{S}AB}}-\frac{2}{3}S_{AB}\Bigr),\\
    &\text{\ding{173}}\text{\ding{174}}: \sum (S_{E_i}-S_{gaps_i}) \le \sum_{C_3^2} \Bigl(\frac{1}{6}S_{ABg_{1AB}}+\frac{1}{3}S_{ABG_{2_{L}AB}}-\frac{1}{2}S_{AB}\Bigr),\notag\\
    &\text{\ding{174}}\text{\ding{175}}: \sum (S_{E_i}-S_{gaps_i}) \le \sum_{C_3^2} \Bigl(\frac{1}{2}S_{ABg_{1AB}}+S_{ABg_{2_{S}AB}}-\frac{3}{2}S_{AB}\Bigr),\notag\\
    &\text{\ding{174}}\text{\ding{176}}: \sum (S_{E_i}-S_{gaps_i}) \le \sum_{C_3^2} \Bigl(\frac{1}{2}S_{ABg_{1AB}}+\frac{1}{3}S_{ABg_{2_{L}AB}}-\frac{5}{6}S_{AB}\Bigr).\notag
\end{align}
{The six inequalities above are all of the valid, independent, permutation symmetric constructions from the inequalities ((\ref{G2SI4}) and (\ref{other4I4})) for every possible choice of $G$. This results in six generally different upper bounds, and each of them could be saturated at the critical point of an entanglement phase transition where the fake RT surface substitutes the real one during the process of tuning  $E$.} {At the saturation point, the phase transition represented by the two indexes (\eg\ \ding{172}\ding{173}), \ie\ in the gap region denoted by each index for all $AB$, $BC$ and $AC$, should happen simultaneously\footnote{In these six inequalities, these are the phase transitions between $AB$ and the $E$ in $G_{2_{L}AB}$, $AB$ and the $E$ in $G_{2_{S}AB}$; between $AC$ and the $E$ in $G_{2_{L}AC}$, $AC$ and the $E$ in $G_{2_{S}AC}$; between $BC$ and the $E$ in $G_{2_{L}BC}$, and finally $BC$ and the $E$ in $G_{2_{S}BC}$. }}.
Among the above inequalities, we aim to determine the one that gives the tightest bound on \(I_4\).

Although all the inequalities are correct, the lack of tightness in the other five inequalities indicates that their phase transition conditions cannot be satisfied at the same time. 
To illustrate this more clearly, we can briefly consider the example {of upper-bounding a completely different entropy combination} where{, without loss of generality,} the disconnectivity condition requires \(EW(AE)\) to be precisely connected. 
In this case, when \(A\) undergoes an RT surface phase transition with strips of $E$ in \(g_{2AB}\) and in \(g_{2BC}\) at the same time, {\ie  ~the entanglement wedges of $A$ and the respective $E$ in those gaps become precisely fully connected as transitioned from the fully disconnected phase.} {At that point,} another phase transition between \(A\) and strips in \(g_{2AB}\cup g_{2BC}\) cannot simultaneously occur {because the lack of independence of the first two conditions prevents the third condition from being the simple union of the former two. Restrictions on lengths of certain RT surface segments set by the former two is so strong that satisfying the final phase transition condition for $E$ in \(g_{2AB}\cup g_{2BC}\) is impossible.} However, when we replace \(g_{2BC}\) with \(g_{1A}\), then the three phase transitions concerning \(g_{1A}\), \(g_{2AB}\) and \(g_{2AB}\cup g_{1A}\) can happen {simultaneously, because these gaps, residing respectively inside and outside $A$, present independent phase transition conditions. The phase transition condition of \(g_{2AB}\cup g_{1A}\) is the union of the conditions for \(g_{2AB}\) and \(g_{1A}\), and is naturally satisfied together with these conditions.} Due to the notable complexity of analyzing whether the conditions can be simultaneously met, we can focus on the simpler task of seeking the tightest inequality {using a different method: to compare the difference in all upper bounds utilizing some known inequalities on the entanglement entropy}.

{Let us instead analyze the tightest upper bound from examining the difference between each two of the inequalities in (\ref{combinationI4}) as follows.} First of all, it is obvious that the difference between the RHS of \ding{174}\ding{175} and \ding{172}\ding{175} is simply the sum of conditional mutual information $\sum_{C_3^2} \frac{1}{2}I(g_1:g_{2_{S}AB}|AB)$. A similar observation holds for \ding{174}\ding{176}, whose RHS is larger than that of \ding{173}\ding{174} by $\sum_{C_3^2} \frac{1}{3}I(g_1:g_{2_{L}AB}|AB)$. Due to the non-negativity of CMI as a result of the strong subadditivity, neither \ding{174}\ding{175} nor \ding{174}\ding{176} could be the tightest among the six in (\ref{combinationI4}), {and we do not consider them anymore.}

For the remaining four inequalities, the differences in their respective right-hand side upper bounds are presented below
\begin{align}\label{gapchoice}
    \text{\ding{172}}\text{\ding{173}}-\text{\ding{172}}\text{\ding{175}}&: \sum_{C_3^2} (\frac{1}{4}S_{ABG_{2_{L}AB}}-\frac{1}{4}S_{ABG_{2_{S}AB}}+\frac{1}{2}S_{AB}-\frac{1}{2}S_{ABg_{2_{S}AB}})\notag \\
    & =\sum_{C_3^2}\frac{1}{4}(S_{ABG_{2_{L}AB}}-S_{ABG_{2_{S}AB}})+\sum_{C_3^2}\frac{1}{2}(S_{AB}-S_{ABg_{2_{S}AB}}),\notag \\
    \text{\ding{172}}\text{\ding{173}}-\text{\ding{172}}\text{\ding{176}}&: \sum_{C_3^2} (\frac{1}{4}S_{ABG_{2_{L}AB}}-\frac{1}{4}S_{ABG_{2_{S}AB}}+\frac{1}{6}S_{AB}-\frac{1}{6}S_{ABCg_{2_{L}AB}})\notag\\
    & =\sum_{C_3^2}\frac{1}{4}(S_{ABG_{2_{L}AB}}-S_{ABG_{2_{S}AB}})+\sum_{C_3^2}\frac{1}{6}(S_{AB}-S_{ABg_{2_{L}AB}}),\notag\\
    \text{\ding{172}}\text{\ding{173}}-\text{\ding{173}}\text{\ding{174}}&: \sum_{C_3^2} (\frac{1}{4}S_{ABG_{2_{S}AB}}-\frac{1}{6}S_{ABg_{1AB}}-\frac{1}{12}S_{ABG_{2_{L}AB}})\notag\\
    & =\sum_{C_3^2}\frac{1}{6}(S_{ABG_{2_{S}AB}}-S_{ABg_{1AB}})+\sum_{C_3^2}\frac{1}{12}(S_{ABG_{2_{S}AB}}-S_{ABG_{2_{L}AB}}),\notag\\
    \text{\ding{172}}\text{\ding{175}}-\text{\ding{172}}\text{\ding{176}}&: \sum_{C_3^2} (\frac{1}{2}S_{ABg_{2_{S}AB}}-\frac{1}{3}S_{AB}-\frac{1}{6}S_{ABg_{2_{L}AB}})\\
    & =\sum_{C_3^2}\frac{1}{3}(S_{ABg_{2_{S}AB}}-S_{AB})+\sum_{C_3^2}\frac{1}{6}(S_{ABg_{2_{S}AB}}-S_{ABg_{2_{L}AB}}),\notag\\
    \text{\ding{172}}\text{\ding{175}}-\text{\ding{173}}\text{\ding{174}}&:
    \sum_{C_3^2} (\frac{1}{2}S_{ABg_{2_{S}AB}}-\frac{1}{2}S_{AB}-\frac{1}{6}S_{ABg_{1AB}}+\frac{1}{2}S_{ABG_{2_{S}AB}}-\frac{1}{3}S_{ABG_{2_{L}AB}})\notag\\
    & =\sum_{C_3^2}\frac{1}{2}(S_{ABg_{2_{S}AB}}-S_{AB})+\sum_{C_3^2}\frac{1}{6}(S_{ABG_{2_{S}AB}}-S_{ABg_{1AB}})\notag\\
    &+\sum_{C_3^2}\frac{1}{3}(S_{ABG_{2_{S}AB}}-S_{ABG_{2_{L}AB}}),\notag\\
    \text{\ding{172}}\text{\ding{176}}-\text{\ding{173}}\text{\ding{174}}&:
    \sum_{C_3^2} (\frac{1}{6}S_{ABg_{2_{L}AB}}-\frac{1}{6}S_{AB}-\frac{1}{6}S_{ABg_{1AB}}+\frac{1}{2}S_{ABG_{2_{S}AB}}-\frac{1}{3}S_{ABG_{2_{L}AB}})\notag\\
    & =\sum_{C_3^2}\frac{1}{6}(S_{ABg_{2_{L}AB}}-S_{AB})+\sum_{C_3^2}\frac{1}{6}(S_{ABG_{2_{L}AB}}-S_{ABg_{1AB}})\notag\\
    &+\sum_{C_3^2}\frac{1}{2}(S_{ABG_{2_{S}AB}}-S_{ABG_{2_{L}AB}}).\notag
\end{align}
With no known evident entanglement inequalities available, the signs of these value differences in different bounds can be examined by analyzing the UV divergence.

{In order to simplify the discussion, we can first break the long results into entropy subtraction pairs as presented in (\ref{gapchoice}). There are two kinds of such entropy pairs, and their distinction lies in the property of the non-overlapping subregion between the corresponding boundary subregions for the two entropy terms in each pair. This leads to different UV behaviours.}          
{The first kind refers to the case where the non-overlapping boundary subregion between the pair is simultaneously adjacent to the smaller subregion in the pair and the complement of the larger one. In these cases, one term of the pair in AdS$_3$/CFT$_2$ can be regarded as produced by the other term through translation of endpoints. Thus the boundary subregions of them have the same number of endpoints and the subtraction is not UV divergent. In the second kind the non-overlapping boundary subregion is only adjacent to the smaller subregion in the pair, and the larger term can be seen as obtained by the smaller one absorbing certain gaps. Thus the larger term contain less endpoints and the subtraction has a UV divergence with the same sign as the smaller term.}

{For clearer illustration, without loss of generality, we analyze terms with subscript $AB$ in (\ref{gapchoice}) that originate from the disconnectivity condition of $EW(ABE)$.} First, it should be noted that in AdS$_3$/CFT$_2$, boundary gap subregions $g_{2_{AC}}$ and $g_{2_{BC}}$ are respectively adjacent to $A${ or $B$} on one side and $C$ on the other side. {Consequently, we can consider subregions $ABG_{2_{L}AB}$ and $ABg_{2_{L}AB}$ on the boundary as being formed by the inclusion of $g_{2_{AC}}$ and $g_{2_{BC}}$ from $ABG_{2_{S}AB}$ and $ABg_{2_{S}AB}$.} Then such inclusions can be regarded as the expansion of boundary subregions $A$ and $B$ through translation of endpoints. {This results in no elimination of the existing endpoints or addition of new endpoints and cannot lead to UV divergence in $S_{ABG_{2_{S}AB}}-S_{ABG_{2_{L}AB}}$ and $S_{ABg_{2_{S}AB}}-S_{ABg_{2_{L}AB}}$ as well as in similar terms for $AC$ and $BC$.}

On the other hand, the inclusion/exclusion of gap region $g_{2_{AB}}$ which is adjacent to $A$ and $B$ on both sides can lead to UV divergence in terms $S_{AB}-S_{ABg_{2_{S}AB}}$, $S_{AB}-S_{ABg_{2_{L}AB}}$, $S_{ABG_{2_{S}AB}}-S_{ABg_{1AB}}$, and $S_{ABG_{2_{L}AB}}-S_{ABg_{1AB}}$. In these terms, the larger boundary subregion can be formed by absorbing these gaps through the smaller region, thereby causing disconnected boundary regions to merge,{ and the corresponding endpoints for those gaps vanish. Thus, such terms can contribute to a UV divergence with the same sign as the entropy of the smaller region.} 

Therefore, \ding{172}\ding{173}-\ding{173}\ding{174}, \ding{172}\ding{175}-\ding{172}\ding{176}, \ding{172}\ding{175}-\ding{173}\ding{174} and \ding{172}\ding{176}-\ding{173}\ding{174} have negative UV divergence, while \ding{172}\ding{173}-\ding{172}\ding{175} and \ding{172}\ding{173}-\ding{172}\ding{176} have positive UV divergence, and it can be concluded that \ding{172}\ding{175} gives the tightest bound among the six. 

\begin{figure}[H]
	\centering
	\includegraphics[scale=0.5]{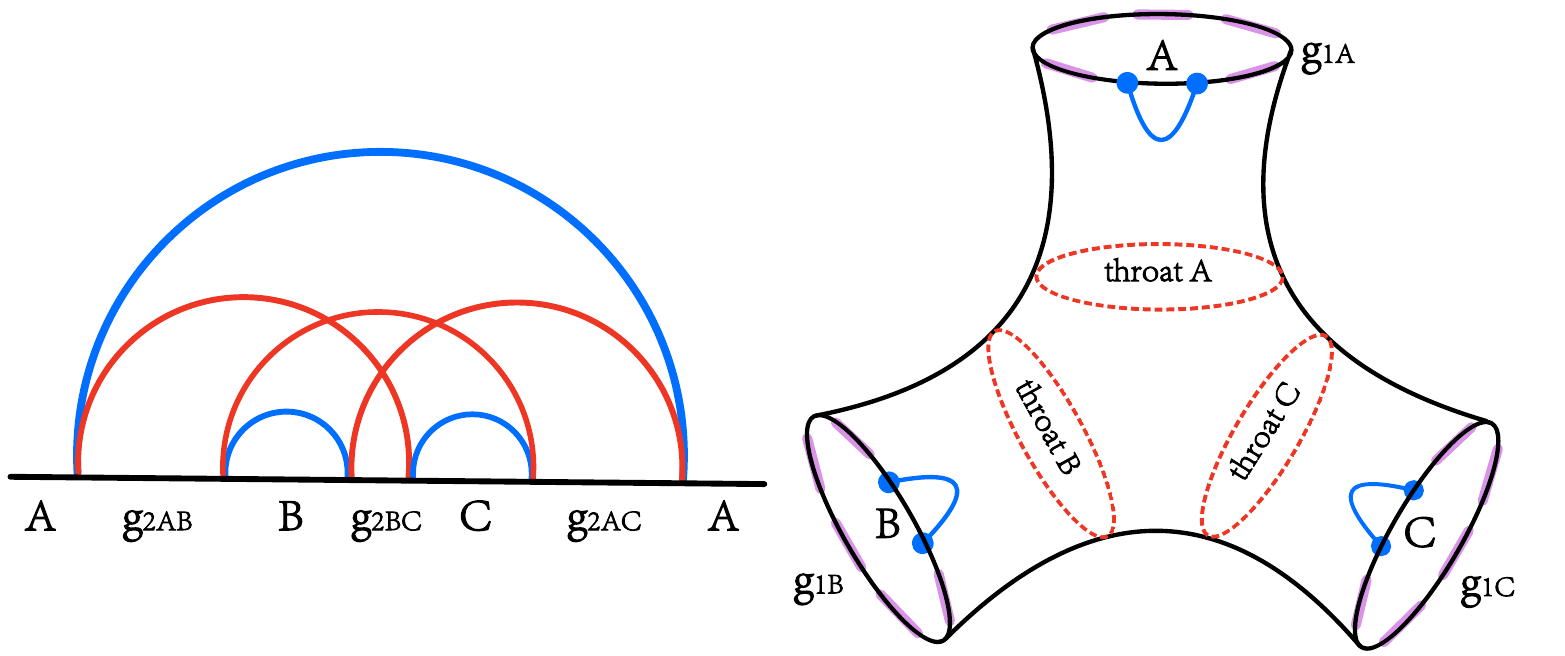}
		\caption{{The upper bound for $I_4(A:B:C:E)$ when $A$, $B$ and $C$ are single boundary intervals in AdS$_3$/CFT$_2$ (left), and when they reside respectively on the three boundaries of a three-mouth wormhole (right). These two examples can be regarded as special cases of our result (\ref{I4generalbound}), where only certain types of gap regions exist. In the first case, the upper bound is given by the difference between the total length of the blue and red curves, while the upper bound is related to the throat area for the second case. Our universal upper bound can successfully unify the two results.}}\label{I4example}
\end{figure}

Finally, substituting \ding{172}\ding{175} into (\ref{I4def}), we can obtain the inequality
\begin{equation}\label{I4generalbound}
    \begin{aligned}   
    I_4 \le & S_{ABC}+\sum_{C_3^2} (\frac{1}{2}S_{ABg_{2_{S}AB}}+\frac{1}{2}S_{ABG_{2_{S}AB}}-S_{AB}), \ie\\
    I_4 \le & S_{ABC}+\frac{1}{2}(S_{ABg_{2AB}}+S_{ACg_{2AC}}+S_{BCg_{2BC}})\\
    &+\frac{1}{2}(S_{ABG_{2_{S}AB}}+S_{ACG_{2_{S}AC}}+S_{BCG_{2_{S}BC}})-(S_{AB}+S_{AC}+S_{BC}),
    \end{aligned}
\end{equation}
which gives the universal upper bound for $I_4$ in AdS$_3$/CFT$_2$.
Now we can show that such a universal upper bound can be reduced to the simple results when $A$ $B$ and $C$ are single intervals in AdS$_3$/CFT$_2$ as discussed in \cite{Ju:2024hba}, as well as when $A$ $B$ and $C$ are single intervals residing respectively on one boundary of the three-mouth wormhole, as shown in Figure  \ref{I4example}. Therefore, we have found a way to unify the upper bounds of $I_4$ in the two cases. 

In the left figure of Figure \ref{I4example}, the case of $A$, $B$ and $C$ being single intervals in AdS$_3$/CFT$_2$ is shown, and only $g_2$ exists. Meanwhile, in the right figure of the three sided wormhole with $A$, $B$ and $C$ being again single intervals, only $g_1$ exists. We can directly use the universal formula in (\ref{I4generalbound}) to evaluate the upper bound in these two specific cases as follows.
In the first case in AdS$_3$/CFT$_2$, we have
\begin{equation}
\begin{aligned}
    I_4(A:B:C:E)&\leq S_{ABC}-S_{AB}-S_{BC}-S_{AC}+S_{ABg_{2AB}}+S_{ACg_{2AC}}+S_{BCg_{2BC}}\\
    &=S_{ABg_{2AB}}+S_{ACg_{2AC}}+S_{BCg_{2BC}}-S_{A}-S_{B}-S_{C}+I_3(A:B:C),
\end{aligned}
\end{equation}
\ie, the length of the red curves minus the length of the blue curves plus a term without UV divergence ($I(A:B:C)$). This is exactly the result presented in \cite{Ju:2024hba}, and one can observe that all UV divergent terms cancel out
so that the upper bound of $I_4$ is a finite term. 

In the three-mouth wormhole case, as only $g_1$ exists, the upper bound of $I_4$ would be
\begin{equation}
    I_4(A:B:C:E)\leq S_{ABC} -\frac{1}{2}(S_{AB}+S_{BC}+S_{AC})+\frac{1}{2}\text{Area}(\text{throat}_{A}+\text{throat}_{B}+\text{throat}_{C}),
\end{equation}
where $S_{Ag_{1A}}=S_{BCg_{1B}g_{1C}}={\text{Area}(\text{throat}_A)}$ (we take $4G_N=1$ for convenience here). Specifically, when $A,B,C$ are small regions so that $EW(ABC)$ is totally disconnected, the upper bound of $I_4$ would simply be half of the summation of three throat areas.

{Now, given that $I_4$ in AdS$_3$/CFT$_2$ has been successfully upper-bounded, we are interested in obtaining a general bound for $I_4$ in higher dimensions.} However, in higher dimensions (\eg, AdS$_4$/CFT$_3$), $g_{3ABC}$ might exist, and this will result in a divergence behavior of the upper bound of $I_4$, which makes it approach its information theoretical upper bound \cite{Ju:2024kuc}. Note that instead of the union of intervals, we choose $E$ to be curved strips in higher dimensional holography.  

We first take the simplest case where only one kind of gap $g_{3ABC}$ exists, {while all other $g_{2...}$ and $g_{1...}$ do not exist},  as an example. Using the disconnectivity condition of $EW(ABE)$ \ie, the disconnected real RT surface has a smaller area compared to the connected fake RT surface, we can get
\begin{equation}
    S_{AB}+\sum S_{strip}\leq S_{ABg_{3ABC}}+\sum S_{gap},
\end{equation}
{It should be noted that, because now there is only one kind of gap, then for $AB$, $AC$ and $BC$, the $G$ where we put $E$ is always $g_{3ABC}$. Moreover, $S_{ABg_{3ABC}}=S_C$, $S_{ACg_{3ABC}}=S_B$, and $S_{BCg_{3ABC}}=S_A$.}
 Using the expression of $I_4$ in (\ref{I4def}), we can finally get the upper bound of $I_4$ in this higher dimensional case
\begin{equation}
    I_4\leq \min(2S_A-I(A:BC),2S_B-I(B:AC),2S_C-I(C:AB)),
\end{equation}
which is UV divergent because the UV divergence of $I(A:BC)$ must be smaller than the UV divergence of $2S_A$. This reveals the fundamental difference of the four-partite entanglement structure in AdS$_3$/CFT$_2$ and higher dimensional holography.

For more complicated cases where $g_2$ and $g_1$ also exist, we can use the previous procedure to list all combinations of choices of the gap regions that we should choose and perform the same calculations to find the tightest upper bound. However, there are so many possible choices and trying to find the tightest one is a very tedious task. Therefore, we present the calculation in Appendix D.

\subsection{Difference of five-partite entanglement structures between AdS$_4$/CFT$_3$ and even higher dimensional holography}

\noindent 
The difference of four-partite entanglement structure between AdS$_3$/CFT$_2$ and higher dimensional holography, though non-trivial, is not very surprising, because AdS$_3$/CFT$_2$ is quite ``special" in that the spatial part of CFT$_2$ is only one-dimensional. That makes us wonder if there in principle exist further differences in more-partite entanglement structure in AdS$_4$/CFT$_3$ and even higher dimensional holography. 

A very famous theorem came to our mind, which is the four-color theorem. It states that on a plane or a spherical world map, no more than four colors are required to color the regions so that no two adjacent regions have the same color. This theorem is, of course, invalid in higher dimensions, because we can simply use thin threads to connect every two {distant} regions so that any two regions among them could be adjacent, and no number of colors less than the number of regions is enough to fill in them. 

{Therefore, with the aid of the four-color theorem, we aim to find the difference in the entanglement structure between AdS$_4$/CFT$_3$ and higher dimensional holography by analyzing the distinctive existence of certain gap regions in the two cases.} Note that the premise of our proof below is that each of \(A,B,C,D\) is a connected region. This condition is also a meaningful condition in quantum information theory since for any spatial partition of connected region \(A\) into \(A_1\) and \(A_2\), \(I(A_1:A_2)\) will always have UV divergence. When \(n=3\), in AdS$_4$/CFT$_3$, it is easy to construct a configuration of connected \(A, B, C\) where all types of gap regions \(g_1, g_2, g_3\) exist, so no difference in the four-partite entanglement structure in AdS$_4$/CFT$_3$ and higher dimensional holography could be found following the procedure in the previous sections.

However, for \(n=4\), there are fifteen categories of gap regions among \(A,B,C,D\) as follows
\begin{equation}\label{Gcha_n4d3}
    \begin{aligned}
        &g_{1A},\,\,\, \,\quad g_{1B},\,\,\,\, \quad g_{1C},\,\,\, \quad g_{1D}, \quad\\
        &g_{2AB}, \quad g_{2AC}, \,\quad g_{2AD}, \quad g_{2BC},\quad g_{2BD}, \quad g_{2CD},\quad\\
        &g_{3ABC},\,\,\, g_{3ABD},\,\,\,\,  g_{3ACD},\,\,\,g_{3BCD}, \\
        &g_{4ABCD}.
    \end{aligned}
\end{equation}
Now we give a simple proof that in AdS$_4$/CFT$_3$, all those six \(g_2\) gap regions and \(g_{4ABCD}\) cannot simultaneously exist using the four-color theorem. More precisely, when \(g_{2AB}\) exists, \(g_{2AB}\) would be the only gap region that entirely separates \(A\) and \(B\) if \(A\) is not adjacent to \(B\), so that \(g_{4ABCD}\) cannot exist. As a result, the existence of \(g_{2AB}\) and \(g_{4ABCD}\) requires that $A$ and $B$ must be adjacent. Due to this fact, if all \(g_2\) gap regions and \(g_{4ABCD}\) exist, any two regions among \(A,B,C,D,g_{4ABCD}\) has to be adjacent to each other. Then we have to use at least five different colors to fill in \(A,B,C,D, g_{4ABCD}\) respectively, which violates the four-color theorem. Similarly, we also have the fact that four \(g_3\) gaps and \(g_{4ABCD}\) cannot simultaneously exist.

This lack of coexistence of certain gap regions in AdS$_4$/CFT$_3$ results in the fact that when we are dealing with the upper bound of entropy combinatoins with \(n=4\) fixed subregions and one tunable region, we should not assume that all gap regions exist {at the same time}, and therefore the formula for the {tightest} upper bound would {depend on the specific configuration of fixed subregions and gap regions that could exist}. However, in even higher dimensional holographic CFTs where all gap regions could exist simultaneously, we could in principle write down more combinations of the inequalities and the upper bound in those configurations could in principle be larger than the case in AdS$_4$/CFT$_3$. 

{Furthermore, if the spatial geometry of the CFT under consideration has non-zero genus \cite{Klebanov:2007ws,Horowitz:2022hlz}—such as a torus—the classical four-color theorem must be replaced by a corresponding higher-color theorem (for example, a 7-color theorem). In principle, this adjustment allows for more combinations of gap regions to coexist, and in a high-genus CFT, the five-partite (or even more-partite) entanglement structure might closely resemble that of its higher-dimensional counterpart. In this sense, the gap classification method could potentially reveal differences in multipartite entanglement structures in CFT\(_3\) with varying spatial topologies.
}

Overall, we can state that we have actually found a difference in the five-partite entanglement structure between AdS$_4$/CFT$_3$ and even higher dimensional holographic CFTs in principle, and there could exist various and more intricate five-partite entanglement structures in higher dimensions that do not exist in AdS$_4$/CFT$_3$.

\section{Several examples of explicit calculations}\label{sec4}

\noindent {In Section \ref{sec3.2}, a symmetric, universal upper bound for $I_4$ has been derived utilising the lamp diagram, the disconnectivity condition, and classification of gap regions. In this section, we aim to use the same formalism to obtain the upper bounds for other explicit examples of entropy combinations. It should be noted that our formalism is actually very effective and can provide the correct upper bounds for a wide range of entropy combinations. Eventually, we can calculate the general upper bounds for families of combinations that have similar structures.} In this section, we will first analyze the case of $n$-partite conditional mutual information ($n$-CMI), a CMI-type combination with only one maximum configuration specified by the disconnectivity condition for each entanglement wedge, in Section \ref{sec4.1}, while in Section \ref{sec4.2}, two $I_4$-type combinations with one local maximum in addition to the one satisfying {the specific} disconnectivity condition are discussed. These combinations can be eventually generalized to arbitrary $n$, resulting in the knowledge of the upper bounds for three families of entropy combinations.  {These examples can reveal the participation of $n$ fixed subsystems in $n+1$ partite entanglement structures, and thus provide a deeper understanding of multipartite entanglement.}
\subsection{$n$-partite conditional mutual information}\label{sec4.1}
\noindent We first focus on the case of the $n$-partite conditional mutual information, which is defined as
\begin{equation}\label{n-CMIdef}
    I_n(A:B:...:N|E)=-(n-1)S(E)+S(AE)+S(BE)+...+S(NE)-S(AB...NE).
\end{equation}
{We will search for the upper bound of $n$-CMI using the method that we developed in the previous two sections, which can be summarized into the following steps. Firstly, we study the lamp diagrams for this family of entropy combinations to obtain the correct disconnectivity condition at the maximum configuration, which also preliminarily simplifies the entropy combinations into terms like $\sum_{i} S_{E_{i}}-\sum_{i} S_{gaps_{i}}$ along with constant terms. Secondly, we identify possible gap intervals where strips of $E$ can reside and provide bounds on a certain part of $\sum_{i} S_{E_{i}}-\sum_{i} S_{gaps_{i}}$ by comparing the true RT surfaces with the respective fake RT surfaces. {Again, we should note that the sum is over all single gap intervals. Therefore, these bounds need to be combined to  build an inequality that precisely bounds $\sum_{i} S_{E_{i}}-\sum_{i} S_{gaps_{i}}$. Eventually, among all the upper bounds given by such combinations, we can }obtain the tightest one, which is the general, universal and symmetric upper bound we want for our family of entropy combinations.}

{Before we begin, it should be checked that $n$-CMI is a well-defined quantity to study.} It satisfies the permutation symmetry, while fulfilling the conditions of $E$ being balanced and single parties $A$, $B$,..., $N$ balancing in all terms that contain $E$.

{The splitting process to search for the maximum} of $n$-CMI is explicitly illustrated in the lamp diagrams. According to the definition in (\ref{n-CMIdef}), these lamp diagrams for $n$-CMI contain only three rows of non-black lamps when all lamps are lit:
a red lamp with multiple $n-1$ on the top row, the second row fully lit in blue and the bottom row consisting of only one red lamp. {This is because other terms of entanglement entropy like $S(ABE)$... do not appear in the expression for $n$-CMI in (\ref{n-CMIdef}). Therefore, the lamps corresponding to these terms, even lit, are colored in black, representing that disconnecting the entanglement wedge in such terms cannot change the current rising/falling trend of $n$-CMI. }
\begin{figure}
    \centering
    \includegraphics[width=0.6\linewidth]{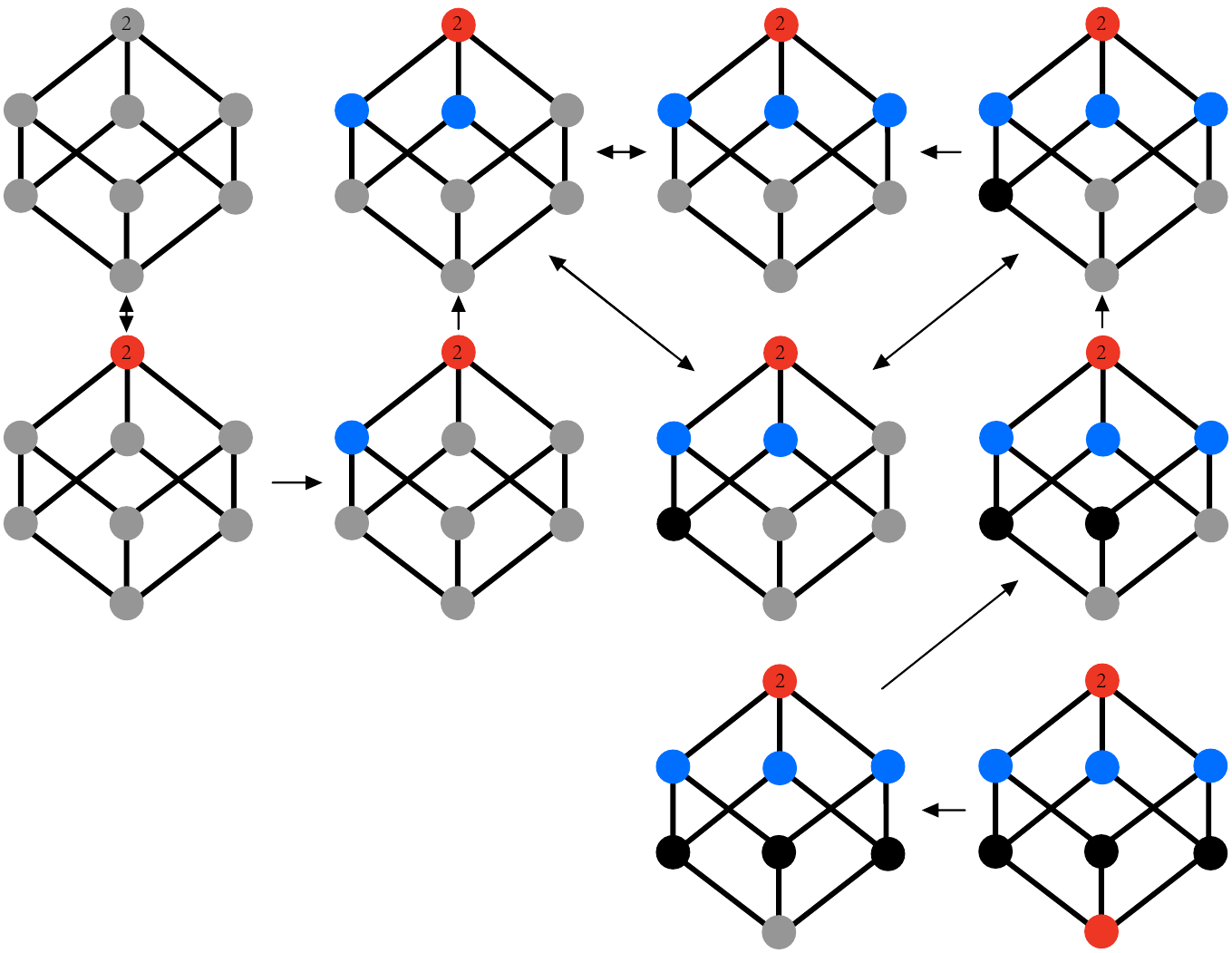}
    \caption{Lamp diagrams {that determine} the disconnectivity conditions for 3-CMI. The number $2$ inside the top lamp in each diagram denotes the coefficient of $S(E)$ in the entropy combination of $3$-CMI. {When we gradually split $E$ from the top left figure where all entanglement wedges containing region $E$ are connected, the maximum value of 3-CMI corresponds to three pictures: the second and third pictures on the first row and the third picture on the second row. However, similar to our discussion for CMI, only the third picture on the first row is unique, {which means that switching among $A$, $B$ and $C$ cannot produce different conditions, thus preventing the difficulty of analyzing cases where for instance, some $E$ intervals are connected with $A$ and $B$ in $AE$ and $BE$, while others are connected with $A$ and $C$ in $AE$ and $CE$... Therefore, we can limit our analyzation to this third picture on the first row.} }}
    \label{3-CMI}
\end{figure}

The maximizing process of $n$-CMI is displayed in Figure  \ref{3-CMI} through an example of $n=3$. First, the coefficient $n-1$ {as well as the minus sign in front of $S(E)$ indicated by the red color} of the lamp for $S(E)$ ensure that during the splitting process, the value of $n$-CMI first increases before all lights on the second row but one are lit. {This is because in this period, the number of red lamps is always not less than that of the blue ones. Now in principle, some of the black lamps below can be illuminated, but this does not contribute to changes in the increasing/decreasing trend of $n$-CMI. On the other hand, the bottom lamp cannot be illuminated unless all $n$ lamps on the second row are lit.} Therefore, further splitting from the $n-1$ blue lamps cannot result in changes of $n$-CMI until the last $n$-th blue lamp on the second row is illuminated. However, with the current blue lamps outnumbering the red lamps by one, any further splitting from this point now leads to the decrease of $n$-CMI.

Therefore, the disconnectivity condition of $n$-CMI is satisfied when the entanglement wedge of $E$ precisely disconnects with that of $n$ single regions: $A$, $B$, ..., $N$, while the entanglement wedge of $AB...NE$ remains connected. As a result,
\begin{equation}
\begin{aligned}
    I_n(A:B:...|E)&\le S(E)-S(AB...NE)+S(A)+S(B)+...+S(N)\\
    &=\sum_{i} S_{E_{i}}-\sum_{i} S_{gaps_{i}}+S(A)+S(B)+...+S(N),
\end{aligned}
\end{equation}
{and now finding the upper bound of $n-$CMI becomes upper-bounding $\sum_{i} S_{E_{i}}-\sum_{i} S_{gaps_{i}}$, similar to the $I_4$ case.}
This can be achieved by first analyzing various choices of gap intervals to place $E$. 

{In order to clarify all the valid bounds, we again classify all possible gap regions in which $E$ could be placed.}
First of all, $g_{1A}$... and $g_{2AB}$... maintain their former definitions as single gap regions adjacent to one and two of the $n$ fixed subregions respectively. Meanwhile, taking various unions of the gap regions into consideration, we first introduce the notation of $G_{2A}$. It is the gap region that contains all the gap intervals adjacent to no more than $2$ parties, one of which is $A$. An example in AdS$_3$/CFT$_2$ when $n=3$ is given by:
\begin{equation}
\begin{aligned}
    &G_{2A}=g_{1A}\cup g_{2{AB}}\cup g_{2{AC}},
\end{aligned}
\end{equation}
and $G_{2B}$, $G_{2C}$... can be similarly defined. Moreover, following the pattern of the $I_4$ case, we also have:
\begin{equation}
\begin{aligned}
    &g_{1A}=g_{1A},\\
    &g_{2A}= g_{2{AB}}\cup g_{2{AC}}.
\end{aligned}
\end{equation}

{Now the upper bounds imposed by the area of fake RT surfaces enclosing a connected entanglement wedge can be calculated {for the intervals of $E$ residing in the above unions of gap regions}.}
We first consider the example of $n=3$ before generalizing the result to arbitrary $n$. The disconnectivity condition $I(A:E)=I(B:E)=I(C:E)=0$ can be applied to $G_{2}$, $g_{1}$ and $g_{2}$. Similar to the $I_4$ case, we have
\begin{equation}
\begin{aligned}
    &G_{2}:\sum_{i\subset g_1} (S_{E_i}-S_{gaps_i})+2\sum_{i\subset g_2}(S_{E_i}-S_{gaps_i}) \le \sum_{C_3^1} (S_{AG_{2A}}-S_{A}),\\
    &g_{1}:\sum_{i\subset g_1} (S_{E_i}-S_{gaps_i}) \le \sum_{C_3^1} (S_{Ag_{1A}}-S_{A}),\\
    &g_{2}:2\sum_{i\subset g_2}(S_{E_i}-S_{gaps_i}) \le \sum_{C_3^1} (S_{Ag_{2A}}-S_{A}),
\end{aligned}
\end{equation}
{where $g_1$ and $g_2$ remain as the union of all $g_{1A}$... and $g_{2AB}$... respectively, and $C_3^1=(A,B,C)$ denotes the summation over the three parties. Again the possible bounds for $\sum_{i} S_{E_{i}}-\sum_{i} S_{gaps_{i}}$ can be obtained through combinations of the inequalities above, so that the terms $\sum_{i\subset g_1} (S_{E_i}-S_{gaps_i})$ and $\sum_{i\subset g_2}(S_{E_i}-S_{gaps_i})$ have identical coefficients. Eventually, there are two such combinations: $g_{1}$ and $G_{2}$, or $g_{1}$ and $g_{2}$.} They give
\begin{equation}\label{g1G2}
    g_{1} G_{2}:\sum (S_{E_i}-S_{gaps_i}) \le \sum_{C_3^1} (\frac{1}{2}S_{Ag_{1A}}+\frac{1}{2}S_{AG_{2A}}-S_{A}),
\end{equation}
and
\begin{equation}\label{g1g2}
    g_{1} g_{2}:\sum (S_{E_i}-S_{gaps_i}) \le \sum_{C_3^1} (S_{Ag_{1A}}+\frac{1}{2}S_{Ag_{2A}}-\frac{3}{2}S_{A}),
\end{equation} respectively. 

Note that the difference in the RHS of (\ref{g1g2}) and (\ref{g1G2}) is the CMI $\sum_{C_3^1} \frac{1}{2}I(g_{1A}:g_{2A}|A)$ again, and the non-negativity of this term results in (\ref{g1G2}) always being the tighter of the two. {The combination $g_{1} g_{2}$ then gives an upper bound which is not tight and we do not consider it any more.} Thus, for 3-CMI, we have eventually derived its upper bound as
\begin{equation}
    I_3(A:B:C|E)\le \frac{1}{2}(S_{Ag_{1A}}+S_{Bg_{1B}}+S_{Cg_{1C}}+S_{AG_{2A}}+S_{BG_{2B}}+S_{CG_{2C}}).
\end{equation}

Furthermore, we have proven that the above discussion remains applicable in the general case of $n$-CMI. Similar to the $I_4$ case, when the disconnectivity condition stipulates that $EW(E)$ must disconnect with $EW(A)$, $EW(B)$,..., the gap regions that we study with a fake RT surface considering strips of $E$ in this gap region should always be closest to $A$, $B$,... In AdS$_3$/CFT$_2$, where gap intervals adjacent to more than two parties cannot exist, the groups under consideration are again $G_2$, $g_1$ and $g_2$, which do not alter with $n$. By simple calculations one can finds that the valid combinations are always $g_{1}G_{2}$ or $g_{1}g_{2}$, with the former consistently tighter by a positive coefficient of CMI. Therefore, we can obtain the general bound with arbitrary $n$, which gives:
\begin{equation}
    I_n(A:B:...:N|E)\le \frac{1}{n-1}\sum_{C_n^1}(S_{AG_{2A}}+(n-2)S_{Ag_{1A}}).
\end{equation}
Therefore, we have successfully derived the general upper bound for a family of CMI-type of entropy combinations, \ie, the $n$-CMI.
\subsection{{More $I_4$ type entropy combinations}}\label{sec4.2}
\noindent {According to our discussion in Section \ref{sec2}, aside from the CMI-type, there also exists another type of entropy combinations whose disconnectivity condition at the maximum configuration is known. These combinations, defined as  $I_4$-type, have two maximums, a local one when no lamps are lit in the lamp diagram, and a true one when all lights but the bottom one are illuminated. The latter displays the picture of all entanglement wedges being disconnected except $EW(ABC...E)$, and gives the correct disconnectivity condition which we utilise to generate upper bounds by introducing fake RT surfaces.} In this subsection, we attempt to find other sets of such $I_4$-type entropy combinations{, and give a general upper bound for two of those series of combinations}.

Before further discussion, we can constrain several {properties} of the desired combinations to narrow down the search, and establish a general pattern of the corresponding lamp diagrams\footnote{Again, all the restrictions on lamp diagrams refer to the case where every lamp has been lit.}. First of all, the permutation symmetry requires all lamps on the same row to have identical colors. These terms should also have the same coefficients in the entropy combinations.

Meanwhile, in order to simplify our search, we limit our consideration to diagrams where only the lamps on four rows are not black. This restriction is not necessary, but given that our goal is merely to find feasible examples, it can be imposed deliberately for convenience. Therefore, for $I_4$-type combinations, according to the disconnectivity theorem (Theorem 2.2), the illuminated rows are the first one (corresponding to $EW(E)$), the last two rows (representing respectively $EW(\underbrace{ABC...}_{n-1\ \text{parties}}E )$ and $EW(\underbrace{ABC...N}_{n\ \text{parties}}E )$), and a randomly selected row distinct from the other three. Then, we can denote the coefficients of terms that belong to these four illuminated rows as $x$, $z$, $w$ and $y$, respectively ($x, y, z, w \in \mathbb{Z}$). In addition, $n$ remains the number of the total fixed parties, and the random row is denoted as the $(k+1)$-th ($1 \le k \le n-2$ and $k,n \in \mathbb{Z}$), in which every entropy term contains $E$, together with $k$ of the fixed parties.

According to Theorem 2.2, the combination value during splitting should first experience a decrease from the all-connected local maximum. This requires that $S(E)$, as the first non-black row is lit in blue ($x\ge0$), so that its contribution during splitting is negative. Then the value must first increase before reaching the global maximum where all rows are lit except the bottom one. It can be seen that this condition can only be satisfied when the random row is lit in red while the second-to-last row is blue ($y\le0$, $z\ge0$). From there, the global maximum demands two conditions. First, the combination value increases when all the lights but one on the second-to-last row is lit, which gives:
\begin{equation}\label{4.2cond1}
        x+C_{n}^{k} y + (C_{n}^{n-1} -1)z\le 0.
\end{equation}
Meanwhile, from the global maximum, splitting $E$ should decrease the combination value. This demands more blue lights illuminated than red ones at the maximum. Combined with the balance requirements which will be discussed below, this is equivalent to requiring that the bottom light, when lit, is red ($w\le0$).
    
In addition, the balance requirements can also be explicitly written as follows. To begin with, $E$ is balanced, \ie, the number of blue and red lamps are the same when all lamps are lit. Therefore, 
    \begin{equation}\label{4.2cond2}
        x+C_{n}^{k} y+C_{n}^{n-1} z+w=0.
    \end{equation}
Moreover, $A$, $B$, $C$ and $D$ are balanced in all terms that contain region $E$. Similarly, this requires:
    \begin{equation}\label{4.2cond3}
        C_{n-1}^{k-1} y+C_{n-1}^{n-2} z+w=0
    \end{equation}
    from counting the corresponding red and blue lamps.

As a result, conditions (\ref{4.2cond1}), (\ref{4.2cond2}) and (\ref{4.2cond3}) have provided various solutions, from which we are particularly interested in two sets of combinations:
\begin{equation}\label{5211multiple}
        k=1; x=n^{2}-3n+1; y=2-n; z=1; w=-1,
    \end{equation}
and 
\begin{equation}
        k=2; x=\frac{1}{2}(n-1)(n^2-4n+2); y=2-n; z=n-1; w=1-n.
    \end{equation}
    
It should be noted that (\ref{4.2cond1}) is a necessary but not sufficient condition for the disconnectivity condition, because it cannot guarantee the absence of additional local maximum other than the desired one and the trivial one. Detailed proof that eliminates such possibilities for the two sets above is provided below when we analyze the splitting process.

\begin{figure}
    \centering
    \includegraphics[width=0.6\linewidth]{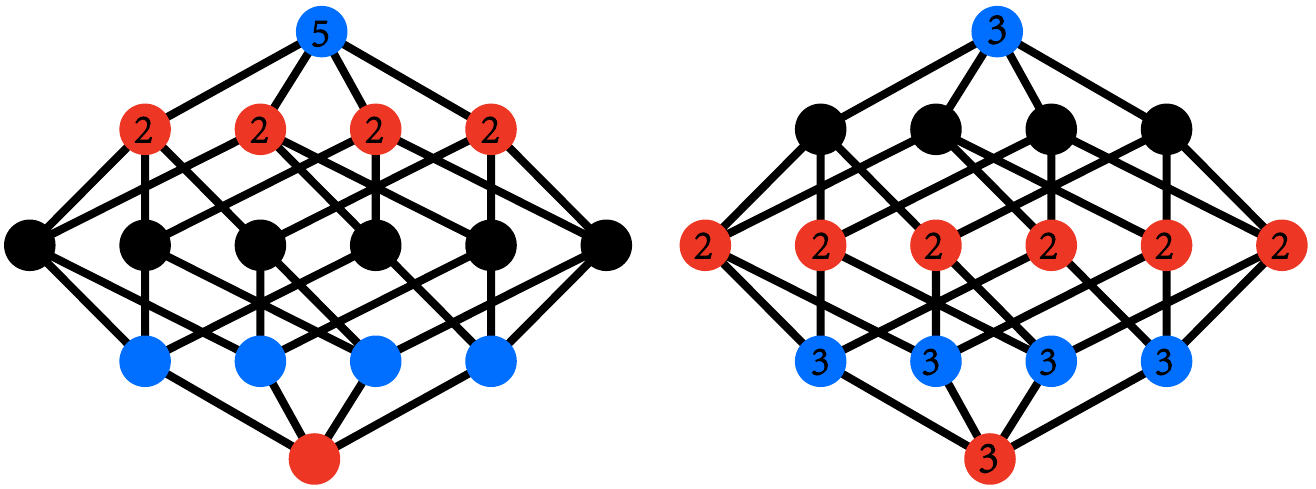}
    \caption{Fully lit lamp diagrams for $n=4, k=1$ and $n=4, k=2$. After the decreasing phase, the number of the blue lamps is always smaller than that of the red lamps until all lights but the bottom one are lit.}
    \label{new combinations}
\end{figure}

In both cases, the corresponding lamp diagrams, when fully lit, include a blue lamp on the top, a lower row of red lamps, the second-to-last row lit in blue and a red lamp at the bottom. Examples corresponding to both cases for $n=4$ are displayed in Figure \ref{new combinations}. Other lamps in the diagrams are black. Therefore, during the splitting process, the values of these combinations first decrease from the trivial maximum before rising to a global maximum corresponding to the complete illumination of the second-to-last row, and drops again if the bottom lamp is to be lit. To exclude other maximums, we must try to light as many lamps on the second-to-last row as possible, and check if this leads to a premature decline of the combination value.

When $k=1$, illuminating one light on the second-to-last row requires at least $n-1$ red lamps on the $k=1$ row to be lit in advance. The difference of the currently lit blue and red lamps is
\begin{equation}
    n^{2}-3n+1+(2-n)(n-1)+1=0,
\end{equation}
which does not affect the general downward or upward trend and cannot introduce additional maximums. Subsequently, illuminating another lamp on the second-to-last row demands that all lights on the $k=1$ row are lit, and the increasing trend still dominates before the disconnectivity condition is satisfied.

Similarly, when $k=2$, lighting one lamp on the second-to-last row indicates that at least $C_{n-1}^{2}$ lamps on the $k=2$ row are lit, and the number of blue lamps again equals that of the red lamps. To light another lamp on the second-to-last row calls for at least $C_{n-1}^{2}+C_{n-2}^{1}$ lamps illuminated on the $k=2$ row, and the combination value rises. Finally, when three lamps on the second-to-last row are lit, all lamps on the $k=2$ row should be illuminated, the overall difference of blue and red lamps no larger than zero. Therefore, we can conclude that both combinations for arbitrary $n$ reach the maximum when $E$ disconnects with every $(n-1)$-party-subsystem while connecting with  the $n$-party-subsystem. Thus, we have attained {two sets of entropy combinations with known disconnectivity condition.}

Now, utilising the above disconnectivity condition, we move on to capture the form of the corresponding upper bounds. We first consider the case where $n=4$ and $k=1$. The entropy combination that is studied is
\begin{equation}
        C_{HEE}=5S_E-2S_{AE}-2S_{BE}-2S_{CE}-2S_{DE}
        +S_{ABCE}+S_{ABDE}+S_{ACDE}+S_{BCDE}-S_{ABCDE}.
\end{equation}
Due to the disconnectivity condition, 
\begin{equation}\label{halfway ineq}
    \begin{aligned}
        C_{HEE}&\le S_E-S_{ABCDE}-2(S_A+S_B+S_C+S_D) +(S_{ABC}+S_{ABD}+S_{ACD}+S_{BCD})\\
          &= \sum_{i} S_{E_i}-\sum_{i} S_{gaps_i}-2(S_A+S_B+S_C+S_D) +(S_{ABC}+S_{ABD}+S_{ACD}+S_{BCD}).    
\end{aligned}
\end{equation}

Similar to the $I_4$ example, the gap unions where we can place $E$ are defined below (taking $ABC$ as an example):
\begin{equation}
\begin{aligned}
    &G_{2_{S}ABC}=g_{1A}\cup g_{1B}\cup g_{1C}\cup g_{2{AB}}\cup g_{2{AC}}\cup g_{2{BC}}\\
    &G_{2_{L}ABC}=g_{1A}\cup g_{1B}\cup g_{1C}\cup g_{2{AB}}\cup g_{2{AC}}\cup g_{2{BC}}\cup g_{2{AD}}\cup g_{2{BD}}\cup g_{2{CD}}\\
    &g_{1ABC}= g_{1{A}}\cup g_{1{B}}\cup g_{1{C}}\\
    &g_{2_{S}ABC}= g_{2{AB}}\cup g_{2{AC}}\cup g_{2{BC}}\\
    &g_{2_{L}ABC}= g_{2{AB}}\cup g_{2{AC}}\cup g_{2{BC}}\cup g_{2{AD}}\cup g_{2{BD}}\cup g_{2{CD}},
\end{aligned}
\end{equation}
with the definition of all the $g_{1...}$s and $g_{2...}$s unchanged. Again, we obtain the following upper bounds by introducing fake RT surfaces:
\begin{align}
    &\text{\ding{172}} G_{2_{S}}: 3\sum_{g_1} (S_{E_i}-S_{gaps_i})+2\sum_{g_2}(S_{E_i}-S_{gaps_i}) \le \sum_{C_4^3} (S_{ABCG_{2_{S}ABC}}-S_{ABC}),\notag\\
    &\text{\ding{173}} G_{2_{L}}: 3\sum_{g_1} (S_{E_i}-S_{gaps_i})+4\sum_{g_2}(S_{E_i}-S_{gaps_i}) \le \sum_{C_4^3} (S_{ABCG_{2_{L}ABC}}-S_{ABC}),\notag\\
    &\text{\ding{174}} g_{1}: 3\sum_{g_1} (S_{E_i}-S_{gaps_i}) \le \sum_{C_4^3} (S_{ABCg_{1ABC}}-S_{ABC}),\\
    &\text{\ding{175}} g_{2_{S}}: 2\sum_{g_2} (S_{E_i}-S_{gaps_i}) \le \sum_{C_4^3} (S_{ABCg_{2_{S}ABC}}-S_{ABC}),\notag\\
    &\text{\ding{175}} g_{2_{L}}: 4\sum_{g_2} (S_{E_i}-S_{gaps_i}) \le \sum_{C_4^3} (S_{ABCg_{2_{L}ABC}}-S_{ABC}).\notag
\end{align}
Now the sum over $C_4^3$ denotes the addition of terms corresponding to regions $ABC$, $ABD$, $ACD$ and $BCD$, \ie{} $C_4^3=(ABC,ABD,ACD,BCD)$. Combining them gives
\begin{equation}
\begin{aligned}\label{gapchoice2}
    &\text{\ding{172}}\text{\ding{173}}: \sum (S_{E_i}-S_{gaps_i}) \le \sum_{C_4^3} (\frac{1}{6}S_{ABCG_{2_{L}ABC}}+\frac{1}{6}S_{ABCG_{2_{S}ABC}}-\frac{1}{3}S_{ABC}),\\
    &\text{\ding{172}}\text{\ding{175}}: \sum (S_{E_i}-S_{gaps_i}) \le \sum_{C_4^3} (\frac{1}{6}S_{ABCg_{2_{S}ABC}}+\frac{1}{3}S_{ABCG_{2_{S}ABC}}-\frac{1}{2}S_{ABC}),\\
    &\text{\ding{172}}\text{\ding{176}}: \sum (S_{E_i}-S_{gaps_i}) \le \sum_{C_4^3} (\frac{1}{12}S_{ABCg_{2_{L}ABC}}+\frac{1}{3}S_{ABCG_{2_{S}ABC}}-\frac{5}{12}S_{ABC}),\\
    &\text{\ding{173}}\text{\ding{174}}: \sum (S_{E_i}-S_{gaps_i}) \le \sum_{C_4^3} (\frac{1}{12}S_{ABCg_{1ABC}}+\frac{1}{4}S_{ABCG_{2_{L}ABC}}-\frac{1}{3}S_{ABC}),\\
    &\text{\ding{174}}\text{\ding{175}}: \sum (S_{E_i}-S_{gaps_i}) \le \sum_{C_4^3} (\frac{1}{3}S_{ABCg_{1ABC}}+\frac{1}{2}S_{ABCg_{2_{S}ABC}}-\frac{5}{6}S_{ABC}),\\
    &\text{\ding{174}}\text{\ding{176}}: \sum (S_{E_i}-S_{gaps_i}) \le \sum_{C_4^3} (\frac{1}{3}S_{ABCg_{1ABC}}+\frac{1}{4}S_{ABCg_{2_{L}ABC}}-\frac{7}{12}S_{ABC}).\\
\end{aligned}
\end{equation}
Once again, the RHS of \ding{174}\ding{176} and \ding{174}\ding{175} are respectively larger than  \ding{173}\ding{174} and \ding{172}\ding{175} by $\sum_{C_4^3} \frac{1}{4}I(g_1:g_{2_{L}ABC}|ABC)$ and $\sum_{C_4^3} \frac{1}{3}I(g_1:g_{2_{S}ABC}|ABC)$, and they fail to provide the tightest bound.

The differences between the remaining four bounds are:
\begin{align}\label{gapchoice3}
    \text{\ding{172}}\text{\ding{173}}-\text{\ding{172}}\text{\ding{175}}&: \sum_{C_4^3} (\frac{1}{6}S_{ABCG_{2_{L}ABC}}-\frac{1}{6}S_{ABCG_{2_{S}ABC}}+\frac{1}{6}S_{ABC}-\frac{1}{6}S_{ABCg_{2_{S}ABC}})\notag\\
    & =\sum_{C_4^3}\frac{1}{6}(S_{ABCG_{2_{L}ABC}}-S_{ABCG_{2_{S}ABC}})+\sum_{C_4^3}\frac{1}{6}(S_{ABC}-S_{ABCg_{2_{S}ABC}})\notag\\
    \text{\ding{172}}\text{\ding{173}}-\text{\ding{172}}\text{\ding{176}}&: \sum_{C_4^3} (\frac{1}{6}S_{ABCG_{2_{L}ABC}}-\frac{1}{6}S_{ABCG_{2_{S}ABC}}+\frac{1}{12}S_{ABC}-\frac{1}{12}S_{ABCg_{2_{L}ABC}})\notag\\
    & =\sum_{C_4^3}\frac{1}{6}(S_{ABCG_{2_{L}ABC}}-S_{ABCG_{2_{S}ABC}})+\sum_{C_4^3}\frac{1}{12}(S_{ABC}-S_{ABCg_{2_{L}ABC}})\notag\\
    \text{\ding{172}}\text{\ding{173}}-\text{\ding{173}}\text{\ding{174}}&: \sum_{C_4^3} (\frac{1}{6}S_{ABCG_{2_{S}ABC}}-\frac{1}{12}S_{ABCg_{1ABC}}-\frac{1}{12}S_{ABCG_{2_{L}ABC}})\notag\\
    & =\sum_{C_4^3}\frac{1}{12}(S_{ABCG_{2_{S}ABC}}-S_{ABCg_{1ABC}})+\sum_{C_4^3}\frac{1}{12}(S_{ABCG_{2_{S}ABC}}-S_{ABCG_{2_{L}ABC}})\notag\\
    \text{\ding{172}}\text{\ding{175}}-\text{\ding{172}}\text{\ding{176}}&: \sum_{C_4^3} (\frac{1}{6}S_{ABCg_{2_{S}ABC}}-\frac{1}{12}S_{ABC}-\frac{1}{12}S_{ABCg_{2_{L}ABC}})\\
    & =\sum_{C_4^3}\frac{1}{12}(S_{ABCg_{2_{S}ABC}}-S_{ABC})+\sum_{C_4^3}\frac{1}{12}(S_{ABCg_{2_{S}ABC}}-S_{ABCg_{2_{L}ABC}})\notag\\
    \text{\ding{172}}\text{\ding{175}}-\text{\ding{173}}\text{\ding{174}}&:
    \sum_{C_4^3} (\frac{1}{6}S_{ABCg_{2_{S}ABC}}-\frac{1}{6}S_{ABC}-\frac{1}{12}S_{ABCg_{1ABC}}+\frac{1}{3}S_{ABCG_{2_{S}ABC}}-\frac{1}{4}S_{ABCG_{2_{L}ABC}})\notag\\
    & =\sum_{C_4^3}\frac{1}{6}(S_{ABCg_{2_{S}ABC}}-S_{ABC})+\sum_{C_4^3}\frac{1}{12}(S_{ABCG_{2_{S}ABC}}-S_{ABCg_{1ABC}})\notag\\
    &+\sum_{C_4^3}\frac{1}{4}(S_{ABCG_{2_{S}ABC}}-S_{ABCG_{2_{L}ABC}})\notag\\
    \text{\ding{172}}\text{\ding{176}}-\text{\ding{173}}\text{\ding{174}}&:
    \sum_{C_4^3} (\frac{1}{12}S_{ABCg_{2_{L}ABC}}-\frac{1}{12}S_{ABC}-\frac{1}{12}S_{ABCg_{1ABC}}+\frac{1}{3}S_{ABCG_{2_{S}ABC}}-\frac{1}{4}S_{ABCG_{2_{L}ABC}})\notag\\
    & =\sum_{C_4^3}\frac{1}{12}(S_{ABCg_{2_{L}ABC}}-S_{ABC})+\sum_{C_4^3}\frac{1}{12}(S_{ABCG_{2_{L}ABC}}-S_{ABCg_{1ABC}})\notag\\
    &+\sum_{C_4^3}\frac{1}{3}(S_{ABCG_{2_{S}ABC}}-S_{ABCG_{2_{L}ABC}}).\notag
\end{align}
Here the divergence behaviors of $S_{ABCG_{2_{S}ABC}}-S_{ABCG_{2_{L}ABC}}$ and $S_{ABCg_{2_{S}ABC}}-S_{ABCg_{2_{L}ABC}}$ as well as $S_{ABC}-S_{ABCg_{2_{S}ABC}}$, $S_{ABC}-S_{ABCg_{2_{L}ABC}}$, $S_{ABCG_{2_{S}ABC}}-S_{ABCg_{1ABC}}$, and $S_{ABCG_{2_{L}ABC}}-S_{ABCg_{1ABC}}$ are similar to the case of $I_4$. In AdS$_3$/CFT$_2$, the terms in each of the former two pairs have the same number of endpoints and the pairs display no divergence, while in the rest pairs, the smaller term has more endpoints and  therefore contributes to UV divergence. \ding{172}\ding{175} is still the tightest bound among the six, and we obtain the inequality
\begin{equation}
    \begin{aligned}   &5S_E-2S_{AE}-2S_{BE}-2S_{CE}-2S_{DE}+S_{ABCE}+S_{ABDE}+S_{ACDE}+S_{BCDE}-S_{ABCDE} \\
    &\le \frac{1}{6}(S_{ABCg_{2_{S}ABC}}+S_{ABDg_{2_{S}ABD}}+S_{ACDg_{2_{S}ACD}}+S_{BCDg_{2_{S}BCD}})\\
    &+\frac{1}{3}(S_{ABCG_{2_{S}ABC}}+S_{ABDG_{2_{S}ABD}}+S_{ACDG_{2_{S}ACD}}+S_{BCDG_{2_{S}BCD}})\\
    &+\frac{1}{2}(S_{ABC}+S_{ABD}+S_{ACD}+S_{BCD})-2(S_{A}+S_{B}+S_{C}+S_{D}).
    \end{aligned}
\end{equation}

The above discussion can then be generalized to the entropy combination of {arbitrary} $n$ with $n\ge 3$, $k=1$, where the UV divergent features do not change, and \ding{172}\ding{175} remains the tightest bound. Here we denote $AB...$ as the subregion that consists of $n-1$ parties and does not include the $n$-th party, $N$. These general {upper bounds are} 
\begin{equation}\label{5211general}
    \begin{aligned}   
    &(n^2-3n+1)S_E-(n-2)\sum_{C_n^1}S_{AE}+\sum_{C_n^{n-1}}S_{AB...E}-S_{AB...NE}\le \\
    &\frac{1}{(n-1)(n-2)}\sum_{C_n^{n-1}}S_{AB...g_{2_{S}AB...}}+\frac{1}{n-1} \sum_{C_n^{n-1}} S_{AB...G_{2_{S}AB...}}\\
    &+\frac{n-3}{n-2} \sum_{C_n^{n-1}} S_{AB...}-(n-2)\sum_{C_n^1}S_{A},
    \end{aligned}
\end{equation}
{where the first line is the general entropy combination with an arbitrary $n$ parameter that we consider.} One can notice that $I_4-I_3(A:B:C)$ is a special case of (\ref{5211multiple}) with $n=3$ and the upper bound of $I_4$ is consistent with (\ref{5211general}).

Similarly, we can derive the set of upper bounds for the $k=2$ case. When $n=4$, using the above method, we can prove that
\begin{equation}
    \begin{aligned}   &3S_E-2S_{ABE}-2S_{ACE}-2S_{ADE}-2S_{BCE}-2S_{BDE}-2S_{CDE}\\
    &+3S_{ABCE}+3S_{ABDE}+3S_{ACDE}+3S_{BCDE}-3S_{ABCDE} \\
    &\le \frac{1}{2}(S_{ABCg_{2_{S}ABC}}+S_{ABDg_{2_{S}ABD}}+S_{ACDg_{2_{S}ACD}}+S_{BCDg_{2_{S}BCD}})\\
    &+(S_{ABCG_{2_{S}ABC}}+S_{ABDG_{2_{S}ABD}}+S_{ACDG_{2_{S}ACD}}+S_{BCDG_{2_{S}BCD}})\\
    &+\frac{3}{2}(S_{ABC}+S_{ABD}+S_{ACD}+S_{BCD})-2(S_{AB}+S_{AC}+S_{AD}+S_{BC}+S_{BD}+S_{CD}).
    \end{aligned}
\end{equation}
{Again, this upper bound can be generalized to the case of arbitrary $n$ with $k=2$:}
\begin{equation}
    \begin{aligned}   
    &\frac{1}{2}(n-1)(n^2-4n+2)\sum_{C_n^{n-1}} S_E-(n-2)\sum_{C_n^2} S_{ABE}+(n-1)\sum_{C_n^{n-1}} S_{AB...E}\\
    &-(n-1)S_{AB...NE} \le \frac{1}{(n-2)}\sum_{C_n^{n-1}}S_{AB...g_{2_{S}AB...}}+\sum_{C_n^{n-1}} S_{AB...G_{2_{S}AB...}}\\
    &+\frac{(n-3)(n-1)}{n-2} \sum_{C_n^{n-1}} S_{AB...}-(n-2)\sum_{C_n^2}S_{AB},
    \end{aligned}
\end{equation}
{where the general form of the set of entropy combinations is given on the left-hand side.}

\section{Conclusion and discussion}\label{sec5}

\noindent In this paper, we build a formalism to evaluate the upper bounds of large classes of holographic entropy combinations, in which \(n\) subsystems are fixed and one additional subsystem is arbitrarily chosen. By tuning the arbitrary subsystem, we obtain a series of upper bounds for entropy combinations (e.g., \(n\)-partite conditional mutual information, etc.). The specific procedure is to first obtain the (dis)connectivity conditions of upper bound configurations through lamp diagrams and then give the upper bounds utilizing fake RT surfaces and classification of gap regions.

The upper bound for many entropy combinations derived in our work depends on the dimension of the holographic theory, a feature that originates from our gap region classification procedure. In AdS\(_3\)/CFT\(_2\), no \(g_3\) type of gap regions exist because every interval has only two endpoints. In AdS\(_4\)/CFT\(_3\), however, certain combinations of gap regions for \(n\geq 4\) are absent due to the four-color theorem. These observations result in differences in the entanglement structure across different dimensions. Specifically, the absence of divergence in \(I_4\) in AdS\(_3\)/CFT\(_2\) implies a lack of four-partite global entanglement involving three arbitrarily fixed regions. For five-partite entanglement—where four fixed and self-connected regions participate—the differences between AdS\(_4\)/CFT\(_3\) and higher-dimensional theories are more complex and subtle, but in principle, different upper bounds can be obtained. Thus, our results suggest that higher-dimensional theories possess more intricate entanglement structures than lower-dimensional ones.

The procedures are highly general, independent of the geometry or topology of the AdS bulk. Moreover, our upper bounds could be verified to remain valid in the HRT formalism \cite{Hubeny:2007xt}. In our prescription, we alter the connectivity of only a single entanglement wedge during the splitting process (as detailed in Appendix A), which allows us to place all HRT surfaces for that wedge on a single bulk time slice. Consequently, the (dis)connectivity conditions and the resulting upper bounds are both remain unchanged in HRT formalism. By contrast, while the holographic entropy cone program has made significant progress in generalizing results to the HRT formalism \cite{Grado-White:2025jci,Grado-White:2024gtx,Czech:2019lps}, a rigorous proof of all holographic entropy inequalities in a general bulk spacetime without any topological or geometrical restrictions remains an open question.

Several open questions remain. First, in principle, we could build holographic entropy inequalities from our upper bound results. However, there is subtle difference between our upper bounds and the standard holographic entropy inequalities. For example, the upper bounds for CMI and \(I_4\) in AdS\(_3\)/CFT\(_2\) can be easily transformed into unbalanced entropy inequalities\footnote{The unbalanced nature arises because the upper bound of a balanced entropy combination may exhibit divergent behavior, \ie, the upper bound itself is an unbalanced entropy combination.}. However, a general method to perform this transformation in more complicated cases is still lacking. Second, an important generalization is to develop a formalism to evaluate the upper bounds of combinations of holographic entanglement entropy with \(n\) regions fixed and \(m\) regions fine-tuned, in order to investigate the \((n+m)\)-partite entanglement structures in which the fixed \(n\) regions participate, which we will report in a future work. Additionally, the upper bounds of multipartite entanglement measures beyond entropy combinations—such as multi-entropy \cite{Gadde:2022cqi}, EWCS \cite{Takayanagi:2017knl}, and multi-EWCS \cite{Umemoto:2018jpc,Bao:2018fso} with modified IR geometry \cite{Ju:2024xcn} are also of significant interest, and we leave further discussion of these topics to future work.

\section*{Acknowledgement}

\noindent We would like to thank Bart{\l}omiej Czech, Joydeep Naskar and Diandian Wang for their valuable discussions. This work was supported by Project 12035016 supported by the National Natural Science Foundation of China.

\appendix

\section{Proof of CMI type disconnectivity condition}
\noindent In this appendix we prove the (dis)connectivity condition~(\ref{I3I4dis}) for conditional mutual information via a splitting procedure, and recast it in the language of lamp diagrams.

\begin{figure}[H]
	\centering
	\includegraphics[scale=0.6]{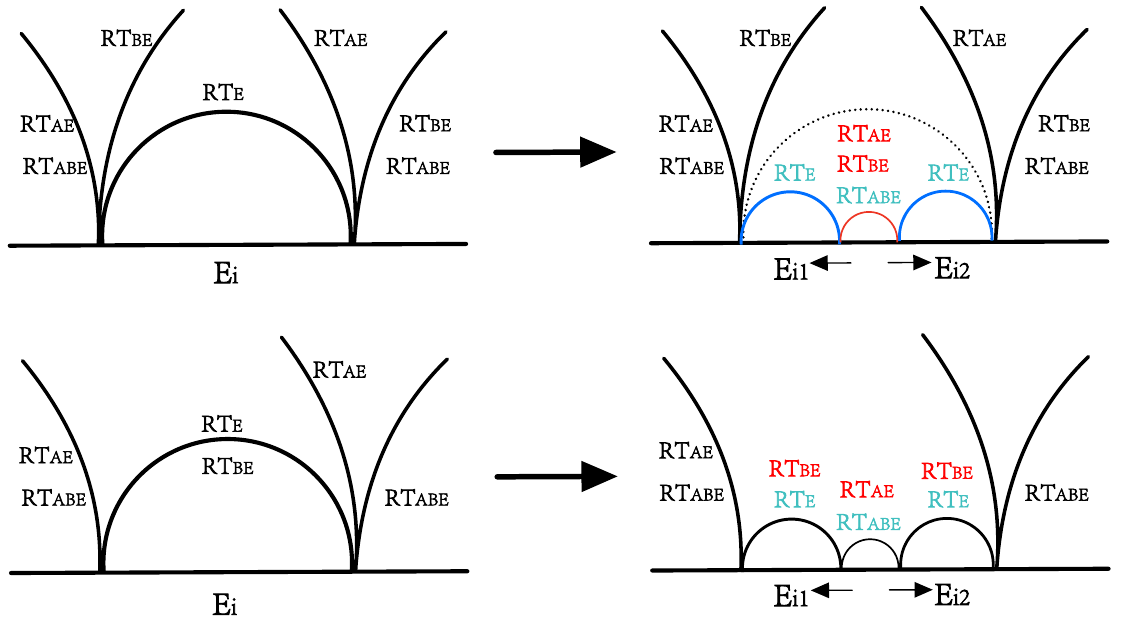}
	\caption{The process of deleting the gap regions from the ``bad" subregion in the calculation of the upper bound of \(I(A:B|E)\). Red (blue) geodesics indicate positive (negative) contributions in \(I(A:B|E)\). The figure is taken from \cite{Ju:2024hba}.}\label{DISCMI}
\end{figure}

As emphasized in the paper, the key step is to split the ``bad'' interval into three pieces—a new gap in the middle and two subregions included in $E'$—and test whether replacing $E$ by $E'$ increases the entropy combination. The detailed steps are shown in Figure \ref{DISCMI}. In subfigure (1.1)\footnote{Throughout this paper, $(p.q)$ denotes the $q$-th subfigure on the $p$-th row.}, the ``bad" subregion \(E_i\) is connected with both \(A\) in \(EW(AE)\) and \(B\) in \(EW(BE)\) while it should be disconnected in both of them according to the (dis)connectivity condition (\ref{I3I4dis}). We first choose a midpoint of the gap region that we are about to delete in \(E_i\), then enlarge this gap region from zero size while preserving the midpoint. During this enlarging process, \(I(A:B|E)\) never changes because all areas of the RT surfaces of the gap region cancel out, until the first phase transition occurs in \(EW(E)\) where \(E_{i1}\) and \(E_{i2}\) become disconnected. 

When we continue to enlarge the gap region, \(S_E\) will decrease while all other entanglement entropies that contain region \(E\) will increase, as shown in the subfigure (1.2). Since \(S_E\) has a negative sign in \(I(A:B|E)\), continually enlarging the gap region will cause \(I(A:B|E)\) to increase, until the next phase transition occurs. In this phase transition, at least one of \(A\) or \(B\) can become disconnected from one of \(E_{i1}\) or \(E_{i2}\) in \(EW(AE)\) or \(EW(BE)\). Without loss of generality, let us assume it is \(B\) that becomes disconnected from \(E_{i1}\). However, we want to have both \(E_{i1}\) and \(E_{i2}\) simultaneously disconnected from \(B\) in \(EW(BE)\). Fortunately, this can always be achieved by fine-tuning the midpoint of the gap region. One can imagine that if the midpoint of the gap region is far to the left so that \(E_{i1}\) is very small, \(B\) must first become disconnected from \(E_{i1}\) in \(EW(BE)\); otherwise, if it is far to the right, \(B\) must first become disconnected from \(E_{i2}\) in \(EW(BE)\). Thus, by the intermediate value theorem, there must exist a fine-tuned midpoint that makes both \(E_{i1}\) and \(E_{i2}\) simultaneously disconnected from \(B\) in \(EW(BE)\)\footnote{It is worth noting that during the enlarging process above, when we fix the connectivity of \(EW(BE)\), \(EW(AE)\) or even \(EW(ABE)\) might also become disconnected. These situations will not affect the result of the entire procedure. We can simply leave the part of the subregion that is disconnected from \(A\) to the second row of the figure, and split the part that connects with both \(A\) and \(B\) once again. This process cannot occur infinitely many times; even if it did, the remaining region would be infinitely small such that its existence would not affect the CMI.}.

Up to now, after the splitting, we can construct \(E'\) with the gap region between \(E_{i1}\) and \(E_{i2}\) deleted, so that \(E_{i1}\) and \(E_{i2}\) only connect with \(A\) in \(EW(AE)\). However, due to the connectivity with $A$ in $EW(AE)$, these \(E_{i1}\) and \(E_{i2}\) are still ``bad" regions that do not satisfy the (dis)connectivity condition (\ref{I3I4dis}). The next step should be to repeat this procedure for each of \(E_{i1}\) and \(E_{i2}\) to remove their connectivity with $A$ in $EW(AE)$.

 To do this, we then move to the second row of the figure and split those ``bad" intervals \(E_i\) that only connect with \(A\) in \(EW(AE)\) once again. The only difference this time is that when we enlarge the gap, \(S_{E}\) and \(S_{BE}\) both decrease while \(S_{AE}\) and \(S_{ABE}\) both increase. Since their signs cancel out in \(I(A:B|E)\), enlarging the gap region will {not} change the value of \(I(A:B|E)\), until the next phase transition occurs between \(A\) and \(E_{i1}\) and \(E_{i2}\) simultaneously. Finally, the (dis)connectivity condition (\ref{I3I4dis}) is satisfied.

 {Therefore, for CMI type entropy combinations, we have confirmed the correspondence between the maximum value of the entropy combination and a valid configuration satisfying a given set of (dis)connectivity conditions of all related entanglement wedges. Note that whether each entanglement wedge should be connected or disconnected at the upper bound configuration is determined by the specific coefficient of each term in the entropy combination. Consequently, when upper-bounding CMI type combinations, it is sufficient to analyze only one configuration that satisfies the set of disconnectivity conditions associated with this entropy combination.}

To extend the previous proof of (dis)connectivity condition (Figure \ref{DISCMI}) {for \(I(A:B|E)\)} to general CMI type entropy combinations more effectively, we now reinterpret it in an alternative language, utilizing the lamp diagram, leading to a general formalism of analyzing the maximum configuration for CMI type combinations. After splitting \(E\) into three pieces and enlarging the gap region in the middle, all connected RT surfaces share the semicircle in the middle and all disconnected RT surfaces share the semicircles on the left and right sides {as shown in Figure \ref{DISCMI}}. Enlarging the gap region will make the length of the former term increase and the latter term decrease. Let us see how this process can be easily represented utilizing the lamp diagram. 

We use lamp diagrams to illustrate the changes in the (dis)connectivity conditions during the splitting processes to reach the upper bound configuration. One lamp diagram represents a set of (dis)connectivity conditions of each entanglement wedge represented by each dot in the lamp diagram. {Grey dots indicate connectivity of the corresponding entanglement wedge while colored dots indicate disconnectivity.} A sequence of lamp diagrams then indicates the splitting process, i.e. each splitting step would lead to the next lamp diagram in the sequence. Specifically, the splitting process always causes one of the connected entanglement wedges to become disconnected; that is, a lamp diagram with fewer lamps turned on will become a diagram with one additional lamp turned on after each splitting. In the end, all lights are turned on, and due to the balance condition of the entropy combination, the number of the red lamps must be equal to the number of the blue lamps. 

Along this splitting procedure, the value of the entropy combination for each lamp diagram might not always increase.
The next task is to determine, throughout the entire splitting process from one lamp diagram to the next, which steps increase the value of the entropy combination and which steps decrease it. This will allow us to identify the lamp diagram that achieves the maximum value of the entropy combination under study. Then from that lamp diagram, we could read out the (dis)connectivity configurations of the entanglement wedges of all regions containing $E$ for that upper bound configuration. 


Here, we provide an explicit example for finding the (dis)connectivity condition in the search for the upper bound of the CMI $I(A:B|E)$ (\ref{I3I4dis}), via the language of lamp diagrams, shown in Figure \ref{CMIlamp}. We start from the first diagram with all entanglement wedges being connected, where all lamps are turned off (grey colored) as shown in the figure.
We now start splitting the region $E$ over and over again, and during this splitting process, more entanglement wedges become disconnected, leading to new lamp diagrams with more lamps turned on. In this procedure, assuming without loss of generality that \(EW(AE)\) becomes disconnected first, five possible lamp diagrams might appear in a sequence. The configuration in each lamp diagram represents the (dis)connectivity of all entanglement wedges.

In the process of each splitting step, we use arrows to denote the direction in which the value of the entropy combination under study increases. Therefore, the diagrams connected by arrows represent configurations that can be transformed by a splitting process. Note that the direction of the arrow does not indicate the direction of the splitting itself; rather, we always split from the diagram with fewer lamps turned on to the one with more lamps turned on, and the arrow shows the direction corresponding to a larger value of the entropy combination. A double-headed arrow indicates that CMI remains unchanged during this splitting step.

As a result, the lamp diagram with all arrows pointing toward it is the rightful diagram with the largest CMI, and we should stipulate the (dis)connectivity condition as read from this diagram in the CMI upper bound configuration.

After having this entire procedure in mind, let us analyze those diagrams and each splitting step in detail one by one. In the first diagram, all lamps are turned off, meaning that \(E\) is at least partially connected in all entanglement wedges. Then we split \(E\) into \(E_{1}\) and \(E_{2}\); the CMI remains unchanged until the first phase transition occurs between \(E_{1}\) and \(E_{2}\) in \(EW(E)\). This leads to the second lamp diagram where the first lamp in this diagram, representing the disconnectivity of \(EW(E)\), is lit. When we continuously enlarge the gap region between \(E_{1}\) and \(E_{2}\), \(S_E\) decreases while each of \(S_{AE}\), \(S_{BE}\), and \(S_{ABE}\) increases for the same amount (shown in Figure \ref{DISCMI}). As the entire entropy combination is balanced, we can focus solely on the sign of \(S_E\) because the sum of the coefficient of those three terms \(S_{AE},S_{BE},S_{ABE}\) in the expression of CMI would simply have the opposite sign. Since \(S_E\) has a negative sign in the expression of CMI $I(A:B|E)$ (represented by red in the lamp diagram), decreasing \(S_E\) will increase the combination. As a result, we can reach the general rule of determining whether each step in the splitting process increases a general entropy combination or not from the lamp diagram:

\begin{itemize}
    \item []
        \textit{If the number of red lamps is greater than the number of blue lamps in the current lamp diagram, then the splitting process to the next step will increase the value of the entropy combination; if the number of blue lamps is greater than the number of red lamps in the current diagram, then the splitting process to the next step will decrease the value of the entropy combination; if the numbers are equal, the next step of the splitting process will not change the value of the entropy combination.}
\end{itemize}

\section{Proof of $I_4$ type disconnectivity condition}

\noindent In this appendix, we prove Theorem 2.2, the $I_4$‑type disconnectivity condition.  
The basic idea is:  
i) {By introducing the complement region $O$ as the purifier of $AB\ldots E$, obtain an equivalent counterpart of fine‑tuning $E$ to seek the upper bound of the entropy combination. In this case, $O$ is fine-tuned.} 
ii) Develop an anti‑splitting process (the merging process) to deal with the $O$‑version lamp diagram.  
iii) Find the maximum configuration of the entropy combination for $O$ with $E=(AB\ldots O)^c$ and rule out the existence of an $E_i$ which is connected in all entanglement wedges containing it.

\textbf{Step One: introducing the complement lamp diagram.}  
We denote the complement of all regions, including \(E\), by \(O\); \ie, \(O\) is the purifier of the union of all systems.  
We can then substitute each entropy term that contains \(E\) in the entropy combination with the entanglement entropy of the region that purifies it, which contains \(O\).  
Thus the task of finding a subsystem \(E\) that maximizes the entropy combination is equivalent to finding an \(O\) that maximizes the counterpart combination.  
For example, finding \(E\) that maximizes the first version of \(3\)-CMI \cite{Wilde_2016} is equivalent to finding \(O\) that maximizes the second version of \(3\)-CMI:
\begin{equation}
\begin{aligned}
    I(A:B:C|E)&=S_{AE}+S_{BE}+S_{CE}-2S_{E}-S_{ABCE}\\
    &=S_{BCO}+S_{ACO}+S_{ABO}-2S_{ABCO}-S_{O}\\
    &=\tilde{I}(A:B:C|O).
\end{aligned}
\end{equation}
Since the two tasks are equivalent, they share the same maximum configuration, with \(O_{\text{max}}\) being the complement that purifies \(ABC\ldots E_{\text{max}}\).

\begin{figure}[H]
	\centering
	\includegraphics[scale=0.5]{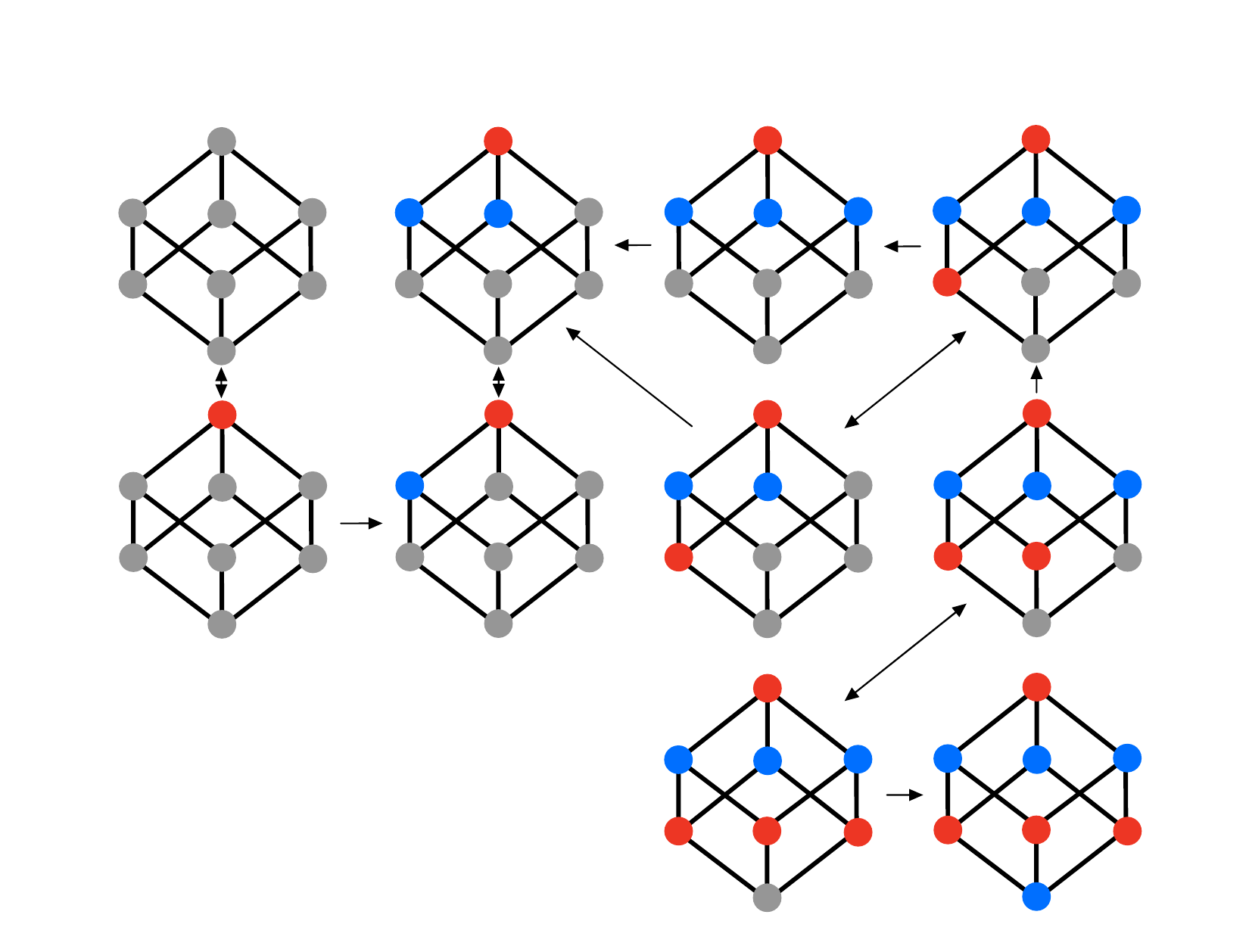}
	\caption{\(O\)-lamp diagram of \(-I_4(A:B:C|O)=I_4(A:B:C|E)\).}
	\label{I4complement}
\end{figure}

Then, what about the complement version of the lamp diagram? Can we observe this process through the lamp diagram? The answer is yes.  
As shown in Figure \ref{I4complement}, the complement \(O\)-lamp diagram is a \(180^\circ\) rotation of the \(E\)-lamp diagram because the lamps corresponding to the connectivity of the complement entanglement wedges are always opposite on the hypercube.  
For example, because the connectivity of \(EW(AE)\) is marked by the lamp on the first row of the second line in the \(E\)-lamp diagram, the connectivity of \(EW(BCO)\) must be marked by the lamp on the last row of the second‑to‑last line in the \(O\)-lamp diagram and indeed it is.

In the \(I_4\) case, for each entanglement wedge, its color in the \(O\)-lamp diagram is the opposite of that in the \(E\)-lamp diagram because \(-I_4(A:B:C|O)=I_4(A:B:C|E)\); we are therefore concerned with the maximum value of \(-I_4(A:B:C|O)\).  
As a result, all arrows in the \(O\)-diagram point in the opposite direction to those in the \(E\)-diagram.  
Now, however, we have a large number of arrows pointing upward toward the diagram with fewer lamps lit, yet the splitting process cannot make a disconnected entanglement wedge become connected again. Therefore, we must develop an anti‑splitting (merging) process to show that the maximum configuration is the second diagram.

\textbf{Step Two: anti‑splitting process (merging process).}  

We now consider two neighbouring subregions $O_i$ and $O_{i+1}$, each of which can exhibit distinct connectivity in the entanglement wedges containing $O$.  
For example, $O_i$ is disconnected from $A$ and $B$ in $EW(AO)$ and $EW(BO)$, respectively, while $O_{i+1}$ is disconnected from $B$ and $C$ in $EW(BO)$ and $EW(CO)$.  
For each of them we can then draw a lamp diagram that denotes the connectivity of the entanglement wedges containing it in the combination. 
We now prove the following statement: as long as the number of blue lights is at least the number of red lights in each diagram, merging\footnote{As the exact opposite of splitting, merging refers to eliminating the gap between $O_i$ and $O_{i+1}$, while preserving their outer two endpoints.} $O_i$ and $O_{i+1}$ into a single subregion will not decrease the entropy combination.

We classify all entanglement wedges into four classes:  
\ding{108}\, Disconnected from both $O_i$ and $O_{i+1}$, \eg, $EW(O)$, $EW(BO)$;  
\ding{52}\, Disconnected from $O_i$ and connected to $O_{i+1}$, \eg, $EW(AO)$;  
\ding{55}\, Disconnected from $O_{i+1}$ and connected to $O_i$, \eg, $EW(CO)$;  
\ding{70}\, Connected to both $O_i$ and $O_{i+1}$.  
Considering the structure of the $I_4$‑type lamp diagram and the blue‑$\ge$‑red condition, we determine the signs of the coefficients as follows:  
$C($\ding{108}$)\ge0$, $C($\ding{70}$)\le0$, $C($\ding{108}$)+C($\ding{52}$)\ge0$, $C($\ding{108}$)+C($\ding{55}$)\ge0$, and $C($\ding{108}$)+C($\ding{52}$)+C($\ding{55}$)+C($\ding{70}$)=0$.

\begin{figure}[H]
	\centering
	\includegraphics[scale=0.45]{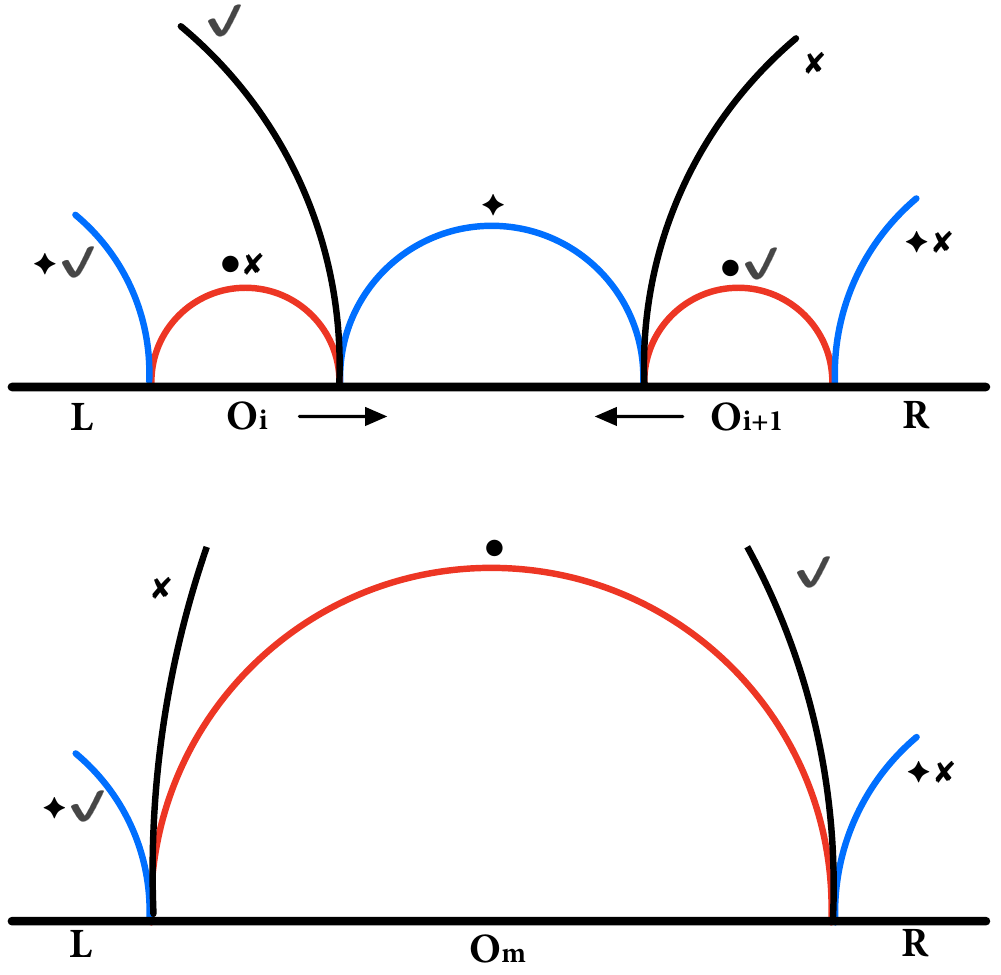}
	\caption{The anti‑splitting process. The upper figure shows $O_i$ and $O_{i+1}$, which exhibit distinct connectivities. The lower figure shows the region $O_m$ after merging. All RT surfaces are labeled by the four classes of entanglement wedges corresponding to them.}\label{antisplit}
\end{figure}

All RT surfaces corresponding to these four types are shown in Figure \ref{antisplit}.  
Although the changes of the RT surfaces appear messy, careful observation shows that the merging process can be viewed as three entanglement‑wedge phase transitions.  
The connectivity of the RT surfaces shifts after merging, and the resulting difference in the entropy combination equals the sum of the area changes of the RT surfaces during these three transitions:  
1. For the RT surface labeled \ding{108}, $O_i$ connects with $O_{i+1}$.  
2. For the RT surface labeled \ding{52}, $O_i$ connects with $O_{i+1}R$.  
3. For the RT surface labeled \ding{55}, $O_{i+1}$ connects with $O_iL$.  
The first transition certainly enlarges the surface labeled \ding{108}; because its coefficient is positive, the entropy combination increases.  
Strong subadditivity shows that the area increases for \ding{52} and \ding{55} are smaller than that for \ding{108}; together with the positive signs of $C($\ding{108}$)+C($\ding{52}$)$ and $C($\ding{108}$)+C($\ding{55}$)$, the total change is therefore an increase.  
In rare cases, the merged region $O_m$ exhibits higher connectivity. One can then shorten the gap between $O_i$ and $O_{i+1}$ continuously; the same analysis shows that the entropy combination still increases before any phase transition occurs.  
This process is repeated until the intervals in $O$ merge into diagram (2.1) in Figure \ref{I4complement}.

One exceptional case occurs when only a single subregion of $O$ lies in a gap region, so no neighbour is available for merging.  
We can “interpolate’’ one nearby region, enlarging it from zero size.  
In a $g_2$ gap, the phase‑transition condition is always degenerate\footnote{For $n\ge5$ $I_4$‑type entropy combinations, this interpolation might in principle fail (though we have found no explicit example). Even if it did, one would merely add a finite correction term to the upper bound whenever $g_2$ gaps exist.}, \ie, the region suddenly exhibits connectivity (2.4) in Figure \ref{I4complement} at a critical point during the enlargement.  
At this point, $I_4$ is unchanged compared with the case without the additional region.  
We then use the merging process as before until the single region inside the gap exhibits connectivity (2.1) in Figure \ref{I4complement}.

\textbf{Step Three: finding the upper‑bound configuration of $O$ and $E$.}  
After finishing the merging process, only two local maximal diagrams remain in the \(O\)-diagram: diagram (2.2)\footnote{Note that (2.2) is obtained from diagram (2.1) by splitting \(O\), where \(EW(AE)\) reaches its critical connection point. We may always choose this critical point, where \(EW(AE)\) is disconnected, as the maximum configuration. In that case, all entanglement wedges are connected except \(EW(O)\).} and the diagram with all lights on.  
Because of the \(180^\circ\) rotation, the maximal \(E\)- and \(O\)-diagrams are exact opposites: the \(E\)-diagram with all lights off becomes the \(O\)-diagram with all lights on, and the true maximum—where every entanglement wedge except \(EW(ABCE)\) is disconnected—becomes the diagram where every wedge is connected except \(EW(O)\).

The configuration with all lamps on cannot be the global maximum: all lights being on means that all entanglement wedges are disconnected, and the mutual information between \(O\) and any region vanishes.  
One can prove that deleting the intervals of \(O\) that are disconnected from all regions does not change the value of the combination.  
We can therefore delete such intervals until no interval of \(O\) connects with \(ABC\) in \(EW(ABCO)\).  
As a result, there is only one true maximum configuration in \(O\), in which all entanglement wedges are connected except \(EW(O)\).

Can we stipulate the opposite connectivity condition in the \(E\)-diagram once its complement \(O\)-diagram has only one true maximum point?  
That is, if \(EW(AO)\), \(EW(BO)\), and \(EW(CO)\) are connected in the maximum \(O\)-configuration, can we stipulate that \(EW(BCE)\), \(EW(ACE)\), and \(EW(ABE)\) are disconnected?  
In most cases, yes, but exceptions occur: the rule fails if an interval of \(E\) is adjacent to \(A\), \(B\), or \(C\); it also fails in higher‑dimensional holography (AdS$_4$/CFT$_3$) where \(O\) and \(E\) sit in a gap region adjacent to \(A\), \(B\), and \(C\) simultaneously.  
Fortunately, we can always find another configuration, with value no less than the former, in which these exceptions do not arise; the details appear in Appendix C.

After all exceptions are ruled out, we may impose the opposite (dis)connectivity condition of the \(O\)-diagram on the \(E\)-diagram.  
Namely, the maximal \(E\)-diagram has $2^{n}-1$ lights on, \ie, all entanglement wedges are disconnected except \(EW(ABC\ldots E)\).  
This concludes the proof of the $I_4$‑type disconnectivity theorem.

\section{Ruling out the exceptions of the opposite (dis)connectivity of $E$-diagram and $O$-diagram in the proof of $I_4$ type (dis)connectivity condition}
\noindent In this appendix, we will rule out all the exceptional cases where $EW(AO)$, $EW(BO)$, and $EW(CO)$ are connected, but there still exist intervals of $E$ (written as $E_i$) that are connected with all entanglement wedges. Note that if $E_i$ is disconnected in one of the entanglement wedges $EW(AE)$, $EW(BE)$, or $EW(CE)$, we can simply split it as we did in Appendix A and Section \ref{sec2.2} until the subregions of $E_i$ are all disconnected in $EW(ABE)$, $EW(BCE)$, and $EW(ACE)$ after splitting. However, if $E_i$ is connected with all entanglement wedges, it reaches the local maximum point in the lamp diagram, and the splitting process might decrease the value of the combination and be inadmissible.

\begin{figure}[H]
	\centering
	\includegraphics[scale=0.6]{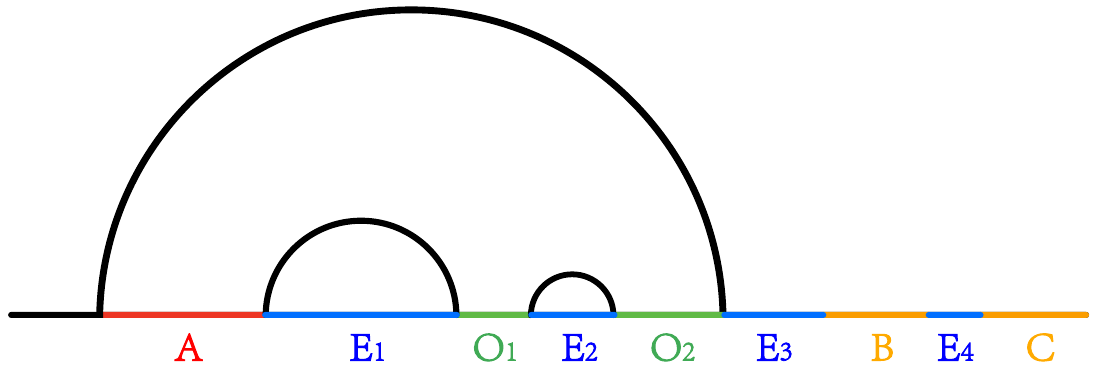}
		\caption{The RT surface of region $AO$ ($RT_{AO}$), or equivalently $RT_{BCE}$, is marked by the black curves. The connectivity of $EW(AO)$ is equivalent to the disconnectivity of $EW(BCE_1E_2)$, but $E_3$ and $E_4$ are adjacent to $B$ or $C$, which makes them connected to $B$ or $C$ in $EW(BCE)$.}\label{adj}
\end{figure}

The first exceptional case occurs when one of the intervals inside $E$ is adjacent to $B$ (w.l.o.g.), such as $E_3$, as shown in Figure \ref{adj}.
\begin{figure}[H]
	\centering
	\includegraphics[scale=0.6]{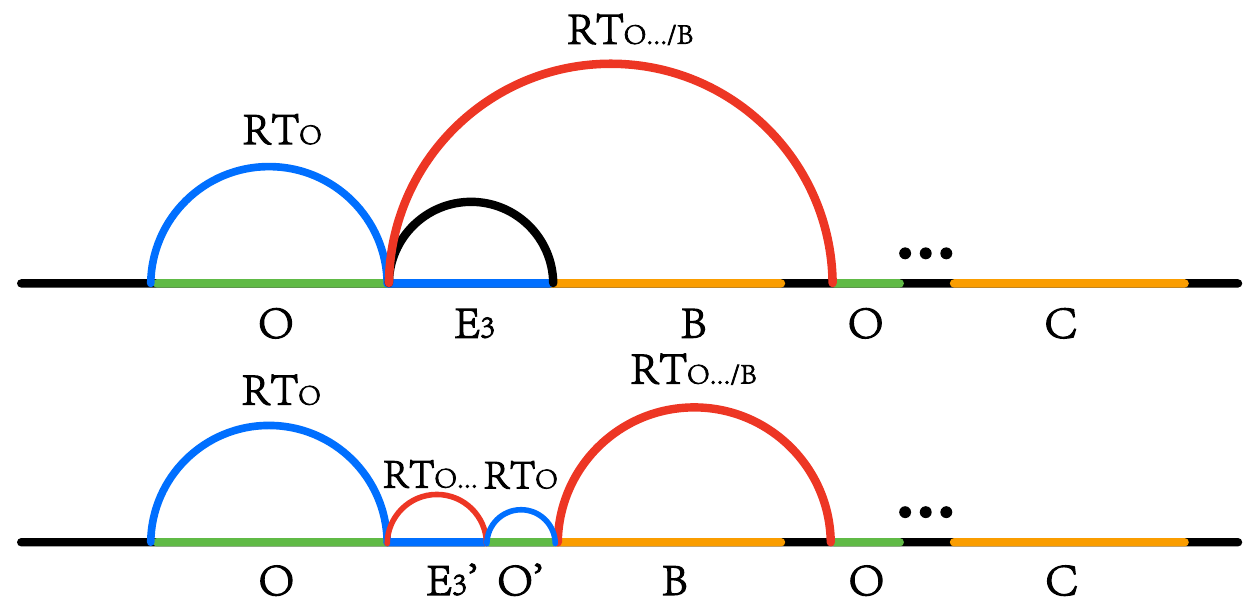}
		\caption{The top figure illustrates the case before $E_3$ is split. In the complement combination, the geodesic labeled $RT_{O.../B}$ is included in the RT surface of every term that contains $O$ and excludes $B$ but is not $S(O)$ itself. Owing to the balance requirements, it is shown in red, while the geodesic enclosing $O$ is blue, which appears only in $RT_O$ regarding the sole disconnectivity of $EW(O)$. In addition, the RT surface enclosing $E_3$ is included in every term that contains $BO$ and is therefore black. In the bottom figure, $E_3$ is partitioned into $E'_3$ and $O'$ with the (dis)connectivity of entanglement wedges unchanged. Thus, the RT surfaces enclosing $O$ and $O'$ are blue, and the component of $RT_{O.../B}$ on the right remains red. Meanwhile, the geodesic encompassing $E'_3$ is part of the RT surface of every term containing $O$, except $S(O)$ itself, and consequently, it is red.
        Therefore, it can be seen that the change in the combination value after splitting is given by the area difference between two pairs of homologous surfaces. The occurrence of an RT surface phase transition ensures that the combination value remains the same.}\label{exc1}
\end{figure}
In this case, as shown in Figure \ref{exc1}, we can always split $E_3$ into two halves, $E'_3$ and $O'$, where the latter is adjacent to $B$ while the former is not. Then, we can enlarge $O'$ until a phase transition occurs in $EW(O.../B)$, where $E_3'$ and $B$ become disconnected. At that time, one can verify that the value of the combination is exactly the same as in the original $E_3$ case. Afterwards, we can repeatedly split $E_3'$ until $EW(ABE)$, $EW(BCE)$, and $EW(ACE)$ are all disconnected.

\begin{figure}[H]
	\centering
	\includegraphics[scale=0.6]{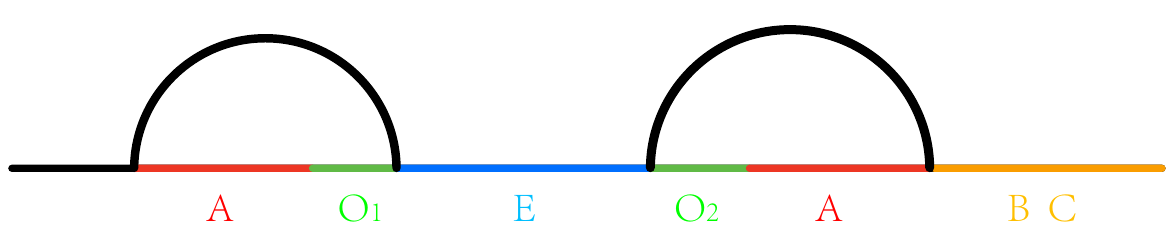}
		\caption{Illustration of the case where $E$ is non-adjacent to $A$, $B$, and $C$. The RT surface of $OA$ is depicted in black. With $O_2A$ sufficiently small, $EW(BE)$ and $EW(CE)$ can be connected, but the connectivity of $EW(AE)$ is then impossible.}\label{exception2}
\end{figure}

After analyzing the case when $E$ is adjacent to at least one of $A$, $B$, or $C$, let us analyze the case where $E$ is not adjacent to any of $A$, $B$, or $C$. As shown in Figure \ref{exception2}, the blue interval of $E$ is not adjacent to $A$, but it might be connected with $B$ in $EW(BE)$ or with $C$ in $EW(CE)$. However, since we demand that $O_1$ and $O_2$ are connected in $EW(AO)$, $EW(BO)$, and $EW(CO)$, the connectivity of $EW(BCO)$ directly implies that $EW(AE)$ cannot be connected. Since $E$ is not connected in all entanglement wedges, we can repeatedly split it until the disconnectivity condition is satisfied.

Note that these two exceptional cases can be ruled out in AdS$_3$/CFT$_2$, and in higher dimensional holography where there does not exist a gap region that is simultaneously adjacent to $A$, $B$, and $C$ (which is equivalent to the lower-dimensional case with one axis compressed). Now we try to deal with the last exceptional case in higher dimensional holography, where there exists a gap region $G_{3ABC}$. Then, if $E$ is inside this region, it could be connected in $EW(AE)$, $EW(BE)$, and $EW(CE)$ without being adjacent to any of $A$, $B$, or $C$, while $O$ is adjacent to $A$, $B$, and $C$. In this case, the splitting method described above will not work. However, since \(I_4\) is an IR term without UV divergence, when we directly delete this problematic \(E\) that is connected with all entanglement wedges, the corresponding change in the combination is always finite. Moreover, as we have found in \cite{Ju:2024hba}, when \(G_{3ABC}\) exists, the upper bound of \(I_4\) obtained by utilizing the disconnectivity condition is infinite when the number of strips of region \(E\) approaches infinity. As a result, for a configuration with some problematic subregions inside \(E\), we can, in principle, simply delete them and construct another configuration that satisfies the disconnectivity condition with the value of the combination not less than that of the original configuration. In conclusion, we have successfully ruled out the last exceptional case.
\section{Gap classification for \(n=3\) in higher dimensional holography}
\noindent In this appendix, we discuss the general case of $I_4$ in AdS$_4$/CFT$_3$ where $g_3$, $g_2$ and $g_1$ simultaneously exist. It should be noted that again we search for general upper bounds satisfying the permutation symmetry across $A$, $B$ and $C$. Similar to the AdS$_3$/CFT$_2$ case, the single gap regions where $E$ can reside in are first listed as follows:
\begin{equation}
    \begin{aligned}
        &g_{1A},\,\, \,\quad g_{1B},\,\,\, \quad g_{1C}, \quad\\
        &g_{2AB}, \quad g_{2AC}, \quad g_{2BC}, \quad\\
        &g_{3ABC}.
    \end{aligned}
\end{equation}
However, now the choice of their unions is more diverse. Again we take $AB$ as an example and introduce the following gap groups:
\begin{equation}
\begin{aligned}
    &G_{123_{S}AB}=g_{1A}\cup g_{1B}\cup g_{2{AB}}\cup g_{3ABC}\\
    &G_{123_{L}AB}=g_{1A}\cup g_{1B}\cup g_{2{AB}}\cup g_{2{AC}}\cup g_{2{BC}}\cup g_{3ABC}\\
    &G_{23_{S}AB}= g_{2{AB}}\cup g_{3ABC}\\
    &G_{23_{L}AB}=g_{2{AB}}\cup g_{2{AC}}\cup g_{2{BC}}\cup g_{3ABC}\\
    &G_{13AB}=g_{1A}\cup g_{1B}\cup g_{3ABC}\\
    &G_{12_{S}AB}=g_{1A}\cup g_{1B}\cup g_{2{AB}}\\
    &G_{12_{L}AB}=g_{1A}\cup g_{1B}\cup g_{2{AB}}\cup g_{2{AC}}\cup g_{2{BC}}\\
    &g_{3AB}=g_{3ABC}\\
    &g_{2_{S}AB}=g_{2{AB}}\\
    &g_{2_{L}AB}=g_{2{AB}}\cup g_{2{AC}}\cup g_{2{BC}}\\
    &g_{1AB}=g_{1A}\cup g_{1B}.
\end{aligned}
\end{equation}
Similar groups can be defined for $AC$ and $BC$. By upper-bounding the ``real" RT surfaces with the corresponding ``fake" ones, and taking the summation over $C_3^2=(AB,AC,BC)$, the following inequalities can be obtained:
\begin{equation}
\begin{aligned}
        &G_{123_{S}}:2\sum_{i \subset g_1} (S_{E_i}-S_{gaps_i})+\sum_{i \subset g_2}(S_{E_i}-S_{gaps_i})+3\sum_{i \subset g_3}(S_{E_i}-S_{gaps_i}) \\
        &\qquad \qquad \qquad \qquad \qquad \qquad \qquad \qquad \qquad \qquad \quad  \le \sum_{C_3^2} (S_{ABG_{123_{S}AB}}-S_{AB})\\
\end{aligned}
\end{equation}
\begin{equation}
\begin{aligned}
        &G_{123_{L}}:2\sum_{i \subset g_1} (S_{E_i}-S_{gaps_i})+3\sum_{i \subset g_2}(S_{E_i}-S_{gaps_i})+3\sum_{i \subset g_3}(S_{E_i}-S_{gaps_i})\\
        &\qquad \qquad \qquad \qquad \qquad \qquad \qquad \qquad \qquad \qquad \quad  \le \sum_{C_3^2} (S_{ABG_{123_{L}AB}}-S_{AB})\\
\end{aligned}
\end{equation}
\begin{equation}
\begin{aligned}
        &G_{23_{S}}:\sum_{i \subset g_2}(S_{E_i}-S_{gaps_i})+3\sum_{i \subset g_3}(S_{E_i}-S_{gaps_i}) \le \sum_{C_3^2} (S_{ABG_{23_{S}AB}}-S_{AB})\\
\end{aligned}
\end{equation}
\begin{equation}
\begin{aligned}
        &G_{23_{L}}:3\sum_{i \subset g_2}(S_{E_i}-S_{gaps_i})+3\sum_{i \subset g_3}(S_{E_i}-S_{gaps_i}) \le \sum_{C_3^2} (S_{ABG_{23_{L}AB}}-S_{AB})\\
\end{aligned}
\end{equation}
\begin{equation}
\begin{aligned}
        &G_{13}:2\sum_{i \subset g_1} (S_{E_i}-S_{gaps_i})+3\sum_{i \subset g_3}(S_{E_i}-S_{gaps_i}) \le \sum_{C_3^2} (S_{ABG_{13AB}}-S_{AB})\\
\end{aligned}
\end{equation}
\begin{equation}
\begin{aligned}
        &G_{12_{S}}:2\sum_{i \subset g_1} (S_{E_i}-S_{gaps_i})+\sum_{i \subset g_2}(S_{E_i}-S_{gaps_i}) \le \sum_{C_3^2} (S_{ABG_{12_{S}AB}}-S_{AB})\\
\end{aligned}
\end{equation}
\begin{equation}
\begin{aligned}
        &G_{12_{L}}:2\sum_{i \subset g_1} (S_{E_i}-S_{gaps_i})+3\sum_{i \subset g_2}(S_{E_i}-S_{gaps_i}) \le \sum_{C_3^2} (S_{ABG_{12_{L}AB}}-S_{AB})\\
\end{aligned}
\end{equation}
\begin{equation}
\begin{aligned}
        &g_{3}:3\sum_{i \subset g_3}(S_{E_i}-S_{gaps_i}) \le \sum_{C_3^2} (S_{ABg_{3AB}}-S_{AB})\\
\end{aligned}
\end{equation}
\begin{equation}
\begin{aligned}
        &g_{2_{S}}:\sum_{i \subset g_2}(S_{E_i}-S_{gaps_i}) \le \sum_{C_3^2} (S_{ABg_{2_{S}AB}}-S_{AB})\\
\end{aligned}
\end{equation}
\begin{equation}
\begin{aligned}
        &g_{2_{L}}:3\sum_{i \subset g_2}(S_{E_i}-S_{gaps_i}) \le \sum_{C_3^2} (S_{ABg_{2_{L}AB}}-S_{AB})\\
\end{aligned}
\end{equation}
\begin{equation}
\begin{aligned}
        &g_{1}:2\sum_{i \subset g_1} (S_{E_i}-S_{gaps_i}) \le \sum_{C_3^2} (S_{ABg_{1AB}}-S_{AB})\\
\end{aligned}
\end{equation}
Once again, no single inequality above can directly give the bound on $\sum_{i} S_{E_i}-\sum_{i} S_{gaps_i}$, and we must consider their combinations. Therefore, the problem of deriving upper bounds with the given inequalities can be transformed into solving linear equations. To each inequality, we assign a vector consisting of the multiples for $\sum_{i \subset g_1}(S_{E_i}-S_{gaps_i})$, $\sum_{i \subset g_2}(S_{E_i}-S_{gaps_i})$ and $\sum_{i \subset g_3}(S_{E_i}-S_{gaps_i})$ labeled by the corresponding gap group. Then, we can identify all 2-combinations and 3-combinations of these 11 vectors that can linearly generate our target vector $(1,1,1)$ while ensuring that all linear combination coefficients are positive. In this case, we eventually obtain 42 possible combinations, which are:
\begin{equation}\label{gapcombsappendix}
\begin{aligned}
        &G_{123_{L}}g_{1},\ G_{23_{L}}g_{1},\ G_{123_{S}}G_{123_{L}}G_{12_{S}},\ G_{123_{S}}G_{123_{L}}G_{12_{L}},\ G_{123_{S}}G_{23_{S}}G_{12_{L}},\\
        &G_{123_{S}}G_{23_{L}}G_{12_{S}},\ G_{123_{S}}G_{23_{L}}G_{12_{L}},\ G_{123_{S}}G_{12_{S}}g_{2_{S}},\ G_{123_{S}}G_{12_{S}}g_{2_{L}},\\
        &G_{123_{S}}G_{12_{L}}g_{3},\ G_{123_{S}}G_{12_{L}}g_{2_{S}},\ G_{123_{S}}G_{12_{L}}g_{2_{L}},\ G_{123_{S}}g_{2_{S}}g_{1},\ G_{123_{S}}g_{2_{L}}g_{1},\\
        &G_{123_{L}}G_{23_{S}}G_{12_{S}},\ G_{123_{L}}G_{13}G_{12_{S}},\ G_{123_{L}}G_{13}G_{12_{L}},\ G_{123_{L}}G_{12_{S}}g_{3},\\
        &G_{23_{S}}G_{23_{L}}G_{12_{S}},\ G_{23_{S}}G_{13}G_{12_{L}},\ G_{23_{S}}G_{12_{S}}G_{12_{L}},\ G_{23_{S}}G_{12_{S}}g_{2_{S}},\ G_{23_{S}}G_{12_{S}}g_{2_{L}},\\
        &G_{23_{S}}G_{12_{L}}g_{1},\ G_{23_{S}}g_{2_{S}}g_{1},\ G_{23_{S}}g_{2_{L}}g_{1},\ G_{23_{L}}G_{13}G_{12_{S}},\ G_{23_{L}}G_{13}G_{12_{L}},\\
        &G_{23_{L}}G_{12_{S}}g_{3},\ G_{13}G_{12_{S}}g_{2_{S}},\ G_{13}G_{12_{S}}g_{2_{L}},\ G_{13}G_{12_{L}}g_{3},\ G_{13}G_{12_{L}}g_{2_{S}},\ G_{13}G_{12_{L}}g_{2_{L}},\\
        &G_{13}g_{2_{S}}g_{1},\ G_{13}g_{2_{L}}g_{1},\ G_{12_{S}}G_{12_{L}}g_{3},\ G_{12_{S}}g_{3}g_{2_{S}},\ G_{12_{S}}g_{3}g_{2_{L}},\ G_{12_{L}}g_{3}g_{1},\\
        &g_{3}g_{2_{S}}g_{1},\ g_{3}g_{2_{L}}g_{1}.
\end{aligned}
\end{equation}
The respective upper bounds of $I_4$ are given in Table \ref{42comb}. Therefore, we can see that seeking the combination that provides the tightest bound is an extremely tedious work. However, similar to the case of \ding{174}\ding{176} and \ding{174}\ding{175} in Section \ref{sec3.2}, some combinations among the 42 are essentially looser than others and can be eliminated. Here we take the former two combinations as an example. The inequalities that they give are:
\begin{equation}
    \begin{aligned}
        &G_{123_{L}}g_{1}:\sum (S_{E_i}-S_{gaps_i}) \le \sum_{C_3^2} \Bigl(\frac{1}{3}S_{ABG_{123_{L}AB}}+\frac{1}{6}S_{ABg_{1AB}}-\frac{1}{2}S_{AB}\Bigr)\\
        &G_{23_{L}}g_{1}:\sum (S_{E_i}-S_{gaps_i}) \le \sum_{C_3^2} \Bigl(\frac{1}{3}S_{ABG_{23_{L}AB}}+\frac{1}{2}S_{ABg_{1AB}}-\frac{5}{6}S_{AB}\Bigr),\\
    \end{aligned}
\end{equation}
and the right-hand sided terms for $G_{23_{L}}g_{1}$ is larger than that for $G_{123_{L}}g_{1}$ by $\sum_{C_3^2}\frac{1}{3}I(G_{23_{L}AB}:g_{1AB}|AB)$. Nevertheless, the tightness of other combinations that cannot be thus excluded may rely on the specific configuration, and the UV divergent behaviors might depend on the particular shapes and positions of $A$, $B$ and $C$.

As a result, when we are provided with an explicit configuration, we can conduct a specific comparison of the bounds corresponding to the 42 combinations in (\ref{gapcombsappendix}) and eventually obtain the tightest one.

\renewcommand\theadfont{\bfseries}

\begin{table}[h!]
\centering
\scriptsize
\rowcolors{2}{pink!20}{cyan!5}  
\begin{tabularx}{\textwidth}{|c|c|X|}
\hline
\rowcolor{white}
\thead{} & \thead{gap groups} & \thead{$I_4$ upper bounds} \\
\hline
1 & $G_{123_{L}}g_{1}$ & \makecell[l]{$I_4\le S_{ABC}+ \sum_{C_3^2} \left(\frac{1}{3}S_{ABG_{123_{L}AB}}+\frac{1}{6}S_{ABg_{1AB}}-\frac{1}{2}S_{AB}\right)$} \\ \hline
2 & $G_{23_{L}}g_{1}$ & \makecell[l]{$I_4\le S_{ABC}+ \sum_{C_3^2} \left(\frac{1}{3}S_{ABG_{23_{L}AB}}+\frac{1}{2}S_{ABg_{1AB}}-\frac{5}{6}S_{AB}\right)$}   \\ \hline
3 & $G_{123_{S}}G_{123_{L}}G_{12_{S}}$ & \makecell[l]{$I_4\le S_{ABC}+ \sum_{C_3^2} \left(\frac{1}{12}S_{ABG_{123_{S}AB}}+\frac{1}{4}S_{ABG_{123_{L}AB}}+\frac{1}{6}S_{ABG_{12_{S}AB}}-\frac{1}{2}S_{AB}\right)$}   \\ \hline
4 & $G_{123_{S}}G_{123_{L}}G_{12_{L}}$ & \makecell[l]{$I_4\le S_{ABC}+ \sum_{C_3^2} \left(\frac{1}{4}S_{ABG_{123_{S}AB}}+\frac{1}{12}S_{ABG_{123_{L}AB}}+\frac{1}{6}S_{ABG_{12_{L}AB}}-\frac{1}{2}S_{AB}\right)$}   \\ \hline
5 & $G_{123_{S}}G_{23_{S}}G_{12_{L}}$ & \makecell[l]{$I_4\le S_{ABC}+ \sum_{C_3^2} \left(\frac{5}{18}S_{ABG_{123_{S}AB}}+\frac{1}{18}S_{ABG_{23_{S}AB}}+\frac{2}{9}S_{ABG_{12_{L}AB}}-\frac{5}{9}S_{AB}\right)$}   \\ \hline
6 & $G_{123_{S}}G_{23_{L}}G_{12_{S}}$ & \makecell[l]{$I_4\le S_{ABC}+ \sum_{C_3^2} \left(\frac{1}{6}S_{ABG_{123_{S}AB}}+\frac{1}{6}S_{ABG_{23_{L}AB}}+\frac{1}{3}S_{ABG_{12_{S}AB}}-\frac{2}{3}S_{AB}\right)$}   \\ \hline
7 & $G_{123_{S}}G_{23_{L}}G_{12_{L}}$ & \makecell[l]{$I_4\le S_{ABC}+ \sum_{C_3^2} \left(\frac{3}{10}S_{ABG_{123_{S}AB}}+\frac{1}{30}S_{ABG_{23_{L}AB}}+\frac{1}{5}S_{ABG_{12_{L}AB}}-\frac{8}{15}S_{AB}\right)$}   \\ \hline
8 & $G_{123_{S}}G_{12_{S}}g_{2_{S}}$ & \makecell[l]{$I_4\le S_{ABC}+ \sum_{C_3^2} \left(\frac{1}{3}S_{ABG_{123_{S}AB}}+\frac{1}{6}S_{ABG_{12_{S}AB}}+\frac{1}{2}S_{ABg_{2_{S}AB}}-S_{AB}\right)$}   \\ \hline
9 & $G_{123_{S}}G_{12_{S}}g_{2_{L}}$ & \makecell[l]{$I_4\le S_{ABC}+ \sum_{C_3^2} \left(\frac{1}{3}S_{ABG_{123_{S}AB}}+\frac{1}{6}S_{ABG_{12_{S}AB}}+\frac{1}{6}S_{ABg_{2_{L}AB}}-\frac{2}{3}S_{AB}\right)$}   \\ \hline
10 & $G_{123_{S}}G_{12_{L}}g_{3}$ & \makecell[l]{$I_4\le S_{ABC}+ \sum_{C_3^2} \left(\frac{1}{4}S_{ABG_{123_{S}AB}}+\frac{1}{4}S_{ABG_{12_{L}AB}}+\frac{1}{12}S_{ABg_{3AB}}-\frac{7}{12}S_{AB}\right)$}   \\ \hline
11 & $G_{123_{S}}G_{12_{L}}g_{2_{S}}$ & \makecell[l]{$I_4\le S_{ABC}+ \sum_{C_3^2} \left(\frac{1}{3}S_{ABG_{123_{S}AB}}+\frac{1}{6}S_{ABG_{12_{L}AB}}+\frac{1}{6}S_{ABg_{2_{S}AB}}-\frac{2}{3}S_{AB}\right)$}   \\ \hline
12 & $G_{123_{S}}G_{12_{L}}g_{2_{L}}$ & \makecell[l]{$I_4\le S_{ABC}+ \sum_{C_3^2} \left(\frac{1}{3}S_{ABG_{123_{S}AB}}+\frac{1}{6}S_{ABG_{12_{L}AB}}+\frac{1}{18}S_{ABg_{2_{L}AB}}-\frac{5}{9}S_{AB}\right)$}   \\ \hline
13 & $G_{123_{S}}g_{2_{S}}g_{1}$ & \makecell[l]{$I_4\le S_{ABC}+ \sum_{C_3^2} \left(\frac{1}{3}S_{ABG_{123_{S}AB}}+\frac{2}{3}S_{ABg_{2_{S}AB}}+\frac{1}{6}S_{ABg_{1AB}}-\frac{7}{6}S_{AB}\right)$}   \\ \hline
14 & $G_{123_{S}}g_{2_{L}}g_{1}$ & \makecell[l]{$I_4\le S_{ABC}+ \sum_{C_3^2} \left(\frac{1}{3}S_{ABG_{123_{S}AB}}+\frac{2}{9}S_{ABg_{2_{L}AB}}+\frac{1}{6}S_{ABg_{1AB}}-\frac{13}{18}S_{AB}\right)$}   \\ \hline
15 & $G_{123_{L}}G_{23_{S}}G_{12_{S}}$ & \makecell[l]{$I_4\le S_{ABC}+ \sum_{C_3^2} \left(\frac{1}{6}S_{ABG_{123_{L}AB}}+\frac{1}{6}S_{ABG_{23_{S}AB}}+\frac{1}{3}S_{ABG_{12_{S}AB}}-\frac{2}{3}S_{AB}\right)$}   \\ \hline
16 & $G_{123_{L}}G_{13}G_{12_{S}}$ & \makecell[l]{$I_4\le S_{ABC}+ \sum_{C_3^2} \left(\frac{5}{18}S_{ABG_{123_{L}AB}}+\frac{1}{18}S_{ABG_{13}AB}+\frac{1}{6}S_{ABG_{12_{S}AB}}-\frac{1}{2}S_{AB}\right)$}   \\ \hline
17 & $G_{123_{L}}G_{13}G_{12_{L}}$ & \makecell[l]{$I_4\le S_{ABC}+ \sum_{C_3^2} \left(\frac{1}{6}S_{ABG_{123_{L}AB}}+\frac{1}{6}S_{ABG_{13}AB}+\frac{1}{6}S_{ABG_{12_{L}AB}}-\frac{1}{2}S_{AB}\right)$}   \\ \hline
18 & $G_{123_{L}}G_{12_{S}}g_{3}$ & \makecell[l]{$I_4\le S_{ABC}+ \sum_{C_3^2} \left(\frac{1}{4}S_{ABG_{123_{L}AB}}+\frac{1}{4}S_{ABG_{12_{S}AB}}+\frac{1}{12}S_{ABg_{3AB}}-\frac{7}{12}S_{AB}\right)$}   \\ \hline
19 & $G_{23_{S}}G_{23_{L}}G_{12_{S}}$ & \makecell[l]{$I_4\le S_{ABC}+ \sum_{C_3^2} \left(\frac{1}{4}S_{ABG_{23_{S}AB}}+\frac{1}{12}S_{ABG_{23_{L}AB}}+\frac{1}{12}S_{ABG_{12_{S}AB}}-\frac{5}{12}S_{AB}\right)$}   \\ \hline
20 & $G_{23_{S}}G_{13}G_{12_{L}}$ & \makecell[l]{$I_4\le S_{ABC}+ \sum_{C_3^2} \left(\frac{1}{8}S_{ABG_{23_{S}AB}}+\frac{5}{24}S_{ABG_{13}AB}+\frac{7}{24}S_{ABG_{12_{L}AB}}-\frac{5}{8}S_{AB}\right)$}   \\ \hline
21 & $G_{23_{S}}G_{12_{S}}G_{12_{L}}$ & \makecell[l]{$I_4\le S_{ABC}+ \sum_{C_3^2} \left(\frac{1}{3}S_{ABG_{23_{S}AB}}+\frac{5}{12}S_{ABG_{12_{S}AB}}+\frac{1}{12}S_{ABG_{12_{L}AB}}-\frac{5}{6}S_{AB}\right)$} \\ \hline
22 & $G_{23_{S}}G_{12_{S}}g_{2_{S}}$ & \makecell[l]{$I_4\le S_{ABC}+ \sum_{C_3^2} \left(\frac{1}{3}S_{ABG_{23_{S}AB}}+\frac{1}{2}S_{ABG_{12_{S}AB}}+\frac{1}{6}S_{ABg_{2_{S}AB}}-S_{AB}\right)$} \\ \hline
23 & $G_{23_{S}}G_{12_{S}}g_{2_{L}}$ & \makecell[l]{$I_4\le S_{ABC}+ \sum_{C_3^2} \left(\frac{1}{3}S_{ABG_{23_{S}AB}}+\frac{1}{2}S_{ABG_{12_{S}AB}}+\frac{1}{18}S_{ABg_{2_{L}AB}}-\frac{8}{9}S_{AB}\right)$} \\ \hline
24 & $G_{23_{S}}G_{12_{L}}g_{1}$ & \makecell[l]{$I_4\le S_{ABC}+ \sum_{C_3^2} \left(\frac{1}{3}S_{ABG_{23_{S}AB}}+\frac{2}{9}S_{ABG_{12_{L}AB}}+\frac{5}{18}S_{ABg_{1AB}}-\frac{5}{6}S_{AB}\right)$} \\ \hline
25 & $G_{23_{S}}g_{2_{S}}g_{1}$ & \makecell[l]{$I_4\le S_{ABC}+ \sum_{C_3^2} \left(\frac{1}{3}S_{ABG_{23_{S}AB}}+\frac{2}{3}S_{ABg_{2_{S}AB}}+\frac{1}{2}S_{ABg_{1AB}}-\frac{3}{2}S_{AB}\right)$} \\ \hline
26 & $G_{23_{S}}g_{2_{L}}g_{1}$ & \makecell[l]{$I_4\le S_{ABC}+ \sum_{C_3^2} \left(\frac{1}{3}S_{ABG_{23_{S}AB}}+\frac{2}{9}S_{ABg_{2_{L}AB}}+\frac{1}{2}S_{ABg_{1AB}}-\frac{19}{18}S_{AB}\right)$} \\ \hline
27 & $G_{23_{L}}G_{13}G_{12_{S}}$ & \makecell[l]{$I_4\le S_{ABC}+ \sum_{C_3^2} \left(\frac{5}{24}S_{ABG_{23_{L}AB}}+\frac{1}{8}S_{ABG_{13}AB}+\frac{3}{8}S_{ABG_{12_{S}AB}}-\frac{17}{24}S_{AB}\right)$} \\ \hline
28 & $G_{23_{L}}G_{13}G_{12_{L}}$ & \makecell[l]{$I_4\le S_{ABC}+ \sum_{C_3^2} \left(\frac{1}{12}S_{ABG_{23_{L}AB}}+\frac{1}{4}S_{ABG_{13}AB}+\frac{1}{4}S_{ABG_{12_{L}AB}}-\frac{7}{12}S_{AB}\right)$} \\ \hline
29 & $G_{23_{L}}G_{12_{S}}g_{3}$ & \makecell[l]{$I_4\le S_{ABC}+ \sum_{C_3^2} \left(\frac{1}{6}S_{ABG_{23_{L}AB}}+\frac{1}{2}S_{ABG_{12_{S}AB}}+\frac{1}{6}S_{ABg_{3AB}}-\frac{5}{6}S_{AB}\right)$} \\ \hline
30 & $G_{13}G_{12_{S}}g_{2_{S}}$ & \makecell[l]{$I_4\le S_{ABC}+ \sum_{C_3^2} \left(\frac{1}{3}S_{ABG_{13}AB}+\frac{1}{6}S_{ABG_{12_{S}AB}}+\frac{5}{6}S_{ABg_{2_{S}AB}}-\frac{4}{3}S_{AB}\right)$} \\ \hline
31 & $G_{13}G_{12_{S}}g_{2_{L}}$ & \makecell[l]{$I_4\le S_{ABC}+ \sum_{C_3^2} \left(\frac{1}{3}S_{ABG_{13}AB}+\frac{1}{6}S_{ABG_{12_{S}AB}}+\frac{5}{18}S_{ABg_{2_{L}AB}}-\frac{7}{9}S_{AB}\right)$} \\ \hline
32 & $G_{13}G_{12_{L}}g_{3}$ & \makecell[l]{$I_4\le S_{ABC}+ \sum_{C_3^2} \left(\frac{1}{6}S_{ABG_{13}AB}+\frac{1}{3}S_{ABG_{12_{L}AB}}+\frac{1}{6}S_{ABg_{3AB}}-\frac{2}{3}S_{AB}\right)$} \\ \hline
33 & $G_{13}G_{12_{L}}g_{2_{S}}$ & \makecell[l]{$I_4\le S_{ABC}+ \sum_{C_3^2} \left(\frac{1}{3}S_{ABG_{13}AB}+\frac{1}{6}S_{ABG_{12_{L}AB}}+\frac{1}{2}S_{ABg_{2_{S}AB}}-S_{AB}\right)$} \\ \hline
34 & $G_{13}G_{12_{L}}g_{2_{L}}$ & \makecell[l]{$I_4\le S_{ABC}+ \sum_{C_3^2} \left(\frac{1}{3}S_{ABG_{13}AB}+\frac{1}{6}S_{ABG_{12_{L}AB}}+\frac{1}{6}S_{ABg_{2_{L}AB}}-\frac{2}{3}S_{AB}\right)$} \\ \hline
\end{tabularx}
\end{table}

\renewcommand\theadfont{\bfseries}
\begin{table}[h!]
\centering
\scriptsize
\rowcolors{2}{pink!20}{cyan!5}  
\begin{tabularx}{\textwidth}{|c|c|X|}
\hline
\rowcolor{white}
\thead{} & \thead{gap groups} & \thead{$I_4$ upper bounds} \\
\hline
35 & $G_{13}g_{2_{S}}g_{1}$ & \makecell[l]{$I_4\le S_{ABC}+ \sum_{C_3^2} \left(\frac{1}{3}S_{ABG_{13}AB}+S_{ABg_{2_{S}AB}}+\frac{1}{6}S_{ABg_{1AB}}-\frac{3}{2}S_{AB}\right)$} \\ \hline
36 & $G_{13}g_{2_{L}}g_{1}$ & \makecell[l]{$I_4\le S_{ABC}+ \sum_{C_3^2} \left(\frac{1}{3}S_{ABG_{13}AB}+\frac{1}{3}S_{ABg_{2_{L}AB}}+\frac{1}{6}S_{ABg_{1AB}}-\frac{5}{6}S_{AB}\right)$} \\ \hline
37 & $G_{12_{S}}G_{12_{L}}g_{3}$ & \makecell[l]{$I_4\le S_{ABC}+ \sum_{C_3^2} \left(\frac{1}{4}S_{ABG_{12_{S}AB}}+\frac{1}{4}S_{ABG_{12_{L}AB}}+\frac{1}{3}S_{ABg_{3AB}}-\frac{5}{6}S_{AB}\right)$} \\ \hline
38 & $G_{12_{S}}g_{3}g_{2_{S}}$ & \makecell[l]{$I_4\le S_{ABC}+ \sum_{C_3^2} \left(\frac{1}{2}S_{ABG_{12_{S}AB}}+\frac{1}{3}S_{ABg_{3}AB}+\frac{1}{2}S_{ABg_{2_{S}AB}}-\frac{4}{3}S_{AB}\right)$} \\ \hline
39 & $G_{12_{S}}g_{3}g_{2_{L}}$ & \makecell[l]{$I_4\le S_{ABC}+ \sum_{C_3^2} \left(\frac{1}{2}S_{ABG_{12_{S}AB}}+\frac{1}{3}S_{ABg_{3}AB}+\frac{1}{6}S_{ABg_{2_{L}AB}}-S_{AB}\right)$} \\ \hline
40 & $G_{12_{L}}g_{3}g_{1}$ & \makecell[l]{$I_4\le S_{ABC}+ \sum_{C_3^2} \left(\frac{1}{3}S_{ABG_{12_{L}AB}}+\frac{1}{3}S_{ABg_{3}AB}+\frac{1}{6}S_{ABg_{1AB}}-\frac{5}{6}S_{AB}\right)$} \\ \hline
41 & $g_{3}g_{2_{S}}g_{1}$ & \makecell[l]{$I_4\le S_{ABC}+ \sum_{C_3^2} \left(\frac{1}{3}S_{ABg_{3}AB}+S_{ABg_{2_{S}AB}}+\frac{1}{2}S_{ABg_{1AB}}-\frac{11}{6}S_{AB}\right)$} \\ \hline
42 & $g_{3}g_{2_{L}}g_{1}$ & \makecell[l]{$I_4\le S_{ABC}+ \sum_{C_3^2} \left(\frac{1}{3}S_{ABg_{3}AB}+\frac{1}{3}S_{ABg_{2_{L}AB}}+\frac{1}{2}S_{ABg_{1AB}}-\frac{7}{6}S_{AB}\right)$} \\ \hline
\end{tabularx}
\caption{The respective upper bounds of $I_4$ corresponding to the 42 gap combinations in (\ref{gapcombsappendix}).} \label{42comb}
\end{table}
\bibliographystyle{elsarticle-num}
\bibliography{reference}

\end{document}